\documentclass[traditabstract]{aa}                                    \usepackage{amsmath}

\usepackage[varg]{txfonts}
\usepackage{graphicx}
\usepackage{natbib}
\usepackage{url}
\usepackage{flafter}
\usepackage{longtable}

\bibpunct{(}{)}{;}{a}{}{,}
\begin{document}
   \title{The discovery and classification of 16 supernovae at high redshifts in ELAIS-S1}

   \subtitle{-- the Stockholm VIMOS Supernova Survey \rm{\bf{II}} 
   \thanks{Based on observations collected at the European Organisation
   for Astronomical Research in the Southern Hemisphere, Chile, under ESO 
   programme ID 167.D-0492.}}
   \author{J. Melinder \inst{1}
          \and
          T. Dahlen \inst{2}
		  \and
		  L. Menc\'ia-Trinchant \inst{1}
		  \and
		  G. \"Ostlin \inst{1}
          \and
		  S. Mattila \inst{1,3}
          \and
          J. Sollerman \inst{1}
		  \and
		  C. Fransson \inst{1}
		  \and
          M. Hayes \inst{4}
          \and 
          S. Nasoudi-Shoar \inst{5}
          }
    \institute{Department of Astronomy, Oskar Klein Centre, Stockholm University, AlbaNova 
               University Centre, SE-106 91 Stockholm, Sweden\\ 
               \email{jens@astro.su.se}
		 \and
		 	 Space Telescope Science Institute, 3700 San Martin Drive, 
             Baltimore, MD 21218, USA
         \and
             Tuorla Observatory, Department of Physics and Astronomy, University of Turku, 
             V\"ais\"al\"antie 20, FI-21500 Piikki\"o, Finland
         \and
			CNRS; Universit\'e de Toulouse, UPS-OMP, IRAP, Toulouse, France
         \and
             Argelander-Institut f\"ur Astronomie, Universit\"at Bonn, 
             Auf dem H\"ugel 71, 53121 Bonn, Germany}
   \date{Received ; accepted }
  \abstract{Supernova surveys can be used to study a variety of
   subjects such as: (i) cosmology through type Ia supernovae (SNe), (ii) 
   star-formation rates through core-collapse SNe, and (iii) supernova
   properties and their connection to host galaxy characteristics.
   The Stockholm VIMOS Supernova Survey (SVISS) is a multi-band imaging
   survey aiming to detect supernovae at redshift $\sim$0.5 and derive
   thermonuclear and core-collapse supernova rates at high redshift. In
   this paper we present the supernovae discovered in the survey
   along with light curves and a photometric classification into
   thermonuclear and core-collapse types.  To detect the supernovae
   in the VLT/VIMOS multi-epoch images, we used difference imaging
   and a combination of automatic and manual source detection to
   minimise the number of spurious detections. Photometry for the
   found variable sources was obtained and careful simulations were made
   to estimate correct errors. The light curves were typed using a
   Bayesian probability method and Monte Carlo simulations were used
   to study misclassification.  We detected 16 supernovae, nine of
   which had a core-collapse origin and seven had a thermonuclear
   origin. The estimated misclassification errors are quite small, 
   in the order of 5\%, but vary with both redshift and type. The
   mean redshift of the supernovae is 0.58. Additionally, we found a
   variable source with a very extended light curve that could
   possibly be a pair instability supernova.}

   \titlerunning{The discovery and classification of 16 supernovae in 
ELAIS-S1}
   \maketitle
\section{Introduction}
During the last couple of decades supernovae have been shown to be
powerful probes of both cosmology and star formation in high-redshift
galaxies.  A number of surveys with various scientific goals have been
conducted and proposed. Most of them have in common that multiple
imaging epochs are used to detect the supernovae. In some cases
the surveys also contain spectroscopic follow-up observations of the
detected candidates. The spectroscopic information makes it possible to
easily characterise the detected supernovae and to measure redshifts,
but this comes at the cost of added telescope time. Whole-sky
supernova surveys have already started, e.g., Pan-STARRS1
\citep{2008A&A...489..359Y} and PTF \citep{2009PASP..121.1334R},
and several more are being planned, e.g., SkyMapper and LSST. With
the large number of expected supernovae it will not be feasible to
obtain spectra for all of them. Photometric techniques will thus be
required to characterise the detected sources.

Core-collapse supernovae (CC SNe) are the result of a massive
star ending its life in an energetic explosion \citep[see, e.g.,]
[]{2009ARA&A..47...63S}. Because the lifetimes of these massive
stars are short, their presence signals that active star-formation
is taking place in the host galaxy. The star-formation rate density 
for a cosmic volume, defined by the redshift range and field size,
can be derived from the core-collapse supernova rates by making an 
assumption on how many stars out of the current star forming 
population that explode. This method provides an independent tracer of
the star-formation history of the universe and has been used by
\citet{2004ApJ...613..189D,2008A&A...479...49B,2009A&A...499..653B}.
While \citet{2011MNRAS.tmp..317L} also provide a local supernova
rate, their big sample of SNe is also used to study the host
galaxy properties and connections to supernova subtype.  Most of
these surveys target low-redshift SNe ($z\lesssim 0.3$), but with
deep enough observations the technique also works at higher $z$
\citep[e.g.][]{2004ApJ...613..189D}. For this kind of projects it is
important that the observations are obtained as a ''blind'' survey
to limit the selection effects.

The use of thermonuclear (Ia) supernovae as
standardisable candles has been instrumental in measuring
cosmological parameters for the $\Lambda$CDM concordance cosmology
\citep[e.g.][]{2006A&A...447...31A,2007ApJ...659...98R,
2010ApJ...716..712A,2007ApJ...666..694W,2009ApJS..185...32K}.
The systematic errors resulting from the determination of the
distance modulus for thermonuclear supernovae have been extensively
studied and minimised. To achieve the precision needed to pinpoint
cosmological parameters, it is necessary to obtain spectra for the
supernovae. Despite the amount of work made on calibrating the
distance modulus relation, the underlying physics of thermonuclear
supernovae is still partly unknown. In particular, the type of
progenitor system that gives rise to the supernovae is uncertain;
see, e.g., \citet{2009ApJ...699.2026R} for a discussion of the
different possibilities. One way of putting constraints on progenitor
models is to study the delay times of thermonuclear supernovae
(i.e., the time between formation of the progenitor star and the
supernova explosion). The rate of thermonuclear supernovae compared
to either the global \citep{2004ApJ...613..189D} or the local
\citep{2006ApJ...648..868S,2008PASJ...60.1327T,2010MNRAS.407.1314M}
star formation history can provide estimates on the delay time
distribution even without observed spectra of the supernovae. A
number of surveys targeting Ia SNe in galaxy clusters have also been
undertaken \citep{2010ApJ...718..876S,2010arXiv1010.5786B}. Recently,
\citet{2010ApJ...713.1026D} presented an accurate measurement of the
$z\sim 0.1$ thermonuclear SN rate from the Sloan Digital Sky Survey
II Supernova Survey.

A major challenge to all intermediate- and high-redshift CC SN
searches is how to take into account the effects of SNe missed owing
to large host galaxy extinctions. \citet{2007ApJ...659L...9M} and
\citet{2008ApJ...689L..97K} present supernovae found in the nuclear
regions of luminous infrared galaxies (LIRGs) by using infrared
adaptive optics imaging. These types of searches are important to
constrain the numbers of supernovae lost in LIRGs, in which most of
the massive stars are formed at intermediate
and higher redshifts \citep{2009A&A...496...57M}.

Typing of supernovae using broad band colours in multiple
epochs has been demonstrated to work by several authors
\citep[e.g.,][]{2004ApJ...613L..21B,2006AJ....132..756J,
2007ApJ...659..530K,2007AJ....134.1285P,2008AJ....135..348S}. The
application of the codes vary, in some cases the codes are used
mainly to reject core-collapse SNe from the follow-up target
lists of cosmological Ia surveys, but in others the codes are
used as the only way of typing SNe without spectra. Most of the
codes use template-fitting methods where a number of k-corrected
thermonuclear and core-collapse SN light curves are compared to the
observed light curve and colour evolution. The simplest way of doing
this fitting is through $\chi^2$ fitting. However, there
are some problems with this approach because it does not allow prior
information on, e.g., redshifts and peak luminosity to be used to full
extent. \citet{2007ApJ...659..530K} introduced a Bayesian approach to
supernova typing where probability distributions for the parameters
could be used as priors. Lately, Bayesian methods have been further
developed by \citet{2009ApJ...707.1064R}, where ''fuzzy'' templates
are used to improve classification of SNe with non-standard light
curves. With this method the templates are assigned an uncertainty
that enters into the likelihood calculations, which improves
the classification quality of SNe with non-standard light curves.
Recently, \citet{2010arXiv1008.1024K} presented the results of a
supernova classification challenge, where a common sample of supernova
light curves was typed by different codes. Their results indicate
that Bayesian typing codes are competitive and give reliable type
determinations.

The Stockholm VIMOS Supernova Survey (SVISS) is a multi-band ($R+I$)
imaging survey aiming to detect supernovae at redshift $\sim$0.5
and derive thermonuclear and core-collapse supernova rates. The
supernova survey data were obtained over a six-month period with
VIMOS/VLT \citep{2003SPIE.4841.1670L}. \citet{2008A&A...490..419M}
describe the supernova search method along with extensive testing
of the image subtraction, supernova detection and photometry. In
this paper we report the discovery of 16 supernovae in one of the
search fields of the survey and provide light curves along with type
classifications for them. We will present the supernova rates and
conclusions to be drawn from them in a future paper (Melinder et al,
in prep.).

The first part of the paper  contains a
description of the data set and the methods used to reduce it. 
In Section~\ref{sec:typing} we describe the supernova typing method
and the template supernova light curves. Section~\ref{sec:tests}
contains a description of how the simulations of supernovae have
been set up and offers the results from extensive testing of the typing
method. Finally, we present the results and possible
implications in Section \ref{sec:cands} and conclude by discussing
and summarising the results in Section~\ref{sec:sum}. The
Vega magnitude system and a standard $\Lambda$CDM cosmology with
$\{h_0,\Omega_{M},\Omega_{\Lambda}\}=\{70,0.3,0.7\}$ have been used
throughout the paper.

\section{The data}
\label{sec:data}
\subsection{Observations}
The data were obtained with the VIMOS instrument
\citep{2003SPIE.4841.1670L} mounted on the ESO Very Large Telescope
(UT3) at several epochs during 2003--2006. The VIMOS instrument
has four CCDs, each 2k$\times$2.4k pixels with a pixel scale of
0.205\arcsec/pxl, covering a total area of $4\times$56 sq. arcmin. The
SVISS observations were obtained in two fields, covering parts of the
Chandra Deep Field-South \citep{2001ApJ...551..624G} and the ELAIS-S1
field \citep{2004AJ....127.3075L}. The observations in the ELAIS-S1
field were obtained in five broad band filters ($U$, $B$, $V$, $R$
and $I$) centred at $\alpha$ = 00:32:13, $\delta$ =
-44:36:27 (J2000). The data used in this paper are only from the
ELAIS-S1 observations, the CDF-S observations will be presented in
a subsequent paper.

The supernova search filters are $R$ and $I$, with roughly twice the
exposure time in $I$ compared to $R$. Observations in these filters
have been divided into one reference epoch (henceforth epoch 0), seven
search epochs and one control epoch. The images used to construct the
reference epoch were obtained in August 2003, the search epoch images
were obtained from July 2004 to January 2005, while the control
epoch was taken in January 2006. The search epochs were separated by
roughly one month.
\begin{table} 
\caption{Overview of the ELAIS-S1 observations}	     \label{table:data} \centering \begin{tabular}{l l c c c} \hline\hline Filter & Ep/Date & Exp. time (s) & 
$m_{lim}$ & Seeing (\arcsec)\\ \hline    R & 0/2003-08 & 8\,640  & -- & 0.65 \\ 
   R & 1/2004-07 & 5\,760  & 26.83\tablefootmark{a} & 0.68 \\ 
   R & 2/2004-08 & 8\,280  & 26.97\tablefootmark{a} & 0.67 \\ 
   R & 3/2004-09 & 5\,520  & 26.71\tablefootmark{a} & 0.69 \\ 
   R & 4/2004-10 & 8\,760  & 26.94\tablefootmark{a} & 0.89 \\ 
   R & 5/2004-11 & 7\,560  & 26.80\tablefootmark{a} & 0.80 \\ 
   R & 6/2004-12 & 7\,320  & 26.75\tablefootmark{a} & 0.80 \\ 
   R & 7/2005-01 & 5\,760  & 26.88\tablefootmark{a} & 0.76 \\ 
   R & C\tablefootmark{b}/2006-jan. & 8\,760  & -- & 0.72 \\
   I & 0/2003-08 & 11\,520 & -- & 0.59 \\ 
   I & 1/2004-07 & 11\,520 & 26.60\tablefootmark{a} & 0.59 \\ 
   I & 2/2004-08 & 18\,240 & 26.41\tablefootmark{a} & 0.64 \\ 
   I & 3/2004-09 & 9\,960  & 26.39\tablefootmark{a} & 0.71 \\ 
   I & 4/2004-10 & 11\,520 & 26.52\tablefootmark{a} & 0.68 \\ 
   I & 5/2004-11 & 15\,960 & 25.85\tablefootmark{a} & 0.76 \\ 
   I & 6/2004-12 & 19\,080 & 26.04\tablefootmark{a} & 0.64 \\ 
   I & 7/2005-01 & 12\,480 & 26.30\tablefootmark{a} & 0.70 \\ 
   I & C\tablefootmark{b}/2006-jan. & 11\,520 & -- & 0.69 \\
\hline					    U & 2003-2006& 37\,740 & 27.44\tablefootmark{c} & 0.73 \\ 
   B & 2003-2006& 34\,940 & 28.53\tablefootmark{c} & 0.71 \\ 
   V & 2003-2006& 20\,060 & 26.93\tablefootmark{c} & 0.71 \\ 
\hline					 \end{tabular} 
\\ 
\tablefoot{The upper part of the table shows the observational 
properties of the $RI$ supernova search epochs, the lower part shows
the properties of the $UBV$ observations.
The entries in the table are mean values over the four VIMOS quadrants.\\
\tablefoottext{a}{3$\sigma$
magnitude limits estimated from the photometric scatter of simulated
sources placed in faint galaxies ($m_I>24.0$) in the individual
subtracted epoch images (see Section~\ref{sec:mcsim}).}
\tablefoottext{b}{Control epochs, taken approximately one year after 
the final SN survey epoch.} 
\tablefoottext{c}{3$\sigma$
magnitude limits using a $4\times$FWHM diameter aperture.}
}
\end{table}

The $UBV$ observations were obtained at several different epochs
during the time period 2004-2006. Details on the data reduction and
calibration of the $UBV$ data along with a galaxy catalogue for the
field will be presented in Menc\'ia-Trinchant et al. (in prep). In the
present paper the $UBV$ observations have only been used to calculate
photometric redshifts for supernova host galaxies.

Table~\ref{table:data} contains the specifications of the different
epochs and parts of the data. It is worth noting that the quality
of data is very good overall with median seeing of 0.76\arcsec and
0.68\arcsec in $R/I$, respectively, and all search epochs have seeing
below 0.9\arcsec. As can be seen in the table, the observations are
also very deep, with mean $3\sigma$ limiting magnitudes of 26.8 and
26.3 in $R$ and $I$, respectively. In \citet{2008ApJ...681..462D}
the 50\% detection efficiency magnitude of the survey in the
F850LP filter is given, converting this to our magnitude system we obtain
$m_I\sim26.5$. This magnitude can be compared to our $3\sigma$
limits, based on the findings in Paper I; the depth is thus quite
similar. The search for variable objects in the Subaru/XMM-Newton
Deep Survey \citep{2008ApJ...676..163M,2008PASJ...60.1327T} has
a limiting magnitude of $m_I\sim 26.0$, again converted to our
magnitude system. Compared to other SN surveys,
our data set is thus among the deepest ever obtained, 
although smaller in field size than others.

\subsection{Data reduction of the $R$ and $I$ observations}
The data were reduced using a data reduction pipeline written in MIDAS
scripting language for SVISS developed by our team. Each image was
bias-subtracted using an epoch master bias frame constructed by 
median-combining $>10$ bias frames. Flatfield calibration data were obtained
for each night in both filters and all science data were flatfielded
using high signal-to-noise stacked frames. Cosmic-ray rejection
routines were used to detect possible cosmic rays, but no automatic
corrections were applied. Parts of images with suspected cosmic rays
were manually inspected and pixels with cosmic ray contamination
were flagged.  Atmospheric extinction corrections were applied to the images
using tabulated extinction coefficients from the ESO quality control
web pages. The images were also normalised to counts per second. A bad
pixel mask was produced for each of the images, containing vignetted
($<5$\% in most cases), saturated and cosmic ray flagged pixels.

The VIMOS $I$ band suffers from quite severe effects of fringing
\citep[see][for a detailed description of fringing for the VIMOS
instrument]{2008A&A...488..533B}, and care has to be taken to
successfully remove these. We constructed a fringe map for each
observation night by median-combining the non-aligned science frames
and rejecting bright pixels by a standard sigma-rejection routine. To
make this possible, the observations were made with a dithering scheme
that was optimised to avoid having individual science frames target the
same area on sky. For some nights it was necessary to make two fringe
maps owing to sky-brightness variations during the night. We did not need
to use any masking or smoothing on the resulting fringe maps, there
were very few extremely bright sources in our field \citep[causing
optical ghosts, as mentioned by][]{2008A&A...488..533B} and the
noise levels were insignificant compared to the individual frames. The
fringe map was then subtracted from each of the flatfielded science
frames. Finally, all of the science frames were manually inspected,
some frames were removed because of vignetting and/or poor seeing (see
Table~\ref{table:data} for the total exposure time used in each epoch).

The good quality frames for each epoch were registered to a common
pixel coordinate system using a shift- and rotate-transform
(yielding a typical rms of $\sim0.1$ pixels or 0.02\arcsec). The
registered frames were then median combined. The bad pixel maps for each
individual image were also registered and summed, yielding an exposure
time map for each epoch. Some pixels ($< 0.2 \%$ in the reference
epoch) had an exposure time of zero, i.e., none of the individual frames
contained useful data for that pixel. These pixels were flagged in
the combined science frames.

Each of the combined epoch images were photometrically calibrated
using $\sim 50$ secondary photometric standards in the field. The calibrated
photometry for these stars was obtained from stacked one-night images
from nights where standard-star observations were available. The
secondary standard sources were selected by requiring a signal-to-noise
ratio of at least 100 and a stellar-like point spread function (PSF). 
The sources were also
required to be isolated in the single-night frame with no visible
neighbours within a $10\times$FWHM (full width at half maximum) radius. 
The photometry of the
primary and secondary standards was obtained with \texttt{IRAF/phot}
using an aperture size of $10\times$FWHM (with aperture correction).
The resulting zeropoint errors for each epoch are $\lesssim 0.02$
magnitudes, but it should be noted that all epoch images are scaled
to the reference epoch zeropoint in the subtraction procedure.

\subsection{Supernova detections and photometry}
In \citet{2008A&A...490..419M}, henceforth Paper \rm{I}, we presented
our supernova detection method, which is explained there in detail. 
We used IRAF/PYRAF scripts
that were run in sequence and proceeded as follows: (i)
accurate image alignment over the entire frame; (ii) convolving the
better seeing image to the same PSF size and shape as the poorer
seeing image, using a spatially varying kernel; (iii) subtracting
the images; (iv) detection of sources in the subtracted frames,
using both source detection software and naked-eye detection; (v)
photometry and construction of light curves of the detected transients.

All search epochs were aligned to the pixel coordinate system of
the reference epoch using the \texttt{geomap/geotran} tasks in
IRAF/PyRAF. We found that the shift- and rotate-transform used when
registering the individual frames was not good enough for registering
the search images to the reference image. This is likely caused by
changes in the geometry (e.g., differential refraction, changes in
flexure of the telescope) of the frames over the time period our
observations were obtained. To perform the registration at sub--pixel
accuracy, we used a general transform that allows for shifts, rotations,
shear, and pixel scale changes in the image being aligned. The reference
sources for registration were bright, point-like objects in the field
(approximately 100 sources were used).  The resulting standard deviations
for the geometrical transforms were smaller than 0.1 pixels for all
epochs.

Convolution to a common PSF was made with the ISIS 2.2 code
\citep{1998ApJ...503..325A,2000A&AS..144..363A}. A convolution kernel
was computed by comparing a number of reference sources in the two
images. The better seeing image (i.e., the image with smaller PSF
width) was then convolved and scaled to match the photometry of the
unconvolved image; the convolved frame was either the reference
or the search frame. A background variation between the frames
was also computed and compensated for. The selection of suitable
image subtraction parameters for our dataset was investigated in
\citet{2008A&A...490..419M}. Finally, the reference epoch was subtracted
from the search epoch and exposure time maps for the images were 
combined to make a weight map for the subtracted frame.

\subsubsection{Source detection}
Source detection was made in the subtracted frames using SExtractor
\citep[SE, ][]{1996A&AS..117..393B} to obtain an initial source
list. Separate detection was made in each epoch and filter, except in
the very last epoch (sources discovered in the last epoch would only
have one point on the light curve and thus automatically fail our
selection criterion, see Section~\ref{sec:phot}). The detection
parameters for SE were set at a quite liberal level to make sure
that no SN candidates were missed (i.e., this detection threshold accepts
sources at a lower signal-to-noise level than the rejection criteria,
see Section~\ref{sec:phot}). Furthermore, the weight maps for the
subtracted frames were used in source detection to lower the number
of spurious detections. Using the weight maps reduced this number by 
about 20\%, primarily close to the edges of the frames.

The number of detections in a single subtracted frame is
typically in the order of 100 times the expected number
of true varying sources. This is consistent with the
findings of other supernova surveys using similar techniques
\citep[e.g.][]{2007ApJ...666..674M}. Most of these spurious detections
are spurious subtraction residuals. About 50\% of the spurious sources
were then rejected, most of these by requiring that true sources must
be present in both the $R$ and $I$ filter at the given epoch. Spurious
sources close to saturated stars and image defects were also removed;
the total number of candidates for all the epochs remaining after this
initial rejection procedure was $\sim$1500. Most of them are spurious
detections related to subtraction residuals of bright galaxies that
are present in both filters.  In Section~\ref{sec:cands} we show how
constraints on the light curve and supernova typing can be used to
safely reject the remaining spurious detections.

\subsubsection{Photometry}
\label{sec:phot}
Photometry on the detected sources was made using the IRAF
\texttt{daophot} package. The PSF photometry was performed on all
detected candidates using the task \texttt{allstar}. We also tried
using aperture photometry with aperture corrections computed from the
original worse seeing image, using the IRAF task \texttt{phot}. This
does give fairly good results, but seems to be more susceptible to
residual flux from the background galaxies than the PSF photometry,
thus giving larger errors (both statistical and systematic) in general.

In Paper \rm{I} we found that the photometric uncertainties estimated
by \texttt{daophot/phot} were underestimating the true noise by a
factor of two or more. The two main reasons for this are that (i)
the pixels in the subtracted image will be positively correlated
owing to the convolution made in the image matching step; (ii) the sky
noise is estimated in a region outside of the host galaxy, thus not
taking subtraction residuals properly into account. Therefore, we used
simulations to obtain reliable error estimates. By simulating SNe at
different brightness in each epoch and finding the scatter in their
measured magnitudes, we obtained an estimate of the true photometric
error for each point on the SN light curve.

The simulations also allowed us to check the photometry for possible
systematic errors. In Paper \rm{I} we discuss the discovery of a
small systematic flux offset ($\sim$10\% at the $3\sigma$ limiting
magnitudes, lower for brighter sources) in subtracted frames. This
offset is not present in all epochs/filters and changes between
epochs. We can calculate this offset from the simulations and correct
the photometry for this effect in the epochs where it is needed.

$R$ and $I$ light curves are then put together for the 1459 sources
remaining after automated detection in the subtracted frames. We 
rejected spurious detections by requiring the candidates to be brighter
than the $3\sigma$ limiting magnitudes (see Table~\ref{table:data}) in
(i) the detection epoch and the subsequent epoch and (ii) both the $R$
and $I$ filters. With this rejection criterion the number of supernova 
candidates decreased to 115. At this point we are reasonably certain 
that no true supernova candidates have been dropped (unless they were 
too faint to fulfil the rejection criterion), but there is likely
still a number of spurious detections remaining. In some cases subtraction 
residuals will appear at the same location in both filters and in multiple 
epochs (e.g., very bright galaxies).

\subsection{Host galaxy redshifts}
\label{sec:photz}
Using the full set of $UBVRI$ observations, we calculated
photometric redshifts for the supernova host galaxies.  We used a
template-fitting method with the mean galaxy luminosity
function as a Bayesian prior, see \citet{2010ApJ...724..425D}
and \citet{2004MNRAS.350..253D}. We used 16 SEDs (spectral energy distributions) that were
constructed by interpolating between four empirical templates (E,
Sbc, Scd and Im galaxy types) from \citet{1980ApJS...43..393C} and
two starburst-galaxy templates from \citet{1996ApJ...467...38K}. For
the supernova host galaxies we used photometry from the stacked $UBV$
images, but excluding frames obtained during the time period when
the supernova is detectable. For the $I$ and $R$ images we used the
reference epoch images. These considerations ensure that the
supernova light does not affect the determination of the photometric
redshifts.

In our observed subsection of the ELAIS-S1 field only two galaxies have
known spectroscopic redshifts. This means that calibration and
validating of the photometric redshifts is not straightforward. We 
obtained archive $UBVRI$ data of the Hubble Deep Field-South (HDF-S)
observed with the VIMOS instrument. The HDF-S has been observed
extensively with spectroscopy and we found 280 sources with
spectroscopic redshifts ranging from z$\sim$0 to z$\sim$3.5 that are
also found in the VIMOS HDF-S observations
\citep{2002A&A...396..847V,2005A&A...440...61R,2006AJ....131.2383G}.
We reduced and analysed the HDF-S photometric data in the
same way as the ELAIS-S1 data set. The photometric calibration was also
made in the same way (including corrections for galaxtic
extinction) as for the ELAIS-S1 observations.

Comparing the spectroscopic redshifts to the photometric redshifts
obtained from the imaging data, we find a redshift uncertainty of
$\delta z= 0.085*(1+z)$ and a frequency of catastrophic failures
(defined as $|z_{phot}-z_{spec}|>0.3$) of 9\%. Because the instrument
and filters are the same, the analysis is made in exactly the same
way and the depth of the imaging data is comparable, we assume
that the uncertainties of the SVISS redshifts are the same as the
uncertainties determined for this dataset. The uncertainty of the
redshift determination was taken into account in the supernova
typing, see Section~\ref{sec:mcsim}. The photometric redshift technique
and calibrations of it as applied to our data set will be presented
in greater detail in Menc\'ia-Trinchant et al. (in prep).

At first, the host galaxies were identified as the closest galaxies to
the SNe in terms of angular separation. But when the redshift
information was also considered, it became clear that some of the
identified hosts were at unrealistical redshifts, either making any SNe
in them too faint to be detectable or too bright compared to our
templates. Instead of angular separation we therefore used the photometric
redshift information and chose the closest galaxy in terms of physical
distance.  With this method we were able to identify hosts for 13 of
the SNe. For one of these (SN309) the chosen host was not the
closest galaxy in angular distance.  Identifying hosts for the SNe in
this way is of course not unproblematic, there is a risk of selecting
an incorrect host galaxy. We did not try to estimate the percentage of
misidentifications, but we reran the typing for all 13 SNe
using a flat redshift prior and, in relevant cases, also with redshift
priors based on other closeby galaxies. The result of these tests is that
none of these SNe change main type and that the changes in redshift are
small. We conclude that host misidentification has only little effect on the
main results of this paper.

For the remaining three supernovae (SN-14, SN-31 and SN-261) there are
no nearby galaxies with redshifts that are consistent with hosting a SNe
with the observed light curve. For these three SNe we ran the typing
code without photometric redshifts supplied, thus using a flat prior on
the redshift. This is also discussed in more detail in
Section~\ref{sec:ccresults}.  More information on the host galaxy
identification and properties of these galaxies will be presented in an
upcoming paper (Menc\'ia Trinchant et al, in prep.).

\section{Supernova typing method}
\label{sec:typing}
Our typing method relies on a Bayesian template-fitting algorithm. The
use of prior probabilities and Bayesian marginalisation makes it
possible to include previously known information on the different
supernova types in the fitting technique while avoiding over-fitting.
The goal of the method is to find the most likely supernova
type ($\mathcal{T}$) given an observed light curve ($\{F\}$) of a
supernova candidate. The description of the Bayesian method follows
\citet{2007ApJ...659..530K}, although the notation and scope is
slightly different. All calculations were made in a parallelised 
FORTRAN 90 code using double precision.

Formally we want to find the type that maximises the probability
$P(\mathcal{T}_j|\{F_i\})$, where $j=[1,...,N_{\mathcal{T}}]$
refers to the different types and $i=[1,...,N_{dp}]$ to the points
on the observed light curve. Nine supernova types (Ia, Ia$_{faint}$,
Ia$_{91bg-like}$ Ia$_{91t-like}$, Ibc$_{normal}$,
Ibc$_{bright}$, IIn, IIL and IIP) were considered. The total
number of data points ($N_{dp}$) is given by the product of the
number of filters ($N_{filters}$) and the number of observation
epochs ($N_{ep}$). The SVISS observations were made in two filters
($R$ and $I$) and in seven epochs, thus $N_{dp}=14$ in this work.
The probability $P(\mathcal{T}_j|\{F_i\})$ cannot be calculated
directly; using Bayes' theorem we can rewrite the probability:
\begin{equation}
\label{eq:bayes}
P(\mathcal{T}_j|\{F_i\})=\frac{P(\{F_i\}|\mathcal{T}_j) P(\mathcal{T}_j)}{\sum_{j}
P(\{F_i\}|\mathcal{T}_j) P(\mathcal{T}_j)}.
\end{equation}
$P(\{F_i\}|\mathcal{T}_j)$ is the probability of obtaining the light
curve data $\{F_i\}$ for a given supernova type $\mathcal{T}_j$.
$P(\mathcal{T}_j)$ contains all prior information on the supernova
model light curve for the specific type. When Bayes' theorem is used in
this fashion, we implicitly assume that the set of supernova types is
complete, i.e., all supernova candidates must be one of the considered
types. For sufficiently peculiar supernovae, and non-supernovae,
Equation~\ref{eq:bayes} will not give valid probabilities.
In Section~\ref{sec:errors} we describe how we use measures of
goodness-of-fit to purge candidates with light curves that are
too dissimilar from the model curves.

For each of the nine supernova types we created template light
curves using absolute magnitude ($M_B$) light curves and SEDs,
which are described in more detail in Sections~\ref{sec:subt}
and \ref{sec:kcor}. We used four parameters that uniquely define
the light curve for a given type: (i) $M_B$ the absolute
rest-frame B band magnitude at peak, (ii) $z$, the redshift
of the supernova, (iii) $t$, the time difference between
the explosion date for the model and the first observational
epoch,(iv) $\{R_V,E(B-V)\}=\vec{\eta}$; extinction in the host
galaxy.  We denote the template light curves by $\{f_{i,j}\}$
\begin{equation} 
\{f_{i,j}\}= \{f_{i,j}(M_B,z,t,\vec{\eta})\},
\end{equation} 
and because these parameters uniquely define the light
curve for a specific type, we may rewrite $P(\{F_i\}|\mathcal{T}_j)$
as $P(\{F_i\}|\{f_{i,j}(M_B,z,t,\vec{\eta})\})$ and $P(\mathcal{T}_j)$
as $P(M_B,z,t,\vec{\eta},\mathcal{T}_j)$.

We define the likelihood function for each supernova type:
\begin{equation}
\label{eq:likf}
\begin{split}
l_j(\{F_i\}|M_B,z,t,\vec{\eta}) & \equiv P(\{F_i\}|\mathcal{T}_j) 
P(\mathcal{T}_j) \\ 
&= P(\{F_i\}|\{f_{i,j}(M_B,z,t,\vec{\eta})\}) 
P(M_B,z,t,\vec{\eta},\mathcal{T}_j).
\end{split}
\end{equation}
The probability that a supernova with a template light curve
${f_{i,j}}$ will have an observed light curve, $\{F_i\}$, is
\begin{equation}
\label{eq:obsprob}
P(\{F_i\}|\{f_{i,j}(M_B,z,t,\vec{\eta})\})= \prod_i^{N_{dp}} 
\frac{e^{-(F_i-f_{i,j})^2/(2 \delta F_i^2)}}{\sqrt{2 \pi} \delta F_i },
\end{equation} 
where each observational data point is allowed to fluctuate according
to Gaussian statistics (the widths of the distributions are given
by the observational errors, $\delta F_i$). It should be noted that
the photometric quantities in this expression are in units of flux
(counts/sec). Non-detections are included in the analysis as data
points with zero flux and with an error given by the $1\sigma$
limiting fluxes.

The $P(M_B,z,t,\vec{\eta},\mathcal{T}_j)$, or equivalently
$P(M_B,z,t,\vec{\eta}|\mathcal{T}_j)$, prior contains all prior
information on the parameters for a given template. We assume that
the parameters are independent and can thus factorise the prior:
\begin{equation}
\label{eq:priors}
\begin{split}
P(M_B,z,t,\vec{\eta}|\mathcal{T}_j)=P(M_B|\mathcal{T}_j) P(z|\mathcal{T}_j)
P(t|\mathcal{T}_j) \\
\times P(\vec{\eta}|\mathcal{T}_j) P(\mathcal{T}_j).
\end{split}
\end{equation}
The individual parameter priors are described in the next section. The
remaining prior, $P(\mathcal{T}_j)$ contains information on whether
a specific supernova subtype is more likely than others. The
relative rates of supernova subtypes are not well constrained at high
redshifts. Therefore, we assume that the prior probability for a 
supernova candidate to be of a certain subtype is equal for all the types 
(i.e., a flat prior),
\begin{equation}
P(\mathcal{T}_j)=\frac{1}{N_{\mathcal{T}}}.
\end{equation}  

There are more priors/parameters that could have been used to construct the
template prior. Including more prior information will of course
affect the typing and can make it more accurate. For this to work, the
parameter in question needs to have a known probability distribution
(or at least a valid range of values). It should be noted that
adding more priors will cause the computation time of the typing
to increase by a factor equal to the number of steps used for the
probability distribution. Our choice of four priors is thus based
on a compromise between limiting computing time and
choosing parameters with well-known distribution on one hand and
typing accuracy on the other. Other authors have used additional
or a different set of priors. \citet{2007ApJ...659..530K} include
prior information on stretch in their typing code, using a Gaussian
distribution with mean and width from observations. In our code
we use thermonuclear supernova templates with different stretch to
account for this variance. The inclusion of a colour uncertainty to
the templates can also be described in terms of an additional prior,
this has been used by, e.g., \citet{2007AJ....134.1285P}. We do not
have colour uncertainty in our typing, instead we choose to use more
templates that have different colour evolution.

\subsection{The parameter prior distributions}
\begin{table} 
\caption{Properties of the supernova photometric templates}	     \label{table:templ} \centering \begin{tabular}{l c c c c c} \hline\hline Type & $\langle M_B\rangle$ & $\sigma_M$ & 
Stretch & Fraction & References \\ \hline Ia -- 91T    & -19.64 & 0.30 & 1.04 & 0.125 & 1-3\\ 
Ia -- normal & -19.34 & 0.50 & 1.00 & 0.615& 1, 3-4\\ 
Ia -- faint  & -18.96 & 0.50 & 0.80 & 0.196 & 1, 3-6\\ 
Ia -- 91bg   & -17.84 & 0.50 & 0.49 & 0.064& 1, 3-4\\ 
\hline					 Ibc -- bright & -19.34 & 0.46 & N/A & 0.057& 1, 7-9\\ 
Ibc -- normal & -17.03 & 0.49 & N/A & 0.259 & 1, 7-9\\ 
IIL           & -17.23 & 0.38 & N/A & 0.028& 1, 10-12\\ 
IIn           & -18.82 & 0.92 & N/A & 0.248& 1, 12-13\\ 
IIP           & -16.66 & 1.12 & N/A & 0.407& 1, 12, 14-21\\ 
\hline					 \end{tabular} 
\\
\tablefoot{$\langle M_B\rangle$ is the absolute magnitude in the
Johnson-B filter at peak (assuming standard $\Lambda$CDM cosmology
$\{h_0,\Omega_{M},\Omega_{\Lambda}\}=\{70,0.3,0.7\}$), $\sigma_M$
is the dispersion in the peak magnitude. Fraction refers to
the expected fraction in a magnitude limited sample.}\\
\tablebib{(1)~\citet{2007NugentMISC}; (2)~\citet{2004ApJ...612..690S};
 (3)~\citet{2003ApJ...594....1T}; (4)~\citet{2002PASP..114..803N};
 (5)~\citet{1999AJ....118.1766P}; (6)~\citet{2006AJ....131..527J};
 (7)~\citet{2005ApJ...624..880L}; (8)~\citet{2002AJ....124..417H};
 (9)~\citet{2006AJ....131.2233R}; (10)~\citet{1999ApJ...521...30G};
 (11)~\citet{1997A&A...322..431C}; (12)~\citet{2002AJ....123..745R};
 (13)~\citet{2002ApJ...573..144D}; (14)~\citet{2008ApJ...675..644D};
 (15)~\citet{2004ApJ...616L..91B}; (16)~\citet{2003MNRAS.338..939E};
 (17)~\citet{2002AJ....124.2490L}; (18)~\citet{2005MNRAS.359..906H};
 (19)~\citet{2006MNRAS.372.1315S}; (20)~\citet{2006MNRAS.370.1752P};
 (21)~\citet{2007ApJ...666.1093Q}}
\end{table}
Each of the four parameters has an associated probability
distribution function. The parameter priors on the right hand side of
equation~\ref{eq:priors} are given by the respective probabilities. 

The prior on the peak absolute $B$ magnitude is given by
\begin{equation}
\label{eq:mprior}
P(M_B|\mathcal{T}) = \frac{e^{-(M_B-\langle M_B\rangle)^2/(2\sigma_M^2)}}
{\sqrt{2\pi}\sigma_M}\Delta M_B,
\end{equation}
where $\langle M_B\rangle$ is the mean peak magnitude for a given
supernova type (see Table~\ref{table:templ} and the discussion
in Section~\ref{sec:subt}), $\sigma_M$ is the dispersion of the
mean magnitude and $\Delta M$ is the numerical step size. Assuming
the peak magnitudes to be normally distributed around the mean is
consistent with observations, c.f.  \citet{2002AJ....123..745R} and
\citet{2006AJ....131.2233R}. The range in peak absolute magnitudes
considered is given by $\left[ \langle M_B\rangle - 2 \sigma_M,\langle
M_B \rangle + 2 \sigma_M \right]$ for each subtype and the numerical step size
used is $\sim 0.02-0.04$.

The redshift prior used is either a Gaussian or a flat distribution,
depending on whether the host galaxy has a valid photometric redshift
available or not. The mean, $z_p$ and width, $\delta z$ of the Gaussian
distribution is given by the photometric redshift fitting of the host
galaxy (see sec.~\ref{sec:photz} and Menc\'ia-Trinchant et al., in
prep.). The redshift of the supernova is independent of the supernova
type, thus we may write $P(z|\mathcal{T}) = P(z)$ and obtain
\begin{equation}
\label{eq:zprior}
P(z) = \frac{e^{-(z-z_p)^2/(2\delta z^2)}}
{\sqrt{2\pi}\delta z}\Delta z.
\end{equation}
$\Delta z$ is the numerical step size, similar to $\Delta M$ in
eq.~\ref{eq:mprior}. The flat redshift prior is given by
\begin{equation}
\label{eq:zpriorfl}
P(z) = \frac{\Delta z}{z_u-z_l},
\end{equation}
where $\left[z_l,z_u\right]$ is the lower and upper limits of the
redshift interval considered.  For both distributions we used a
numerical step size of 0.02 in redshift and a redshift interval given
by $\left[z_l,z_u\right]= \left[0.0,2.0\right]$.

The probability of a given time difference is assumed to be equal
for all time differences between $t_1$ to $t_2$. The time difference 
probabilities do no depend on the type, $P(t|\mathcal{T}_j)=P(t)$,
and the prior is
\begin{equation}
\label{eq:tprior}
P(t) = \frac{\Delta t}{t_2-t_1} \equiv \frac{1}{N_t},
\end{equation}
where $\Delta t$ is the numerical step size and $N_t$ the
number of steps. We used $\Delta t=1$ day throughout
this work. The time difference interval considered is given by
$\left[t_1,t_2\right]=\left[-136,142\right]$, this interval is
chosen based on the time difference between our search epochs and
that we require supernova candidates to be observed in at least two
subsequent epochs.

The extinction models considered are two Cardelli laws
\citep{1989ApJ...345..245C} with $R_V=2.1$ and 3.1 and the Calzetti
law \citep{2000ApJ...533..682C} with $R_V=4.05$. Some authors
have reported indications of steep extinction laws for supernovae
\citep[e.g.][]{2008ApJ...686L.103G}, this motivates the inclusion
of the $R_V=2.1$ law. The Calzetti law is normally used in 
star-forming galaxies and is therefore a reasonable assumption for core-collapse
supernovae. However, all extinction laws are considered for any given
type. The second parameter for the extinction is the colour excess,
$E(B-V)$, the range of excess considered is $0.0\leq E(B-V)\leq0.6$
with step size of 0.1. The total number of combinations of the
different extinction laws and values of $E(B-V)$ ($\vec{\eta}$) is
$N_{ext}=21$. We assume a flat prior that does not depend on the
type under consideration and arrive at
\begin{equation} 
\label{eq:extprior}
P(\vec{\eta}|\mathcal{T}_j) = \frac{1}{N_{ext}}. 
\end{equation}

We performed tests on simulated and real data that indicate
that the typing is fairly insensitive to changes of the extinction
prior range. Using a higher maximum extinction or a different
step size does not change the type of any of the observed SNe (see
Section~\ref{sec:cands}). 
\subsection{Determining the most likely supernova type}
\label{sec:typingend} 
When trying to find the most
likely supernova type, the four parameters are all nuisance
parameters, we marginalize the likelihood function over the
four parameters to get the summed likelihood function, or in
Bayesian terms, the evidence: 
\begin{equation} 
\label{eq:evid}
\mathcal{L}_j = \sum_{M_B}\,\sum_{z}\,\sum_{t}\,\sum_{\vec{\eta}}
P(\{F_i\}|\{f_{i,j}(M_B,z,t,\vec{\eta})\}) P(M_B,z,t,\vec{\eta},
\mathcal{T}_j).  
\end{equation} 
Combining this, we arrive at
the following formula for the evidence (given that photometric redshift
for the host is available): 
\begin{equation} 
\label{eq:evid_full}
\begin{split} 
\mathcal{L}_j = &\frac{1}{N_{\mathcal{T}}}
\sum_{t} \frac{1}{N_{t}} \sum_{\vec{\eta}}\frac{1}{N_{ext}}
\sum_{M_B} \frac{e^{-(M_B-\langle M_B\rangle)^2/(2\sigma_M^2)}}
{\sqrt{2\pi}\sigma_M}\Delta M_B \\ &\times \sum_z
\frac{e^{-(z-z_p)^2/(2\delta z^2)}} {\sqrt{2\pi}\delta z}\Delta z
 \prod_i^{N_{dp}} \frac{e^{-(F_i-f_{i,j})^2/(2 \delta F_i^2)}}
{\sqrt{2 \pi} \delta F_i }.  
\end{split} 
\end{equation}

The probability $P(\mathcal{T}_j|\{F_i\})$ is then equal to the relative
evidence for each supernova type,
\begin{equation}
P(\mathcal{T}_j|\{F_i\}) = \frac{\mathcal{L}_j}{\sum_j \mathcal{L}_j}.
\end{equation}
The most likely type for a given SN candidate is the one
with the highest $P(\mathcal{T}_j|\{F_i\})$. We co-add the probabilities
for the subtypes belonging to either of the two main types
(thermonuclear and core collapse types) and obtain the probabilities
$P(TN)$ and $P(CC)$ for each typed supernovae. Our tests of
the typing using simulated data of the same quality and type
(see Section~\ref{sec:mcsim}) indicate that the resulting subtype
classifications within these two main types are prone to quite large
errors (in the order of 10--20\%), while the errors on the main type
classifications are significantly smaller (in the order of 5--10\%).

In Section~\ref{sec:tests}
we discuss how to set constraints on the evidence and
normalised probability to reliably reject both mis-classified and
non-supernova objects.

We also constructed a "best-fit" light-curve for each supernova. 
This was made by fixing the supernova type and then
finding the most likely value for each parameter one at a time,
by marginalising over the remaining three parameters. As noted
by \citet{2007ApJ...659..530K}, Bayesian techniques do not always
give the best estimate for the individual fitted parameters. To obtain
trustworthy estimates of individual parameters, dedicated simulations
have to be run, which is beyond the scope of this paper. We merely
used the fitted light curves as a measure of the overall quality of fit (a
method that is studied extensively by Monte-Carlo-simulated light
curves, see Section~\ref{sec:mcsim}).

\subsection{Supernova subtype templates} 
\begin{figure}
\centering
\includegraphics[width=8cm]{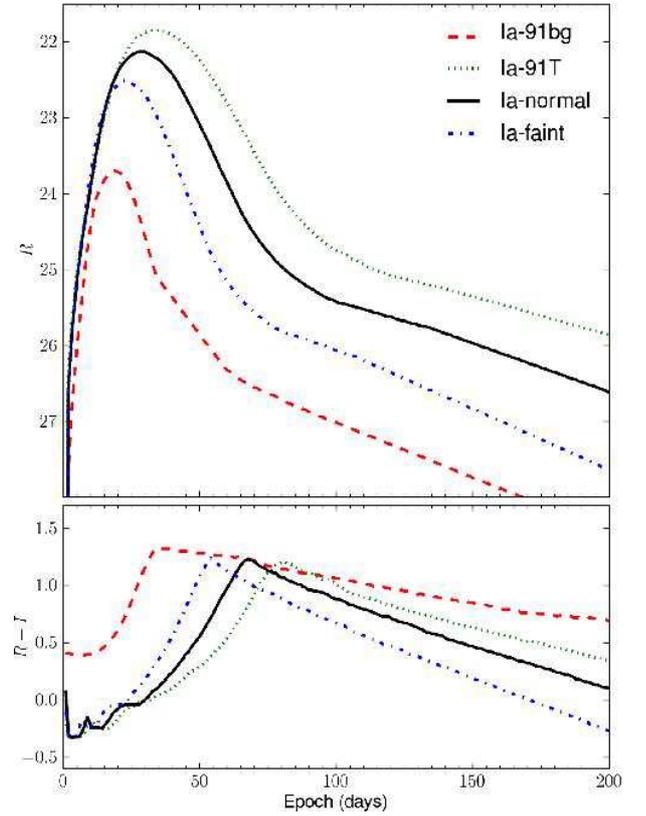}
\caption{Template light curves for the four thermonuclear supernova
types at $z=0.5$. The upper panel shows the $R$ light curve for
the Ia--normal subtype as solid (black), the Ia--faint subtype as
dash-dotted (blue), the 91T-like subtype as dotted (green) and the
91bg-like subtype as dashed (cyan). In the lower panel the $R-I$ colour
evolution for the four subtypes is shown using the same line styles
(this figure is available in colour in the electronic version of the article).}
\label{fig:TNtempl}
\end{figure}
\begin{figure}
\centering
\includegraphics[width=8cm]{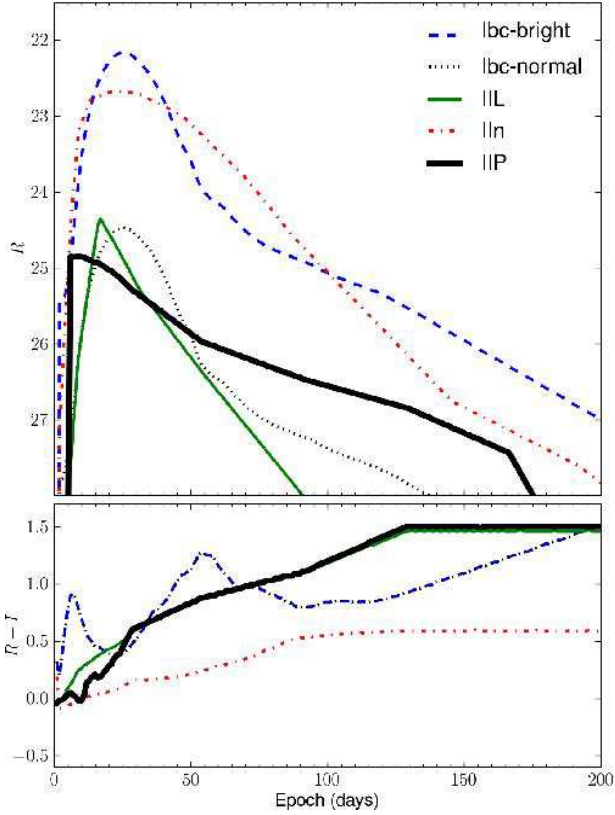}
\caption{Template light curves for the five core-collapse supernova
types at $z=0.5$. The upper panel shows the $R$ light curve for the
IIP subtype as thick solid (black), the IIn subtype as dash-dotted
(red), the IIL subtype as solid (green), the Ibc--normal subtype as
dotted (black) and the Ibc--bright subtype as dashed (blue). In the
lower panel the $R-I$ colour evolution for the five subtypes are shown
using the same line styles (this figure is available in colour in
the electronic version of the article).}
\label{fig:CCtempl}
\end{figure}
\label{sec:subt} 
We used supernova spectra and absolute magnitude light curves
to construct light curve templates for nine different supernova
subtypes. The spectra and $M_B$ light curves for all these types
were (with a few exceptions) obtained from \citet{2007NugentMISC}.
The spectra we used extend from 1\,000~\AA to 25\,000~\AA and cover
epochs from day $\sim$1--2 to $\sim$100 after explosion. The
light curves were scaled to the absolute peak magnitudes
given in Table~\ref{table:templ} and were obtained from
\citet{2004ApJ...613..189D} (and references therein) unless otherwise
noted. The table contains a complete list of the references used in
the construction of each template.

The output light curve template library consists of observer-frame
light curves for the redshift range 0.01--2.0 (with a step size of
0.01 in $z$) in the VIMOS $R$ and $I$ filters for each of the nine
supernova subtypes. As an example, Figures~\ref{fig:TNtempl} and
\ref{fig:CCtempl} show the $R$ band light curve and the $R-I$ colour
evolution for the nine templates at $z=0.5$ (roughly corresponding to
rest frame $B$ and $B-V$). It should be noted that the similarities
in colour between the Ibc--bright and Ibc--normal templates and for
the IIP and IIL templates, respectively, arise because of the spectral
templates used (see relevant sections below).

The subtype fractions listed in Table~\ref{table:templ} are
based on the measurements of \citet{2011MNRAS.tmp..413L} and the
compilation by \citet{2004ApJ...613..189D} and were calculated for our
magnitude limited sample using the limiting magnitudes from the light
curve rejection step. These fractions were not used in the actual typing,
but were only used to study the misclassification ratios, see
Section~\ref{sec:errors}.

\subsubsection{Type Ia supernovae} 
We used four different light curve templates for the thermonuclear
supernovae.  The four templates (91bg--like, Ia--faint, Ia--normal,
91T--like, from fainter to brighter) have different absolute peak
magnitudes and consequently different decline rates as well. The decline
rates were parameterised using the stretch parameter, $s$, first
introduced by \citet{1997ApJ...483..565P}. The range in stretch
($s=0.49$--$1.04$) for the four templates provides a sparsely sampled
grid that covers the observed stretch distribution of type Ia supernovae
\citep[e.g.,][]{2006ApJ...648..868S}.

The spectra and rest-frame $B$ light curve used to construct
the Ia--normal and Ia--faint templates were first presented in
\citet{2002PASP..114..803N}. The absolute peak $B$ magnitude
used for the Ia--normal template, $-19.34$, was obtained from
\citet{2003ApJ...594....1T}. The peak magnitude for the Ia--faint
subtype was calculated by starting from the Ia--normal peak magnitude
and scaling to a $s=0.8$ peak magnitude using the luminosity-to-decline
rate relationship from \citet{1999AJ....118.1766P} and the
decline-rate-to-stretch relation from \citet{2006AJ....131..527J}. The
resulting absolute peak $B$ magnitude is $-18.96$.

The light curve and spectra for the 91T--like template were first
presented in \citet{2004ApJ...612..690S}. These input data were
corrected for extinction assuming $\{R_V,E(B-V)\}=\{3.1/0.2\}$
\citep{2007NugentMISC}. The 91bg--like template is based
on light curves and spectra of SN~1991bg and SN~1999by
\citep{2002PASP..114..803N}. The peak absolute $B$ magnitudes are,
respectively, $-19.64$ and $-17.84$ \citep{2003ApJ...594....1T}. The
templates have stretch values according to their namesakes, 1.04 for
the 91T--like template and 0.49 for the 91bg--like template.

\subsubsection{Type Ib/c supernovae} 
We used two Ib/c supernova templates, the Ibc--bright and
Ibc--normal templates. The spectral evolution is based on the work
of \citet{2005ApJ...624..880L} and the light curve is based on
observations of SN~1999ex from \citet{2002AJ....124..417H}. The
observed data were corrected for extinction because the used SN
have notable extinction, $\{R_V,E(B-V)\}=\{3.1,0.4\}$ is assumed
\citep{2007NugentMISC}. \citet{2006AJ....131.2233R} find that the
absolute peak magnitude distribution of Ib/c SNe can be described
by using one bright and one normal population. We used the mean
absolute peak magnitudes, $-19.34$ and $-17.03$,  and scatter for
the two populations from their paper; converting the magnitudes
to our chosen cosmology and k-correcting to the $B$ band.  The data
available in the SVISS do not allow a refinement of the Ibc typing into
Ib and Ic subtypes, and it is not clear-cut that these two spectral
types correspond to the two different photometric types (the bright
and normal) considered in this work \citep{2002AJ....124..417H,
2011MNRAS.tmp..413L}.

It should be noted that the red colour of the template in pre-peak
epochs (the first 5-7 days in rest-frame) is the result of basing
the template on SN~1999ex, which showed this effect at early epochs
\citep{2002AJ....124..417H}. There are not many Ib/c supernovae that
have been observed before peak brightness, but observations of other
Ib/c's \citep[e.g., SN~2008D from][]{2009ApJ...702..226M} show that
this effect may not be characteristic of the type. We performed
tests using the typing code with a Ib/c template without the early red
phase and found that the choice of template has very little influence
on the resulting probabilities (less than 1\%).

\subsubsection{Type IIL supernovae} 
The IIL spectra and light curves were first used in
\citet{1999ApJ...521...30G}. The light curve is originally from
\citet{1997A&A...322..431C}. \citet{2002AJ....123..745R} presented
the absolute $B$ peak magnitude distribution of IIL SNe, and we
adopt the mean peak magnitude and scatter for their normal IIL
population. Converting the magnitude to our cosmology, we arrive at
a $B$ peak magnitude of $-17.23$.

\subsubsection{Type IIn supernovae} 
The spectra and light curves are based on SN~1999el
\citep{2002ApJ...573..144D}, whereas the absolute $B$ peak magnitude
and scatter are obtained from \citet{2002AJ....123..745R}. After
compensating for the different cosmological parameters, the resulting
peak magnitude is $-18.82$.

\subsubsection{Type IIP supernovae} 
The spectra used to build the IIP supernova template come from
different sources. For the early epochs ($\leq 33$ days after
explosion) we used extinction-corrected spectral models
from \citet{2008ApJ...675..644D}, which in turn are based
on SWIFT UV-optical observations of SN~2005cs and SN~2006bp
\citep{2006MNRAS.370.1752P,2007ApJ...659.1488B,
2007ApJ...664..435I,2007ApJ...666.1093Q}.  The late epoch spectra
are based on SN~1999em from \citet{2004ApJ...616L..91B} and obtained from
\citet{2007NugentMISC}. The reason for not using the SN~1999em spectra
for the early epochs is that the UV part of those spectra are modelled
and extrapolated from optical observations, the more recent modelling based
on early epoch UV/optical observations should therefore provide a more
accurate spectral evolution.

The light curves were constructed using photometric data for
SNe~1999em, 1999gi, 2003gd, 2004et, 2005cs and 2006bp
\citep{2003MNRAS.338..939E,2002AJ....124.2490L,2005MNRAS.359..906H,
2006MNRAS.372.1315S,2006MNRAS.370.1752P,2007ApJ...666.1093Q}. The
light curve of each of these SNe was scaled to a common peak magnitude
and corrected for extinction using the values given in the original
references. Each curve was then re-sampled with a resolution of one day
over the range of observed epochs (using spline interpolation). The
final IIP light curve was then obtained by averaging over the six
interpolated light curves. 
The adopted absolute $B$ peak magnitude, $-16.66$, and scatter
come from \citet{2002AJ....123..745R}, again compensating for the
cosmological parameter difference.

\subsubsection{K-corrections of SN light curve templates}
\label{sec:kcor}
We then calculate VIMOS $R$ and $I$ apparent light curves for
the redshift range $[0,2]$ using the following formula \citep[similar
to][]{2004ApJ...613..189D}:
\begin{equation}
\label{eq:appmag}
\begin{split}
m_y(t,z)= &M_{peak,B} + \Delta M_{B}(t\times(1+z)^{-1}) + \mu(z)\\
&+{\rm A}_B - K^{B}_y(z,t\times(1+z)^{-1}),
\end{split}
\end{equation}
where $y$ refers to the observed filter ($R$ or $I$), $t$ is the
observer frame epoch,  $M_{peak,B}$ is the peak absolute magnitude
in a rest-frame $B$ filter, $\Delta M_{B}$ is the light curve decline
relative to the peak, $\mu(z)$ is the distance modulus and A$_B$ is the
rest frame extinction in the host galaxy. The K-correction $K^{B}_y$,
following the formalism of \citet{1996PASP..108..190K}, is given by
\begin{equation}
\begin{split}
K^{B}_y(z,\tau) = &-2.5\,\log \left(\frac{\int \mathcal{Z}(\lambda) 
S_B(\lambda) d\lambda}{\int \mathcal{Z}(\lambda) S_y(\lambda) 
d\lambda} \right) +2.5\,\log(1+z) \\
&+2.5\,\log \left(\frac{\int F(\lambda,\tau) 
S_B(\lambda) d\lambda}{\int F(\lambda/(1+z),\tau) S_y(\lambda) 
d\lambda} \right).
\end{split}
\end{equation}
In this expression $\mathcal{Z}(\lambda)$ is the spectral energy
distribution of Vega; $S_B(\lambda)$, $S_y(\lambda)$ the Johnson B and
VIMOS $R$/$I$ total filter transmission curves and $F(\lambda,\tau)$
the spectral energy distribution of the supernova at rest frame
epoch $\tau$.

\section{Testing the typing method}
\label{sec:tests}
\subsection{Simulated supernova light curves}
\label{sec:mcsim}
To test the typing accuracy of the code applied to the observed SNe,
we created a simulated mock detection catalogue. This catalogue was created
to resemble the expected characteristics of the detected SN sample
when it comes to peak magnitude distributions, light curve shapes,
redshift distribution, and dust extinction. The catalogue also reflects
the specifics of the SN search, including the cadence between observing
epochs and the limiting magnitudes in the detection filters.

We first assumed a total SN rate by summing the thermonuclear (TN)
and core-collapse (CC) rates from \citet{2004ApJ...613..189D}. We used
this rate to assign a random redshift to each mock SN in the range
$0<z<2$. The SNe were thereafter given a main type, either TN or CC,
from the relative strength of the rate of these types at the assigned
redshift. The thermonuclear SNe were then subdivided into a faint,
a normal, a 91bg--like and a 91t--like population, while the CC SNe
were subdivided into IIL, IIP, IIn, Ibc--bright or Ibc--normal. It
should be noted that the fractions of each simulated subtype are
equal at this point. We did this to ensure that the number of
simulated SNe for each subtype is sufficiently high for statistics,
even for the rare subtypes. The fractions given for each subtype in
Table~\ref{table:templ} were used when analysing the typing results
for the simulated light curves (see Section~\ref{sec:errors}).

A peak B-band magnitude was assigned to
each SN (Table~\ref{table:templ}). This was then perturbed using
the peak magnitude dispersion values given in the table. Each SN was
assigned an explosion date over the course of one year. We then
used the actual cadence of the SN search and the B-band light curves
for each specific type to calculate the absolute magnitude of the SN
at the different observational epochs. Because the SN explosion dates
are distributed over time, we will have some SNe that are observable
in all epochs (i.e., those exploded before the first observation),
and others that explode during the observational period and thus only
observable in some of the epochs.

In the next step we used the type specific SEDs of the SNe to
calculate the K-corrections that give us the apparent magnitudes in
the observed R and I filters corresponding to the absolute B-band
magnitudes (see Section~\ref{sec:kcor}). The apparent magnitudes
were also corrected for extinction in the SN host galaxies using the
extinction distributions from \citet{2005MNRAS.362..671R}. In the final
step, we added a photometric error to the apparent magnitudes. This
consists of both a statistical part, derived from the expected S/N=3
limits in R and I (see Section~\ref{sec:phot}, and a "systematic"
part, for which we assumed an extra 4\% error in the flux. The
simulated catalogue consists of apparent magnitudes and errors in
R and I at each observational epoch. We treated objects with S/N$<1$
as non-detections. This is also the case for epochs observed prior
to the explosion of a particular SN.

Because we rely on photometric redshifts, we also assigned a simulated
redshift to each SN, $z_{sim}$. We calculated this by adding a random
scatter, $\Delta z_{rnd}$ -- drawn from a Gaussian distribution with
$\sigma_z=0.06$ -- to the true redshift ($z_{true}$) according to
$z_{sim}=z_{true} + (1+z_{true})*\Delta z_{rnd}$. The $\sigma_z$
used here is the photometric redshift accuracy of our host galaxy
catalogue, see Section~\ref{sec:photz}.

When testing our code using the mock catalogue, we culled the sample
by applying additional selection criteria. As our default set-up,
we only included objects with a S/N$>$3 in at least two consecutive
epochs in both R and I (mimicking the set-up used for the observed
supernovae). The total number of remaining simulated supernovae
after the cull is $\sim$18\,000, which corresponds to roughly 2\,000 SNe
per subtype (see Table~\ref{table:mcres}).

\subsection{Rejection of non-supernovae}
\label{sec:rejano}
As mentioned in Section~\ref{sec:typingend}, the Bayesian typing
method only gives valid results when the input light curve
can be modelled by one of the templates. If the variability of
a source has a non-SN origin (or, though less likely,
is a SN with a very irregular light curve), the resulting
likelihoods will be very small. Indeed, the evidence will be
zero (within machine accuracy) in many cases where the observed
light curve is just too different from a template SN light curve. The
evidence, $\mathcal{L_\mathcal{T}}$ (see Equation~\ref{eq:evid_full}),
can therefore be used to reject possible anomalous sources (AGN, variable
stars, complicated subtraction residuals). We followed the suggestion
of \citet{2007ApJ...659..530K} and defined rejection limits for each
template, $\mathcal{L}_{99.9,\mathcal{T}}$. These were calculated by
finding the evidence below which fall 99.9\% of the evidences -- resulting
from fitting of the Monte-Carlo-simulated light curves. The likelihood 
thresholds are in the order of $10^{-100}$.

It should be noted that the likelihood threshold is only really
efficient in removing sources with light curves that are sufficiently
different from a supernova light curve. For example, an active galactic nuclei 
(AGN) with
a light curve very similar to a supernova will of course not be
removed by this cut. Using this rejection scheme allows us to perform
an automatic cut that will only reject extreme outliers (0.1\%)
in the supernova population. Additional manual inspection is needed to
safeguard against possible spurious candidates that have light curves
similar enough to SNe to end up with likelihoods higher than the 99.9\%
threshold. This process is described, along with our final supernova
candidates, in Section~\ref{sec:cands}.

Some of the problems related to supernovae with non-standard light
curves may be avoided by using the technique of ''fuzzy'' templates,
pioneered by \citet{2009ApJ...707.1064R}. This method enables
the templates to cover more of the parameter space by giving the
templates themselves an uncertainty (''fuzziness''). In this work
we did not use this technique but it can be implemented
into our code, and we will include it as an option in future versions
of it. Another way to improve typing of these SNe is to use a more
extended set of templates and to update the existing ones.

\subsection{Systematic and statistical errors of typing}
\label{sec:errors} 
\begin{figure*}
\centering
\includegraphics[width=0.95\textwidth]{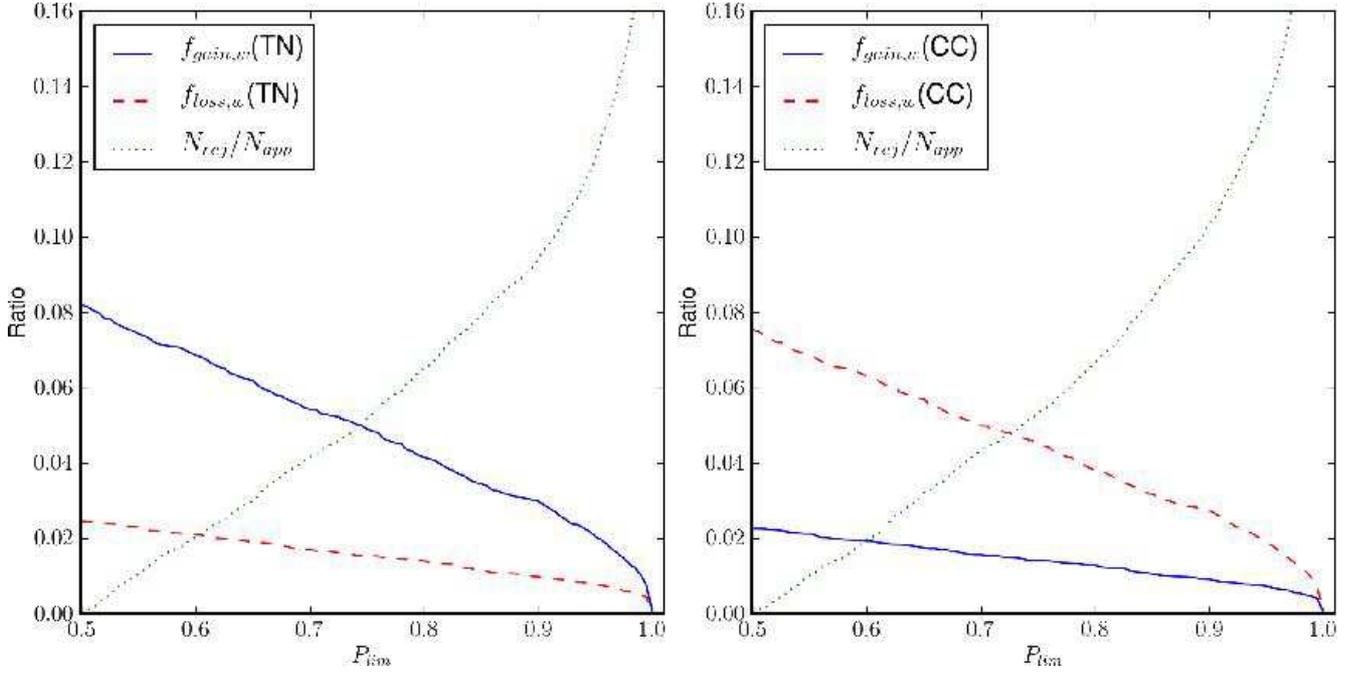}
\caption{Effect on the misclassification ratios of applying
a rejection criterion on the type probability. The left/right
panel shows the weighted misclassification ratios ($f_{loss,w}$
and $f_{gain,w}$) for the thermonuclear/core-collapse SNe along
with the fraction of rejected candidates ($N_{rej}/N_{app}$) for a
given limiting type probability ($P_{lim}$). This information can
be used to select an optimal $P_{lim}$ to avoid misclassifications
without rejecting too many candidates. For details see the text and
Table~\ref{table:mcres} (this figure is available in colour in the
electronic version of the article).}
\label{fig:misratio}
\end{figure*}
\begin{figure*}
\centering
\includegraphics[width=0.95\textwidth]{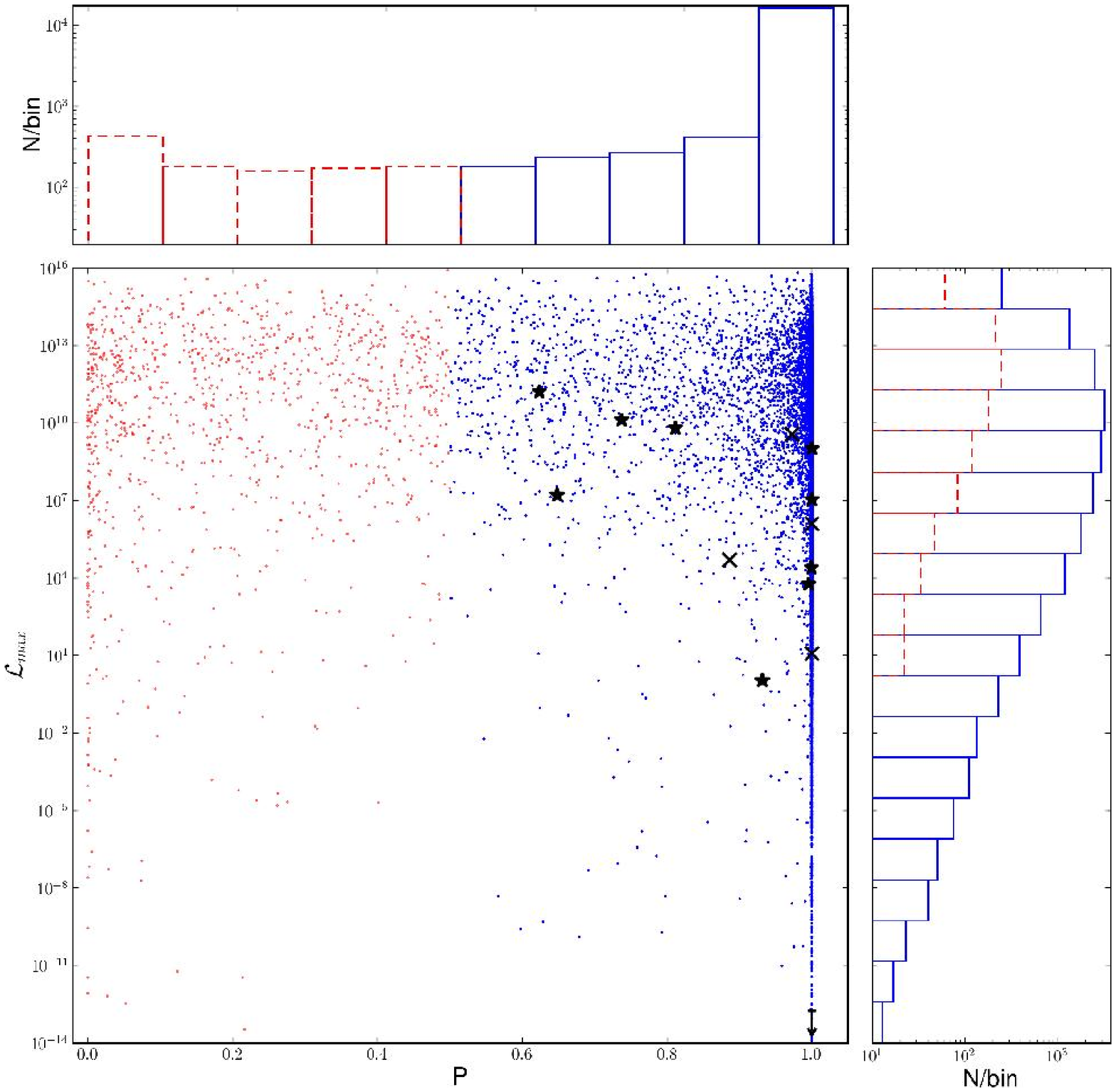}
\caption{Main panel: fit quality as measured by
  the absolute Bayesian likelihood for the best-fitting template
  ($\mathcal{L}_{max}$), versus $P$, the probability of a
  given source being of the simulated type, for the $\sim$18000
  Monte-Carlo-simulated supernovae. The upper panel shows the
  $P$ distribution and the right panel the $\mathcal{L}_{max}$
  distribution. For clarity, large (blue) dots and solid (blue)
  lines marks the correctly typed SNe (with $P>0.5$); small (red)
  dots and dashed (red) lines the incorrectly typed SNe. The cross
  and star symbols in the main panel show $\mathcal{L}_{max}$
  and $P(TN)/P(CC)$, respectively, for the observed supernovae (see
  Section \ref{sec:cands}). Three of the thermonuclear SN candidates
  have $\mathcal{L}_{max}<1\times10^{-14}$, they are marked by a
  caret lower limit symbol at $\mathcal{L}_{max}<1\times10^{-14}$
  (note that they have the same $P(TN)$ values, thus the symbols are
  overlapping) (this figure is available in colour in the electronic
  version of the article).}
\label{fig:SNsim}
\end{figure*}
We used the Monte-Carlo-simulated supernova light curves to find the
expected misclassification ratios for the two main supernova types
at different redshifts. Typing of the simulated light curves was
performed using the same code and assumptions as when typing the observed
supernovae. This approach allows us to estimate the total errors on
the supernova counts including both systematic errors, which arise from
the assumed priors and possible systematic errors in the photometry;
and statistical errors from the photometric statistical errors and
the photometric redshift errors.

We investigated the errors by studying the misclassification ratios, 
$f_{loss}$ and $f_{gain}$ for the simulated thermonuclear and core-collapse 
supernova light curves. The ratios are defined as
\begin{align}
\label{eq:miscratios}
f_{loss}(TN) &\triangleq \frac{N_{TN\rightarrow CC}}{N_{app}(TN)} & 
f_{gain}(TN) &\triangleq \frac{N_{CC\rightarrow TN}}{N_{app}(TN)}\nonumber\\
f_{loss}(CC) &\triangleq \frac{N_{CC\rightarrow TN}}{N_{app}(CC)} &
f_{gain}(CC) &\triangleq \frac{N_{TN\rightarrow CC}}{N_{app}(CC)},
\end{align}
where $N_{TN\rightarrow CC}$ and $N_{CC\rightarrow TN}$ is the
number of SNe that were incorrectly classified (the true type
on the left) and $N_{app}$ is the apparent number of SNe of the given
type. For example, $f_{loss}(TN)$ gives the fraction of thermonuclear
SNe incorrectly classified as CC SNe. Note that using the ratios
as measures of misclassification errors is valid if the underlying
core-collapse to thermonuclear SN ratio in the survey is the same as
the one we assume in Section~\ref{sec:mcsim}.

The different subtypes have quite different misclassification
ratios.  Because we assumed the SN frequency for the different
subtypes to be equal, we also computed weighted misclassification
ratios, $f_{loss,w}(TN)$, $f_{gain,w}(TN)$, $f_{loss,w}(CC)$,
$f_{gain,w}(CC)$, where the number of misclassifications per
subtype (e.g. $N_{IIP\rightarrow TN}$) are weighted by the
relative frequency of that subtype. The subtype fractions are
given in Table~\ref{table:templ}.
The $f_{loss,w}$ and $f_{gain,w}$ ratios were adopted as the final
measure of misclassification errors in our survey.

Figure~\ref{fig:misratio} shows how $f_{gain,w}(TN/CC)$
and $f_{loss,w}(TN/CC)$ decrease when a limit on the type
probabilities, $P_{lim}$, for a source to be included in the analysis
is introduced. The figure also shows that the use of a limiting probability
will decrease the survey efficiency. 
Owing to the low-number statistics nature of our survey we
elected not to use a limiting probability as a rejection criteria, but
it should be noted that the misclassification errors can be somewhat
decreased by only including objects with high probabilities. The
ratios and total number of simulated supernovae for the different
redshift bins can also be found in Table~\ref{table:mcres}.
\begin{table*}
\caption{Simulation results}	     \label{table:mcres} \centering \begin{tabular}{l l l l l l l} \hline\hline $z$ & $f_{gain,w}$ & $f_{loss,w}$ & $f_{gain}$ & $f_{loss}$& $dz$ & N$_{app}$ \\ \hline 
\multicolumn{7}{c}{Thermonuclear Supernovae}\\
\hline
$z\leq0.25$     & 0.019 & 0.025 & 0.039 &0.046  & 0.091 & 540\\   
$0.25<z\leq0.50$& 0.098 & 0.023 & 0.096 &0.043  & 0.12  & 2894 \\
$0.50<z\leq0.75$& 0.096 & 0.032 & 0.10  &0.052  & 0.084 & 3102 \\
$0.75<z\leq1.00$& 0.057 & 0.017 & 0.078 &0.022  & 0.074 & 1759 \\
$1.00<z\leq1.25$& 0.038 & 0.0045& 0.061 &0.0056 & 0.095 & 358 \\
$1.25<z\leq1.50$& 0.26  & 0.0   & 0.50  &0.00   & 0.082 & 20 \\
\hline
$z\leq1.5$      & 0.082 & 0.024 & 0.042 & 0.071 & 0.098 & 8673  \\
\hline
\multicolumn{7}{c}{Core-collapse Supernovae}\\
\hline		
$z\leq0.25$     & 0.0079& 0.0060& 0.015  &0.013 & 0.078 & 1676 \\  
$0.25<z\leq0.50$& 0.020 & 0.085 & 0.038  &0.084 & 0.12  & 3319 \\
$0.50<z\leq0.75$& 0.041 & 0.12  & 0.067  &0.13  & 0.083 & 2428 \\
$0.75<z\leq1.00$& 0.026 & 0.085 & 0.032  &0.12  & 0.094 & 1172 \\
$1.00<z\leq1.25$& 0.0034& 0.029 & 0.0042 &0.047 & 0.12  & 472 \\
$1.25<z\leq1.50$& 0.0   & 0.014 & 0.00   &0.027 & 0.16  & 367 \\
\hline                               
$z\leq1.5$      & 0.022 & 0.075 & 0.084 & 0.037 & 0.11  & 9434 \\
\hline
\end{tabular} 
\tablefoot{$f_{gain,w}$/$f_{loss,w}$ are the misclassification ratios weighted by
subtype frequency and $f_{gain}$/$f_{loss}$ are the unweighted misclassification
ratios. $dz$ is the redshift variance (based on the fitted
redshift from the typing code compared to the true redshift of the
supernova) for the specific bin and N$_{app}$ is the apparent number of simulated
supernovae of the given type in the bin.} 
\end{table*}

There is a caveat with using the misclassification results from
the simulated sample. The errors are only really reliable when the
observed supernova sample has light curves that are similar to the
template light curves on average. If most of the observed supernovae
have light curves different from the templates, the errors
estimated by this method will not be representative. Figure
\ref{fig:SNsim} shows the distribution of Bayesian evidence for most
likely type, $\mathcal{L}$, for the simulated SN sample together with
the observed supernovae (see Section~\ref{sec:cands}). Comparing
the distribution in $\mathcal{L}$ of the simulated SNe to the observed
ones indicate that the samples seems to be fairly different. 

A K--S test applied to the two samples is not conclusive, we cannot
rule out that the two populations are the same with significance
(the K--S p-value is 0.11). Interpreting this result is not trivial,
there is a number of assumptions that go into simulating the SNe
(the choice of specific templates and basically all of the priors)
that can make the distribution different from what you would expect for
a real sample. It is certainly possible that a discovered SNe may have
a light curve that is simply different from any of the templates we
use, which will cause a mismatch like the one we see in the $\mathcal{L}$
distributions. This is also discussed in Section~\ref{sec:rejano},
where we provide some suggestions on how to solve the problem.

The assumption that the the supernovae will be evenly distributed
into the subtypes (see Section~\ref{sec:mcsim}) will influence the
$\mathcal{L}$ distribution and cause it to be slightly different
from the observed distribution. In this particular case we cannot
use the a priori subtype fractions to weight the results (as used in
Section~\ref{sec:errors}) because the numbers of observed SNe in the
different types are too low for subtype K--S calculations.

There is consequently no reason to expect that the the samples would
be perfectly matched (giving p-values close to 1). Nevertheless,
we cannot rule out that the resulting p-values indicate that the
assumption of the simulated light curves are representative of the
sample of real supernovae (at a significance level of 10\%). Note that
the efficient number of data points for the two types are quite low
($\sim 7$ for the thermonuclear and $\sim 9$ for the core-collapse
SNe, see Section~\ref{sec:cands}). A better accuracy for the p-values
requires more real supernovae to compare with, which would also enable
us to look at the subtype statistics.

\subsection{Redshift determination with the typing code}
By fixing the type to the most likely template found in the full
Bayesian fitting run, we tried to find the most likely redshift of
the supernovae (see Section~\ref{sec:typingend} for a more detailed
description). The prior on redshift is based on the probability
distribution for photometric redshift of the host galaxies. By looking
at the most likely redshifts obtained from the typing of the simulated
supernovae, we investigated whether the prior information added
through the supernova light curve changes the redshift accuracy.

For the SVISS host galaxies the normalised photometric redshift
scatter is $\sigma_z=\mbox{rms}((z_{true}-z_{obs})/(1+z_{true}))=0.06$,
the simulated SN redshifts are scattered according to this (see
Section~\ref{sec:mcsim}).  When comparing the most likely redshifts
($z_{fit}$) from the typing code with the true ones ($z_{true}$ as
defined in Section~\ref{sec:mcsim}), we found a normalised redshift
scatter of 0.067 for the thermonuclear supernovae and 0.072 for the
core-collapse with a negligible offset for both types. Overall, a
small number of objects ($<$0.1\%) were assigned a catastrophic redshift,
defined as objects with $(|z_{fit}-z_{true}|/(1+z_{true})>0.2$; these
are all misclassified SNe.  The resulting redshift accuracy is thus
very similar to the input simulated error, the inclusion of prior
information in the form of the supernova data did not improve the
redshift determination.  On the other hand, excluding the misclassified
supernovae, no additional errors to the redshift estimates seem to have
been introduced in the typing code.
\begin{table*}
\caption{Supernovae in the SVISS}	     \label{table:sne} \centering \begin{tabular}{l l l l|l l l c} \hline\hline SVISS ID & Type (Subtype) & $P_t$($P_s$)\tablefootmark{a} & 
$\log \mathcal{L}$& $z$ & $M_{B,peak}$ & $t_e$ (days)\tablefootmark{b} 
& Ext./$E(B-V)$\tablefootmark{c} \\ \hline 
SVISS-SN43  & TN(Ia--normal)& 0.886(0.733)&  4.7 & 0.43 & 
-19.88 &  140 &  --/0.0 \\           
SVISS-SN161 & TN(Ia--faint)& 1.00(0.910)& -19.1 & 0.50 & 
-18.28 &  -47 &  --/0.0 \\          
SVISS-SN115 & TN(Ia--91t)& 1.00(1.00)& -17.8 & 0.40 & 
-19.04 &    1 &  --/0.0 \\          
SVISS-SN116 & TN(Ia--faint)& 1.00(1.00)&  1.1& 0.55 & 
-18.98 &   -6 &  --/0.0 \\          
SVISS-SN309 & TN(Ia--faint)& 1.00(1.00)& -14.8 & 0.47& 
-17.96 &  -77 &  --/0.0 \\          
SVISS-SN402 & TN(Ia--faint)& 1.00(0.929)&  6.1 & 0.22 & 
-18.66 & -127 &  CZ/0.4 \\          
SVISS-SN135 & TN(Ia--normal)& 0.950(0.580)& 9.6 & 0.98& 
-18.9 &  -25 &  C3/0.2 \\          
\hline					 SVISS-SN14 & CC (IIn)& 0.623(0.551)& 11.2  & 0.36\tablefootmark{d} & 
-18.64 &  133 &  --/0.0\\
SVISS-SN51 & CC (IIP) & 1.00 (0.980) & 9.0 & 0.51  & 
-17.56 &   25 &  C2/0.6\\
SVISS-SN54 & CC (IIn) & 0.812 (0.733)& 9.8  & 0.77  & 
-19.04 &   72 &  C2/0.2\\
SVISS-SN261& CC (Ibc--normal) & 0.734 (0.364)& 10.1 & 0.37\tablefootmark{d}& 
-17.48 & -110 &  C2/0.4\\
SVISS-SN55 & CC (IIP) & 0.995 (0.854)& 3.8   & 0.83  & 
-17.91 &  -17 &  --/0.0\\
SVISS-SN31 & CC (IIL) & 0.999 (0.795)& 4.4  & 0.12\tablefootmark{d}& 
-16.52 &   64 &  --/0.0\\
SVISS-SN56 & CC (Ibc--bright) & 0.930 (0.912)& 0.037& 0.57  & 
-19.27 &  -74 &  C1/0.6\\
SVISS-SN357& CC (IIn) & 1.00 (1.00) & 6.0  & 1.4  & 
-19.08 &  -91 &  --/0.0\\
SVISS-SN24 & CC (Ibc--bright) & 0.643 (0.643)& 7.2  & 0.81  & 
-19.56 &   25 &  C2/0.5\\
\hline					 
\end{tabular} 
\tablefoot{
\tablefoottext{a}{$P_t$ refers to the co-added probability for the best-fitting 
main type (TN/CC), $P_s$ refers to the subtype probability from the typing 
code.}
\tablefoottext{b}{This is the time since explosion in the observers frame.}
\tablefoottext{c}{Extinction models are as follows: C3, Cardelli
with $R_V$=3.1; C2, Cardelli with $R_V$=2.1; CZ, Calzetti. For the
SNe with a most likely extinction of zero, no best-fit model is given.}
\tablefoottext{d}{``Hostless'' SNe, for these a flat redshift prior was 
used, see text for details.}
}
\end{table*}

In Table~\ref{table:mcres} we also present the redshift scatter for
the simulated supernovae in the redshift bins used to study
misclassification. Note that the scatter values in the table are not
normalised (i.e. $dz =\mbox{stdev}(z_{true}-z_{obs}))$), as opposed
the the $\sigma_z$ discussed in the previous section. The overall
redshift accuracy is consistent with the simulated input accuracy. The
$0.25<z\leq0.5$ redshift bin has a somewhat increased scatter,
this is because of a higher contribution of catastrophic redshifts (or,
equivalently, misclassified SNe) in this bin. The actual percentage
of outliers in this bin is still quite low, 3\% for the thermonuclear
and 4\% for the core-collapse SNe.

\subsection{SDSSII supernovae}
\label{sec:sdsstest}
We also tested our typing code on a small sample of
spectroscopically confirmed supernovae from the SDSS supernova
survey \citep{2008AJ....135..338F}. The sample contains 55 Ia
supernovae at $z$=0.001--0.2 and 32 IIP supernovae at $z$=0.001-0.2
\citep{2008AJ....135..338F,2009ApJS..185...32K,2010ApJ...717...40K,
2010ApJ...708..661D}. The observations were obtained with a cadence
(1--10 days) and total survey length ($\sim$90 days). 
To be able to compare the typing results from
this sample with the observations and simulations for the SVISS, we
re-sampled the SDSS supernova light curves to a cadence of $\sim$20
days (yielding 4--5 epochs). The resulting light curves (at $z\sim$0.1)
therefore sample approximately the same rest-frame epochs as the light
curves from SVISS (with a mean redshift of 0.57). We used the SDSS
$g$ and $r$ filters, which at $z\sim$0.1 target similar rest-frame
colours as the VIMOS $R$ and $I$ filters at $z\sim0.5$. Using only
these epochs and filters will of course render the typing less optimal,
but will render the results comparable to the SVISS typing because the
same rest-frame light curve information is used. Note that a maximum
of five epochs can be fitted using this sampling, compared to the
maximum of seven used in the SVISS observations and simulations.

The results of testing performed on the sample of SDSS supernovae do
not show any significant differences with the Monte-Carlo-simulated
sample in terms of misclassification. The total misclassification
percentage for a sample of 55 Ia supernovae (i.e., Ia supernovae
typed as any of the core-collapse subtypes) is 7.3 \%. The total
misclassification percentage for a sample of 32 IIP supernovae is 9.3
\%. These ratios are of the same order of magnitude as the results
obtained for the simulated sample, although slightly higher. The
simulated Ia SNe are misclassified in $\sim$3\% of the cases in
the relevant redshift bin, the corresponding percentage for IIP
SNe is $\sim$5\%. The difference could be caused by the somewhat lower
number of data-points available, on average, for the SDSSII SNe with
resampled light curves. We conclude that the simulations provide
valid estimates on the misclassification ratios for a real sample of
supernova light curves.
\begin{figure*}
\centering
\includegraphics[width=0.91\textwidth]{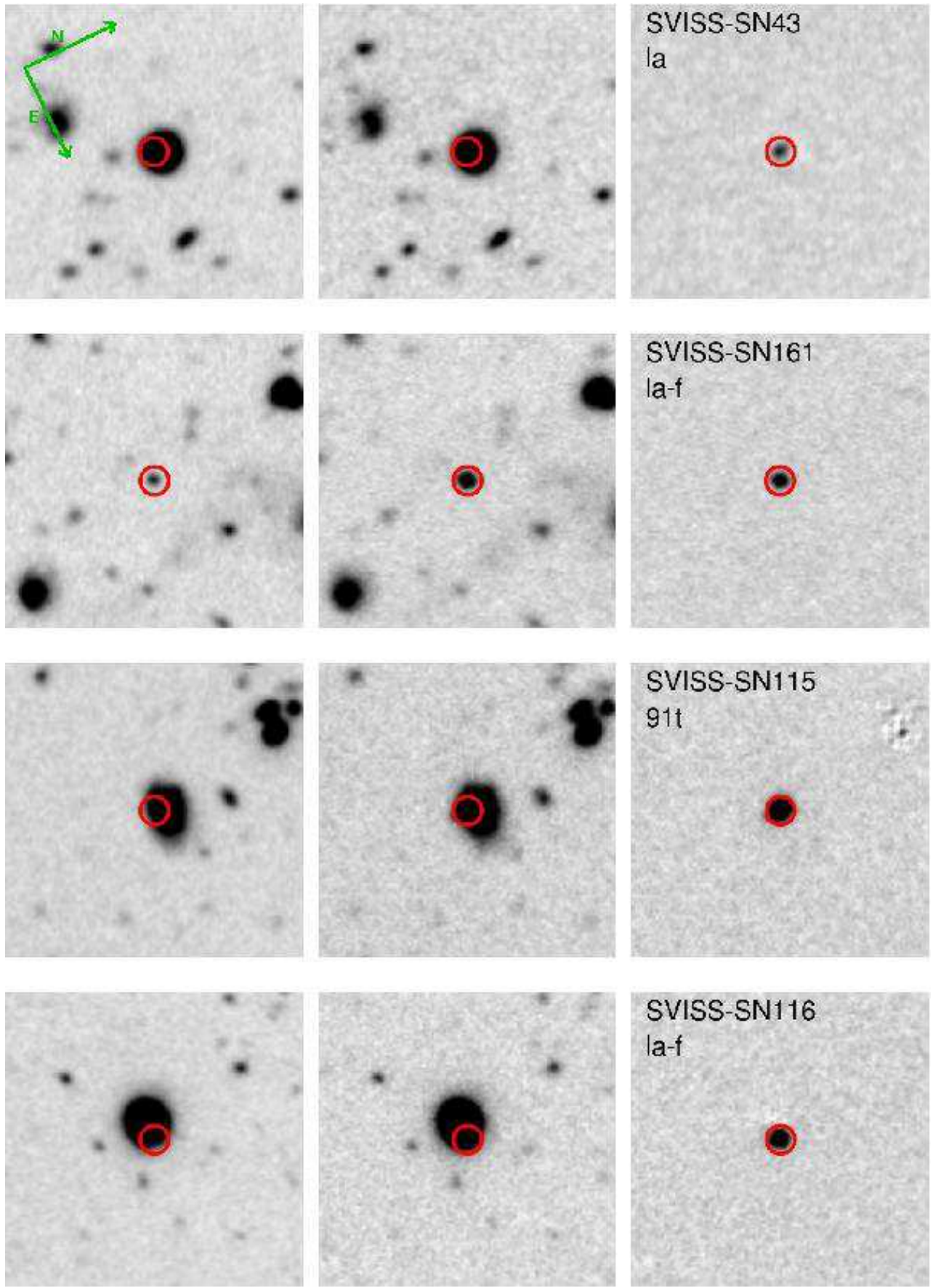}
\caption{R-band image cutouts, approximately $20\times 20$\arcsec large
of the supernova candidates SVISS-SN43, SVISS-SN161, SVISS-SN115
and SVISS-SN116. The leftmost panels show the reference image,
middle panels show the peak brightness (observed) epoch and the
rightmost panels show the subtracted image at peak brightness. The
(red) circle marks the location of the supernova as detected in the
subtracted frame. The most likely redshifts (from the typing code) for 
these SNe are (starting at the top) 0.43, 0.50, 0.40 and 0.55, respectively.}
\label{fig:sta1} 
\end{figure*}
\begin{figure*}
\centering
\includegraphics[width=0.91\textwidth]{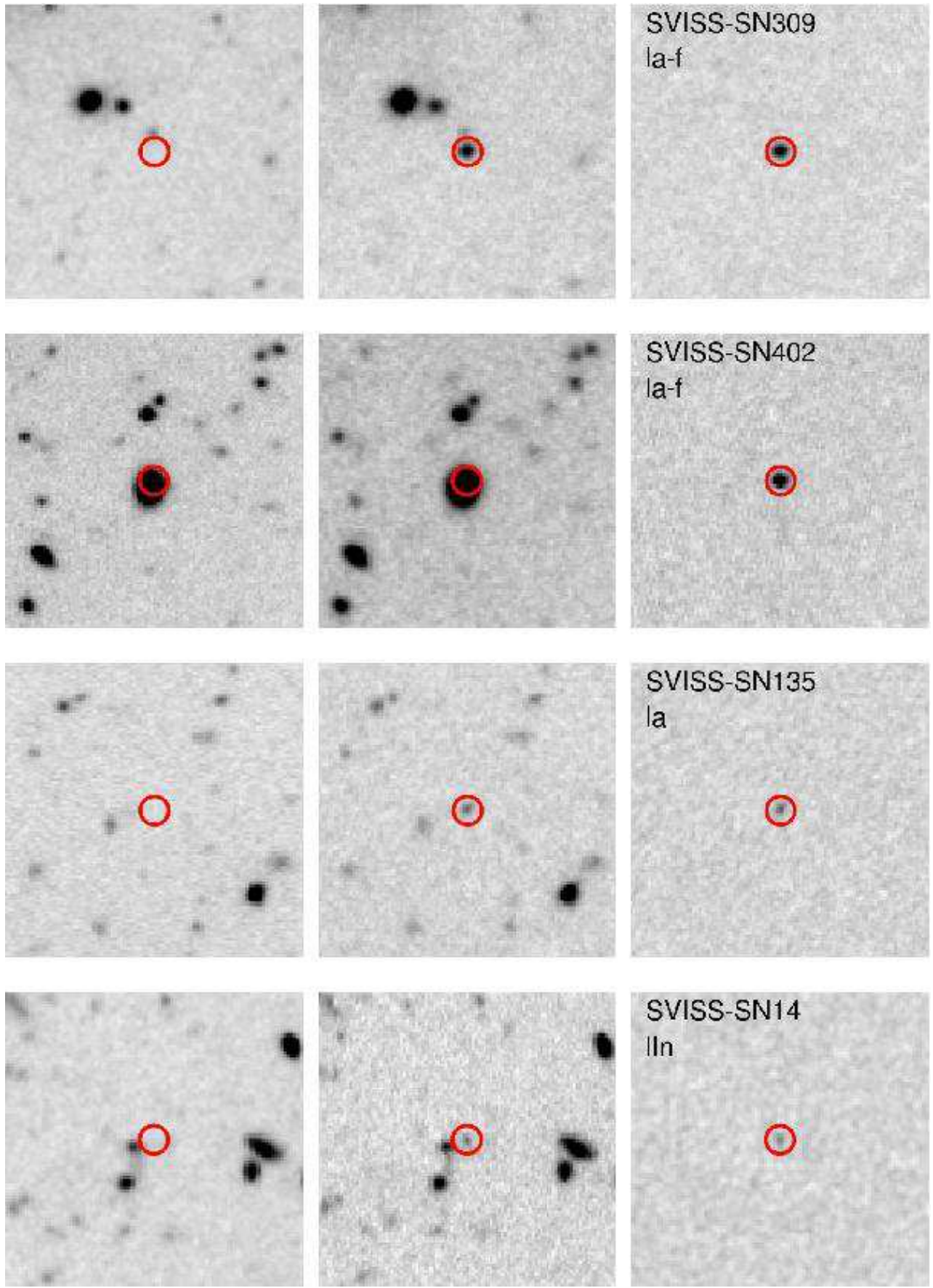}
\caption{R-band image cutouts for the supernova candidates SVISS-SN309, SVISS-SN402,
SVISS-SN135 and SVISS-SN14. See Figure~\ref{fig:sta1} for descriptions on the
different panels and marks. The most likely redshifts (from the typing code) for 
these SNe are (starting at the top) 0.47, 0.22, 0.98 and 0.36, respectively.}
\label{fig:sta2}
\end{figure*}
\begin{figure*}
\centering
\includegraphics[width=0.91\textwidth]{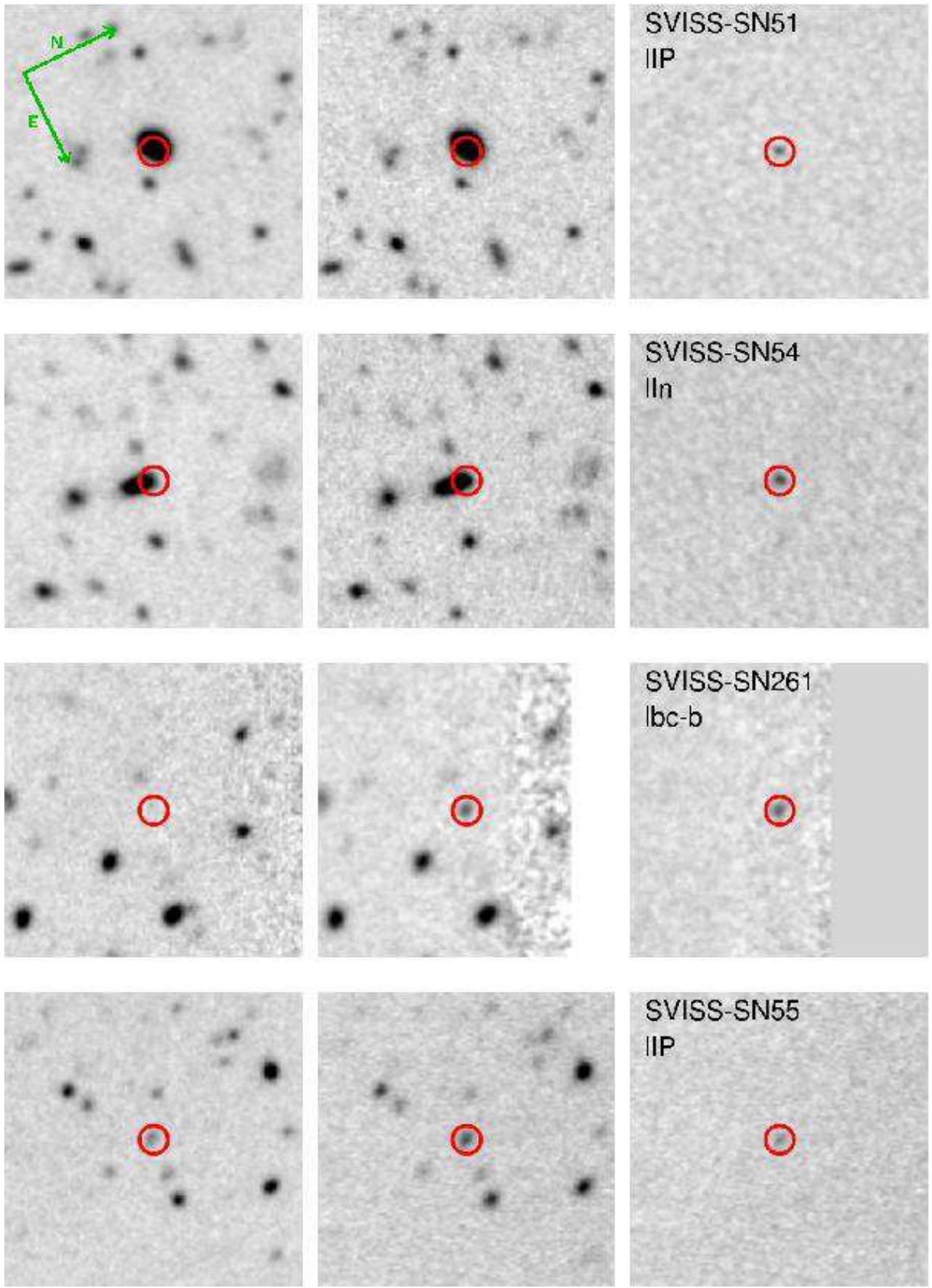}
\caption{R-band image cutouts for the supernova candidates SVISS-SN51, SVISS-SN54,
SVISS-SN261 and SVISS-SN55. See Figure~\ref{fig:sta1} for descriptions on the
different panels and marks. The most likely redshifts (from the typing code) for 
these SNe are (starting at the top) 0.51, 0.77, 0.37 and 0.83, respectively.}
\label{fig:sta3}
\end{figure*}
\begin{figure*}
\centering
\includegraphics[width=0.91\textwidth]{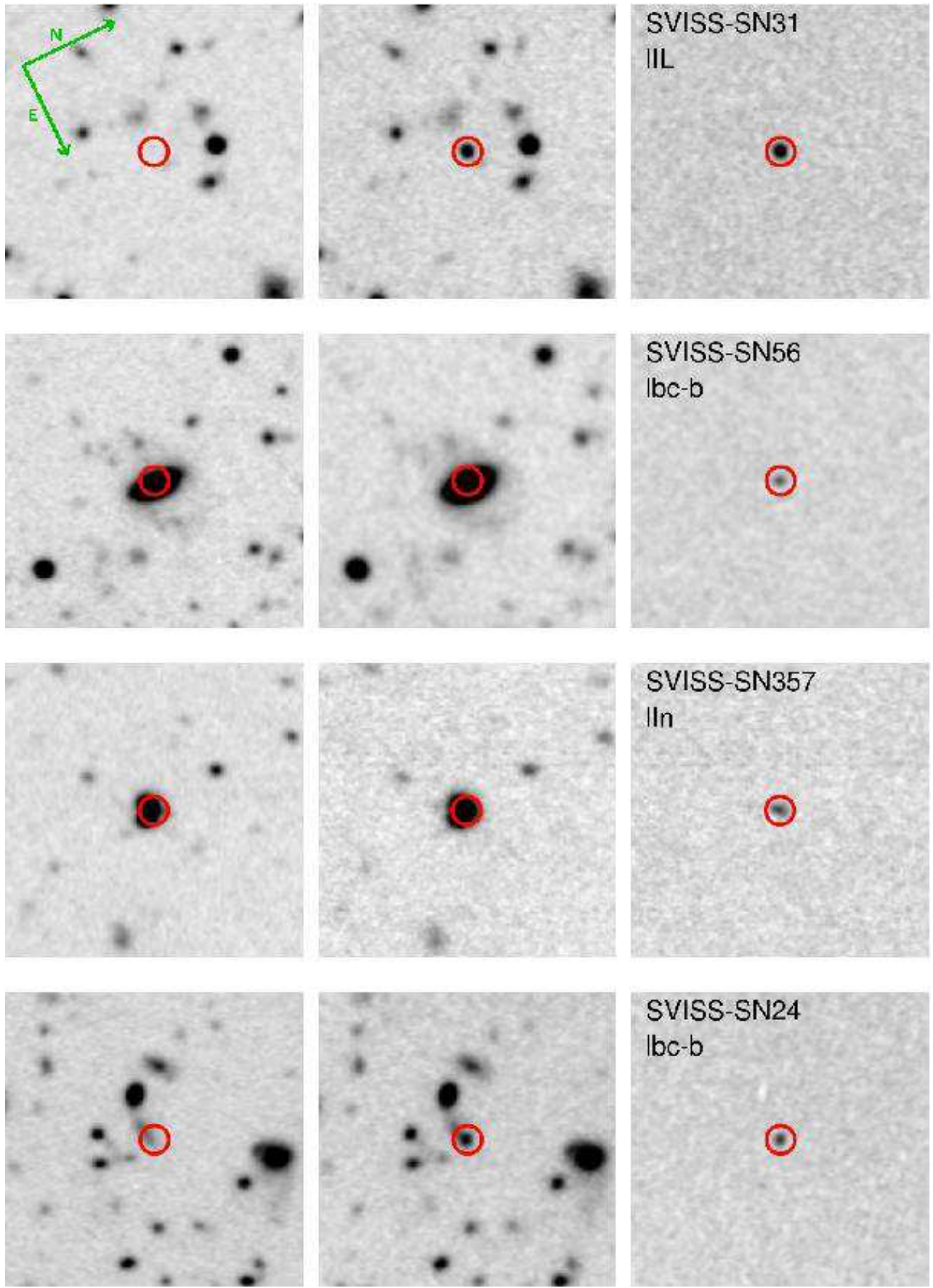}
\caption{R-band image cutouts for the supernova candidates SVISS-SN31, SVISS-SN56,
SVISS-SN357 and SVISS-SN24. See Figure~\ref{fig:sta1} for descriptions on the
different panels and marks. The most likely redshifts (from the typing code) for 
these SNe are (starting at the top) 0.12, 0.57, 1.4 and 0.81, respectively. Note 
that the host galaxy for SN357 is blended with a foreground galaxy, the selected 
host is the lowermost of the two galaxies.}
\label{fig:sta4}
\end{figure*}
\begin{figure*}
\centering
\includegraphics[width=0.91\textwidth]{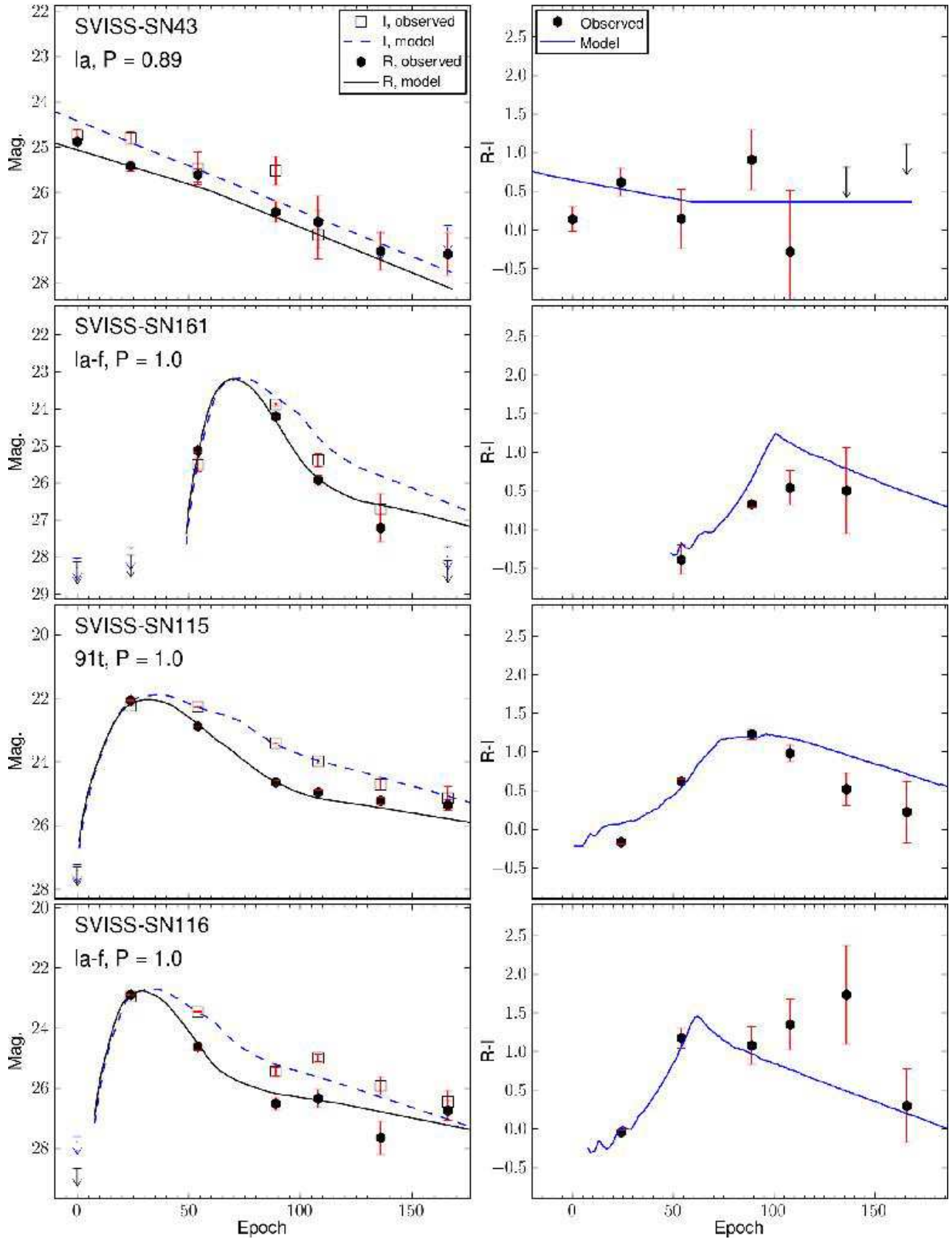}
\caption{Observed and fitted light curves for the supernova candidates
SVISS-SN43, SVISS-SN161, SVISS-SN115, and SVISS-SN116. The left-hand panels show the observed
$R$ (solid hexagons) and $I$ (squares) light curves along with the
best-fit light curve of the most likely supernova subtype, solid
(black) for $R$ and dashed (blue) for I. The error bars given for
the observations are based on the photometric accuracy simulations
described in Section~\ref{sec:phot}. If the source
is non-detected (i.e., has an estimated magnitude error of more than 1),
a magnitude lower limit is given. The right-hand panels show the $R-I$
colour evolution. For the $R-I$ plot, the limit symbols indicate that
the source was only detected in one of the bands and a lower or upper
limit is given (this figure is available in colour in the electronic
version of the article).}
\label{fig:lc1} 
\end{figure*} 
\begin{figure*}
\centering 
\includegraphics[width=0.91\textwidth]{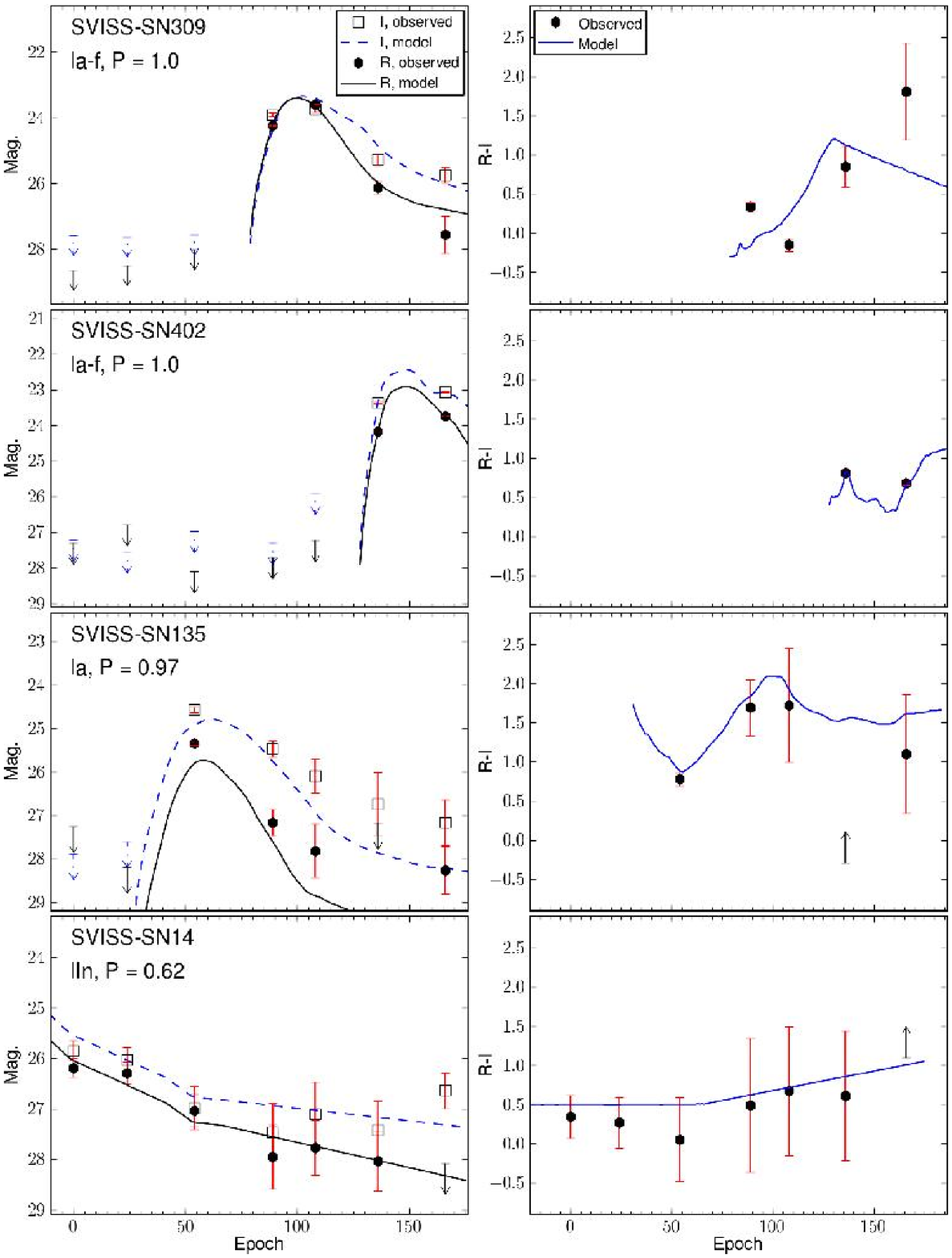}
\caption{Observed and fitted light curves for the supernova candidates
SVISS-SN309, SVISS-SN402, SVISS-SN135, and SVISS-SN14. See Figure~\ref{fig:lc1}
for descriptions on the different panels and marks.}
\label{fig:lc2} 
\vspace{1cm}
\end{figure*} 
\begin{figure*} 
\centering
\includegraphics[width=0.91\textwidth]{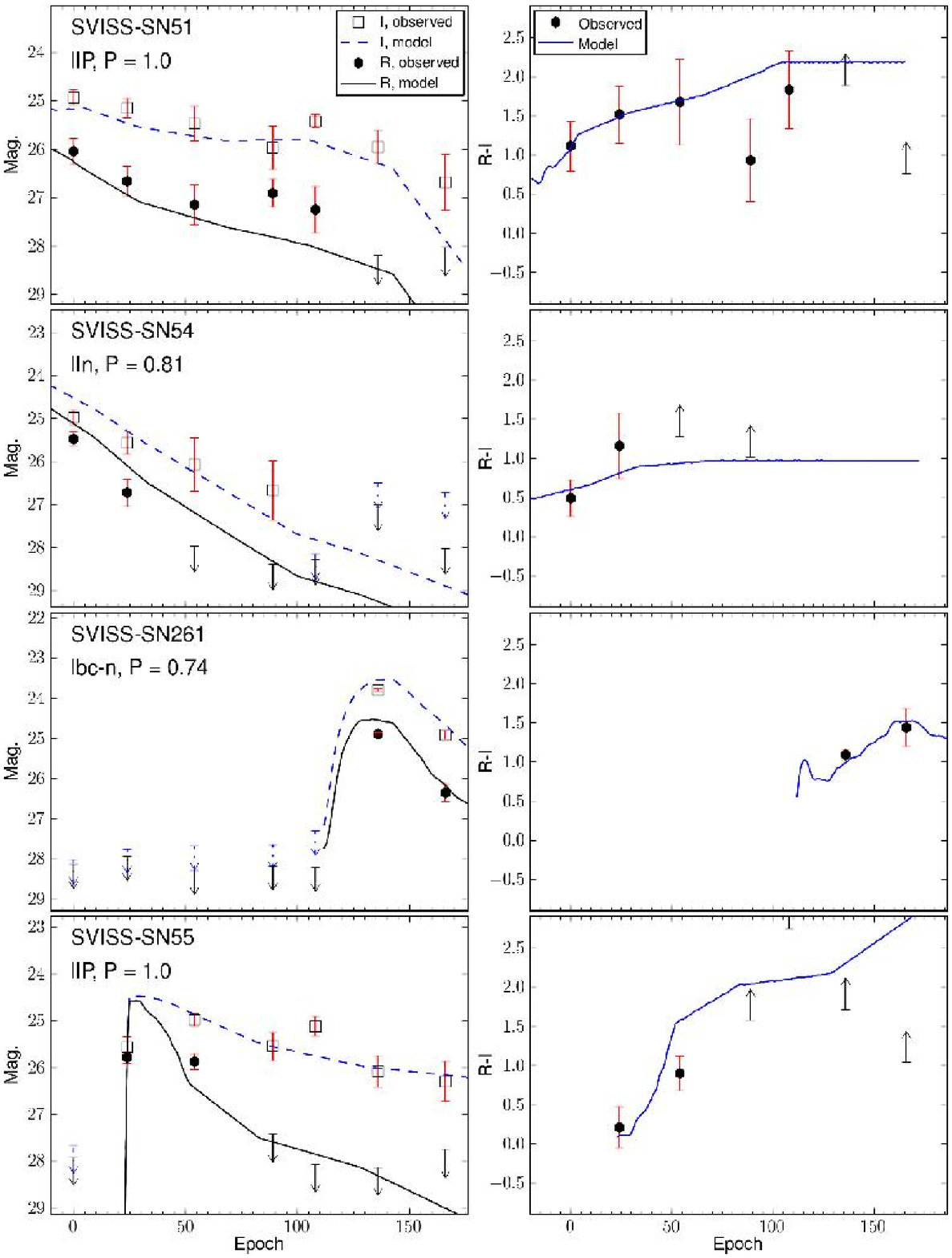}
\caption{Observed and fitted light curves for the
supernova candidates SVISS-SN51, SVISS-SN54, SVISS-SN261, and SVISS-SN55. See
Figure~\ref{fig:sta1} for descriptions on the different panels
and marks.} 
\label{fig:lc3} 
\vspace{1cm}
\end{figure*} 
\begin{figure*}
\centering 
\includegraphics[width=0.91\textwidth]{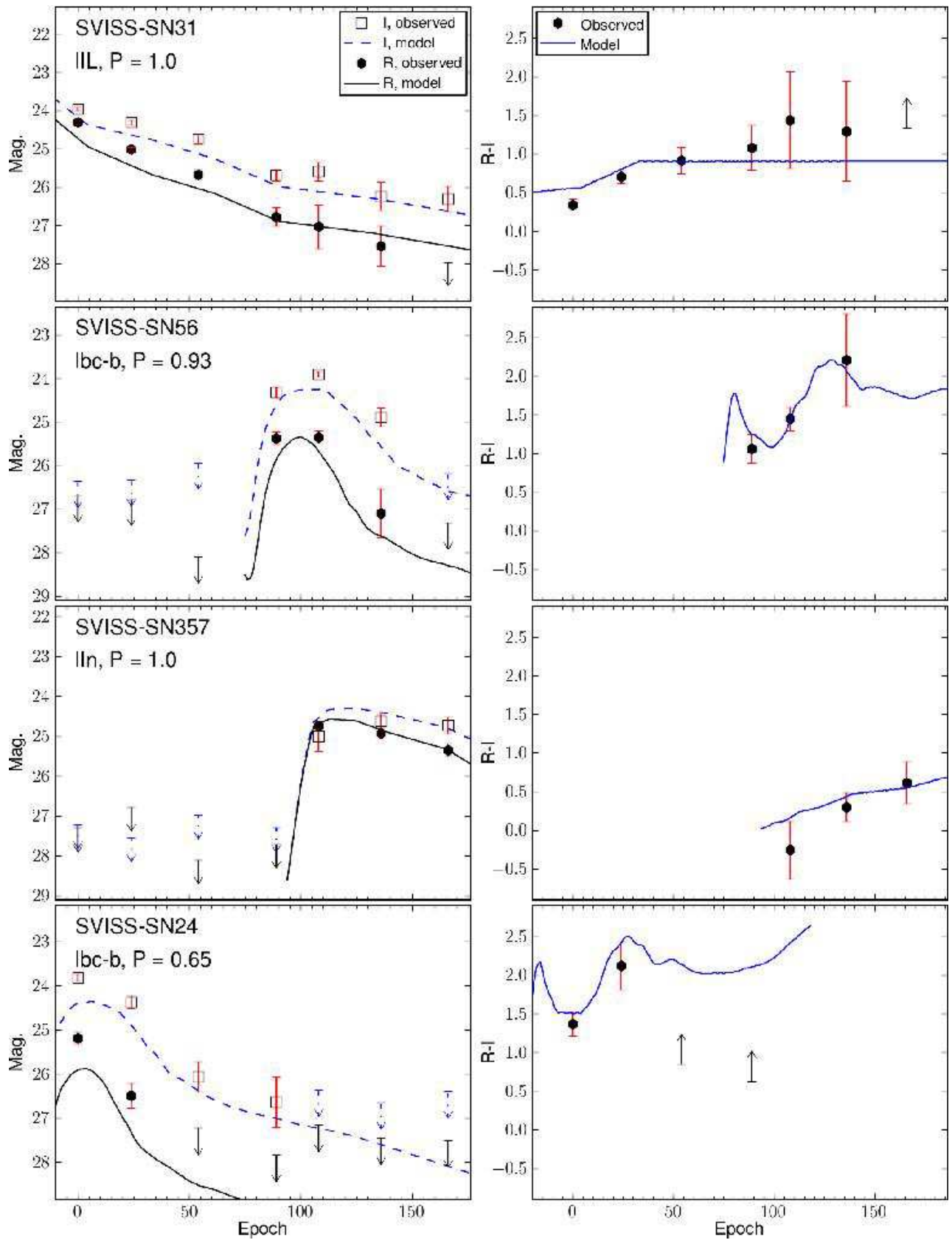}
\caption{Observed and fitted light curves for the supernova
candidates SVISS-SN31, SVISS-SN56, SVISS-SN357, and SVISS-SN24. See Figure~\ref{fig:sta1}
for descriptions on the different panels and marks.} 
\label{fig:lc4}
\vspace{1cm}
\end{figure*} 

\section{Results}
\label{sec:cands}
After applying the photometric rejection criterion described in
Section~\ref{sec:phot}, we ended up with 115 supernova candidates
that are typed using our code.  The subtype probabilities for the
two main types are co-added, yielding $P(TN)$ and $P(CC)$, for each
candidate; the candidate was assigned the main type with the
highest probability. $\mathcal{L}_{max}$ for the candidate is the
maximum evidence among the subtypes belonging to the chosen main
type. The evidence was then compared to the threshold evidence,
$\mathcal{L}_{99.9,\mathcal{T}}$ (see Section~\ref{sec:rejano}),
and candidates with lower $\mathcal{L}_{max}$ than the threshold were
rejected.  The remaining number of candidates at this point was 54,
the rejected objects were also manually checked to ensure that the
technique was working.

We then investigated each of the 54 possible supernovae manually,
making an overall assessment -- using both the images, light curve
and classification -- of whether the source is likely a supernova
 or something else. This investigation was performed by six of us (JM,
TD, G\"O, LMT, JS, and SM), and we individually and separately rated
each candidate. The manual rejection step allows us to remove
spurious transient sources from imperfect image subtraction. These
residuals can be present in both bands and in multiple epochs, which
enabled them to get through the earlier rejection steps. During the
manual inspection of both images and light curves, the residuals
can be discovered. The choice of having several independent inspectors
minimises the risk of errors being made. The rejected residuals have a
number of properties in common: they all have a bright host galaxy,
in general they tend to move (by 0.1-1 pixels) from epoch to epoch,
most of them have low $\mathcal{L}_{max}$ and have erratic light curves
combined with large errors, most of them also show a negative residual
close to the source in at least one epoch/filter. Approximately 20
candidates are considered to be spurious subtraction residuals.
A small number of candidates ($<5$) is found very close to the
edge of the images and are related to the the higher noise level at
the edges. A reanalysis of the photometric errors in these regions
showed that the sources did not fulfil our photometric criteria,
we consequently decided to reject these sources. This decision also
means that the efficient field of view will be somewhat smaller, a
trade-off we are willing to accept to make the detection efficiency
of the survey constant over the full field. After this rejection step
we were left with 31 transient sources that we believe are real. The
world coordinates, $RI$ light curves and errors for the 31 transient
sources are given in the appendix (available in the electronic version
of the article).

At this point we also rejected possible AGN that contaminate our
sample. To do this we used subtracted frames with a two-year difference
in time (the reference epoch from August 2003 and the control epoch
obtained in November 2005 to January 2006). If a candidate showed
variability over this time span, it is very unlikely to be a supernova,
and we rejected it (with the exception of one candidate, discussed in
Section~\ref{sec:ccresults}). Of the 31 candidates 15 were rejected because
of this, but note that some of these are likely not AGN but rather some
other non-SN transient object (at least one of them has a light curve
and colour consistent with a variable star). It should be noted that
the AGN rejection scheme will not allow us to get rid of AGN that show
no variation over the two-year baseline and have a SN-like light curve
during the search period; however, the number of AGN fulfilling this
criterion is estimated to be small (also see Section~\ref{sec:pecand}).

The final sample of 16 supernovae is presented in
Table~\ref{table:sne}, cutout images for each SN are available in
Figures~\ref{fig:sta1} through \ref{fig:sta4}. Figures ~\ref{fig:lc1}
through \ref{fig:lc4} show the observed and best-fit light curves for
all of them. The best-fit light curves are here obtained as described
in Section~\ref{sec:typingend}, it should be noted that the typing
method is not based on finding the best-fitting (in the $\chi^2$
sense) parameter set but rather on finding the type with the closest
matching parameter space given the priors. Nevertheless, the fitted
light curves provide a quick estimate on how good the evidence for
the most probable type is, a fitted light curve far from the observed
curve usually also means that the evidence is low. The probability
($P$) might be close to one even for low $\mathcal{L}_{max}$, this
reflects that the chosen type might be improbable in the absolute
sense but that the type is the best among the available alternatives.

\begin{figure}
\centering
\includegraphics[width=8cm]{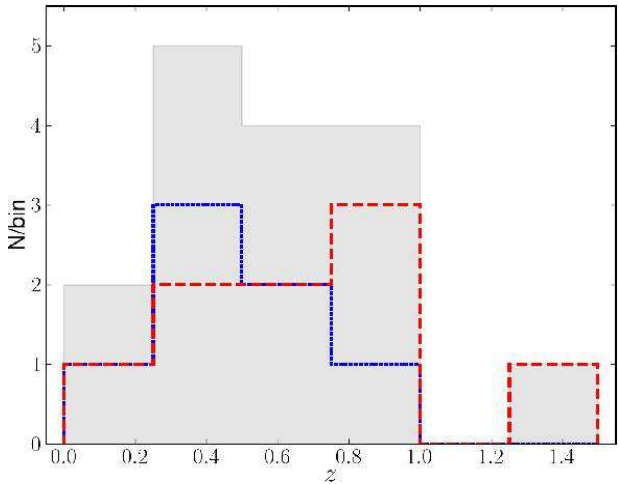}
\caption{Redshift distribution of the supernovae detected in 
the SVISS ELAIS-S1 field using a $z$ bin size of 0.25. 
The redshifts used are the resulting 
best-fit redshifts from the supernova typing code. The grey, 
filled bars show the total supernova counts, the blue (dotted)
lines show the distribution of thermonuclear SNe and the red 
(dashed) lines the distribution of core-collapse SNe
(this figure is available in colour in the electronic
  version of the article).}
\label{fig:zdist}
\end{figure}

The redshift distribution, based on the fitted redshifts from the
typing, of the supernovae is shown in Figure~\ref{fig:zdist}. The joint
distribution (the filled bars in the plot) has a mean of 0.58. The
thermonuclear supernovae are on average found at lower redshifts compared to 
those of the
core-collapse (with a mean of 0.51 compared to 0.64). The requirement
that the supernovae should be detected in both filters makes this
expected because thermonuclear SNe become very faint at redshifts
$\gtrsim 0.8$ in the $R$-band owing to k-corrections.
\subsection{Thermonuclear supernovae} 
\label{sec:tnresults}
Seven of the 16 supernovae in the survey are found to have a
thermonuclear origin. Based on the simulation results, approximately
0.7 (10\% of the total number) could be misclassified core-collapse 
supernovae. Inspecting the light curves and looking at the
fit quality ($P(TN)$ and $\mathcal{L}_{best}$) we can see that one
of the seven supernovae (SVISS-SN43) has a somewhat insecure
typing. The probability, $P(TN)$, is well below 1.0, which means
that it has a significant chance of being a core-collapse supernova.

Three of the supernovae (SVISS-SN161, SVISS-SN115, and SVISS-SN309)
have low evidences ($\mathcal{L}_{best}$). They all have probabilities
close to 1.0, however, indicating that they, while not being perfectly
matched by the available templates and priors, are optimally fit by
the thermonuclear templates. The addition of a more general stretch
prior as described in Section~\ref{sec:typing} would most probably
allow the templates to match the observed light curves better.

\subsection{Core-collapse supernovae} 
\label{sec:ccresults}
Nine of the supernovae are found to be best matched by a core-collapse
template light curve. The simulation results indicate that none
of these are likely to be misclassified thermonuclear supernovae
(2.4\% of the total number). A look at the actual fit qualities
for the supernovae show that three of them (SVISS-SN14, SVISS-SN261,
and SVISS-SN24) have probability values lower than 0.8 and accordingly have a
definite probability of being misclassified thermonuclear SNe. It is
worth noting that for two of them the most likely
subtype is Ibc, the CC subtypes with light curves
most similar to the thermonuclear subtypes. They are also all quite
faint and detected only in 2--3 epochs. However, we
have no reason to mistrust the likelihoods computed for these SNe and
thus conclude that they are likely of core-collapse origin.

One of the CC supernovae,
SVISS-SN56, has a low evidence (also visible in the light curve
fit, see Figure~\ref{fig:lc4}), indicating that none of the supernova
templates fitted the data well. Including more core-collapse supernova
templates, improving the ones used or adding a colour prior could
make it possible to get a better typing result for this
supernova.

As discussed in Section~\ref{sec:photz}, for three CC SNe no photometric
redshift prior was used. In the case of SVISS-SN14 this is because 
the most likely redshift (0.36) for the found subtype, IIn, is
significantly different from the photometric redshift of the host galaxy
(0.76). This type and redshift is found to be the best fit even when the
photometric redshift is used as a prior. The reason for this mismatch
could be either that the host redshift is incorrect (approximately 10\% of
the photometric redshifts are so called catastrophic redshifts, off by
more than 0.3, again see Section~\ref{sec:photz}), or that there is another very
faint and undetected galaxy that is the real host of the SN. A third
possibility is that the supernova is of a peculiar type that is not
well-modelled by the template light curves.

For SVISS-SN261, the closest galaxy is quite far away in angular
distance and it has a redshift of 1.64. Using this as a redshift prior
results in the SNe being placed at a redshift of 0.37 as a Ibc--bright
subtype.  There is just no possible SN type that can give the observed
light curve at such a high redshift. Also, using a flat prior on the redshift
results in the same type and redshift. We believe that the true host of this
SN is undetected, it could possibly be a low-surface brightness, but
star-forming galaxy at intermediate redshift.

The third of the three ``hostless'' SNe is a complicated case.
SVISS-SN31 is located very close to a very faint galaxy with a
photometric redshift of 2.97. At that redshift all of the supernova
types under consideration would be undetectable (owing to distance dimming
and k-corrections). Using a flat prior on the redshift results in a type
IIL SN at $z=0.12$. Other nearby galaxies (in angular distance) have
been considered as hosts, but none of them has a redshift lower than
$\sim$0.6. The second-closest galaxy has a photometric redshift of 1.04
(highly uncertain, the 68\% confidence interval for this measurement is
0.96--1.58), and if this is used as a prior for the supernova typing the
resulting type is a IIn SN at $z=1.92$. However, the evidence for this
type is lower than the IIL solution found with a flat prior. Redshift
priors from the other nearby galaxies result in even worse evidence for
the most likely types. It should also be noted that the main type (CC)
of the supernova is the same for all the tested redshift priors. Based
on the evidence comparisons, host galaxy properties, and angular
distance we conclude that the typing without a redshift prior provides the
most believable solution. The mismatch with the host redshift could be
explained by, as for SVISS-SN14 above, the host galaxy redshift being
catastrophically wrong, a faint undetected host galaxy, or a peculiar SN
type.

\subsection{A peculiar SN candidate}
\begin{figure}
\centering
\label{sec:pecand}
\includegraphics[width=8cm]{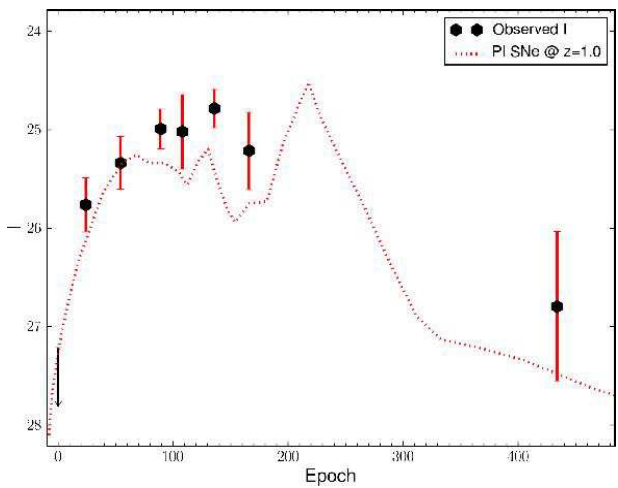}
\caption{$I$-band light curve of SVISS-T167. Also shown
  is a $B$-band light curve for a pair
  instability SNe shifted to $z=1$ (line). Details of how the
  comparison light curve is constructed can be found in the
  text (this figure is available in colour in the electronic version
  of the article).}
\label{fig:hypercand} 
\end{figure}
\begin{figure*}
\centering
\includegraphics[width=0.9\textwidth]{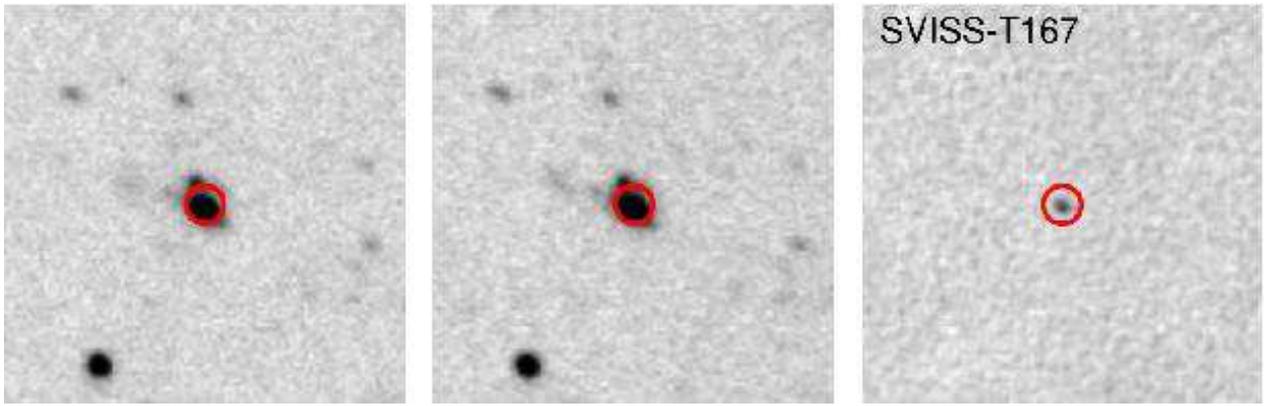}
\caption{$I$-band image cutouts, approximately $20\times 20$\arcsec large
of the transient of unknown type, SVISS-T167. 
The leftmost panel shows the reference image,
middle panel shows the peak brightness (observed) epoch and the
rightmost panel shows the subtracted image at peak brightness. The
(red) circle marks the location of the transient as detected in the
subtracted frame.}
\label{fig:hyperstamp} 
\end{figure*}
In Section~\ref{sec:cands} we described how the AGN that likely contaminate
the sample are dealt with. One of the variable sources showing
variability over a two year period, SVISS-T167 has an intriguing light curve
that could possibly come from a supernova with a very extended light
curve. However, the supernova would in that case have to be of a
peculiar type; our typing code was unable to find the most likely
type (i.e. $\mathcal{L}_{best}<\mathcal{L}_{99.9}$, as defined in
Section~\ref{sec:rejano}). The light curve of this candidate is
shown in Figure~\ref{fig:hypercand} and an image cutout is shown in
Figure~\ref{fig:hyperstamp}. For it to be even remotely similar to a
normal supernova light curve, it needs to be situated at high redshift
where cosmological time dilation will stretch the light curve by a
factor of 2 (or more).

Comparing this to the light curves of bright peculiar supernovae we
find that the light curve  is marginally consistent with that of
a pair instability (PI) supernovae at high redshift, something
that has also been suggested by \citet{2009Natur.462..624G} for
SN~2007bi and by \citet{2007Natur.450..390W} for
SN~2006gy. PI SNe have slowly evolving light curves, sometimes
with a secondary peak; the slow light curve decline of our candidate
indicates that it could be similar to such SNe. The PI light curve
in Figure~\ref{fig:hypercand} was derived by taking a model rest
frame $B$ band light curves for a 110 $M_{\odot}$ progenitor from the
supplemental information available for \citet{2007Natur.450..390W}
and shifting it to $z=1.0$, applying time dilation and distance
modulus. We did not attempt any detailed k-correction, but it should be
noted that the correction term should be fairly small because the rest frame
$B$ central wavelength at $z=1.0$ is close to the central wavelength
of the $I$ filter. We also compared the light curve to that of
SN~2010gx \citep{2010arXiv1008.2674P}, an extremely luminous type
Ic-like supernova. We found that the very extended light curve and
colour evolution of our transient is impossible to match with the
2010gx light curve, even when shifting it to higher redshift.

There is no strong evidence for our candidate being a supernova of
this type, it is still a distinct possibility that the source is
actually an AGN because the position of the source is right in the
centre of its host galaxy. The photometric redshift for the host
galaxy is not constrained, which might point towards it being an AGN
(our photometric redshift code does not include AGN SED templates),
but could also mean that it is a high-redshift galaxy ($z\gtrsim 1$)
where the lack of near infrared data renders the photometric redshift technique
unreliable. We excluded the source from the SN sample presented
in Table~\ref{table:sne}. To discover the true identity of the object,
follow-up observations of its host galaxy are needed.

\section{Summary and discussion}
\label{sec:sum}
We have presented light curves and typing of the 16 supernovae
discovered by SVISS. Using photometric redshifts estimated from host
galaxy photometry, we typed the SNe into thermonuclear or core-collapse
events. The typing uses spectral and light curve templates for
nine different subtypes and is based on a Bayesian fitting method.
From studying Monte-Carlo-simulated SN light curves and a small
sample of SDSS SN light curves, we conclude that the typing code yields
reliable results considering that we only have photometric redshifts
with accuracies in the order of 0.1, only one colour, and a limited
number of epochs. We find that probably not more than  5--10\% of the
supernovae are misclassified.  The misclassification ratios are not
symmetric with respect to the two main types, a CC SN is more likely
to be misclassified into a TN SN than the other way around.

Comparing our typing quality with that of other typing codes is not
straight-forward because the input data and tests of the codes
vary considerably. \citet{2010arXiv1008.1024K} try to remedy this by
offering a test sample of SN light curves and a framework in which
different typing codes can be compared. Currently, our code is written
specifically to deal with the data from SVISS, trying the code out on
the test sample is thus outside the scope of this paper. Nevertheless,
we can compare the typing quality estimated from our Monte-Carlo-simulated 
light curves with the typing qualities reported in
\citet{2010arXiv1008.1024K}. The Ia \emph{efficiency}, which in our
paper is equal to $(N_{app}(TN)-N_{TN\rightarrow CC})/N_{tot}(TN)$,
and Ia \emph{pureness}, $(N_{app}(TN)-N_{TN\rightarrow
CC})/(N_{app}(TN)-N_{TN\rightarrow CC} + N_{CC\rightarrow TN})$ in
our notation, are two of the concepts used to describe the quality
in \citet{2010arXiv1008.1024K}. Our overall efficiency and pureness
values for the thermonuclear SNe, 0.97 and 0.75, respectively, are
comparable to the results obtained by the other typing codes, but with the 
caveat that we did not use the same test sample of supernovae as
these authors.

Out of the 16 SNe, $9^{+0.7}_{-0.2}$ are core-collapse supernovae, and
$7^{+0.2}_{-0.6}$ are thermonuclear; only the systematic typing errors
derived from the misclassification ratios are given here. The mean
redshift for the entire sample is 0.58 with the core-collapse SNe
being at slightly higher redshifts than the thermonuclear ones. In
a supernova survey of this type a large number of spurious sources
will be detected and methods must be devised to safely reject
interlopers without compromising the SN detection efficiency. We used
a combination of automatic (SExtractor, signal-to-noise rejection,
SN typing rejection) and manual (by-eye inspection of image frames
by multiple independent observers, by-eye inspection of light curves
by multiple independent observers) rejection methods to ensure
that the final SN sample is as pure as possible.

In the specific case of AGN interlopers we find that the use
of an additional observing epoch displaced from the search/reference
epochs by at least one year makes it possible to limit the amount of
contamination further. We also find that most of the AGN will also
be rejected in the supernova typing step, because the fitting code is
simply not able to fit the SN templates to their colour and evolution.
At least one of the rejected candidates could possibly be a supernova
of an extremely rare type. This particular source was still detectable,
albeit barely, one year after our search period ended. The extremely
extended light curve is comparable to that of pair instability SNe
\citep{2007Natur.450..390W} at high redshifts.

The SVISS results show that SN imaging searches combined with
photometric typing can be used to find and characterise SNe at high
redshifts. While the basic strategy used in this project does work,
there are a few improvements that could be considered. One of the main
uncertainties in the SN typing is the use of photometric redshifts. 
Securing more precise redshifts for the host galaxies would make the typing 
considerably
more accurate. These redshifts could be in the form of well-calibrated
photometric or spectroscopic redshifts. Future pencil beam surveys
targeting high-redshift SNe should thus be aimed at regions
with good spectral and multiwavelength coverage. We found the cadence of
the search to be more or less optimal, given the constraint on total
observational time, but with more frequent observations it should be
possible to detect the UV breakout phase of core collapse SNe as well
\citep[e.g.,][]{2011ApJS..193...20T}. Observations in more than one filter is
mandatory to be able to do photometric typing, but we
also found it to be quite important for rejecting spurious objects in
the detection step. Adding more filters will break some of the
degeneracies in the typing, but the typing accuracy is also limited by
the quality of the templates.

A full analysis of supernova and star-formation rates based on our
observations will be presented in a forthcoming paper (Melinder et al.,
in prep). The work presented in this paper will help us estimate
the uncertainties of the found rates.  To calculate proper rates,
information about the detection efficiency and the enforced photometric
limits has to be taken into account.  Nevertheless, a quick look at
the raw numbers of discovered SNe of the two types compared to 
Monte-Carlo-simulations of detectable SNe using supernova rates based on
local observations and star-formation rates \citep{1999A&A...350..349D}
shows that our numbers are consistent with the expected rates. We
can also compare the range and distribution in redshift of our
supernovae to that of other surveys. Our sample of thermonuclear
SNe are not really unique in this respect \citep[see, e.g.,
][]{2006AJ....132.1126N,2007MNRAS.382.1169P,2008ApJ...681..462D}, but
it should be noted that the survey methodology we used is different from
most other surveys where the requirement of spectra means that some of
the fainter events may be missed. For the core-collapse SNe our sample
is similar to the observations presented in \citet{2004ApJ...613..189D}
with 17 CC SNe at $z<0.7$. We detect 9 CC SNe out to $z=1.4$, with
most of them at $z\sim$0.8, and this is therefore one of the deepest survey to date
for CC SNe.

\begin{acknowledgements} We thank
A.~Pastorello for comments on the peculiar SN candidate and the SDSS
supernova team for the use of the unpublished SN~Ia data. We also thank
A.~Jerkstrand for helpful discussions regarding parallelisation of
the typing code. We thank the anonymous referee for some helpful comments
that helped us improve the paper. The observations were made in service mode at the
VLT/VIMOS instrument at Paranal Observatories.

We are grateful for financial support from the Swedish Research 
Council. S.M. and J.M. acknowledge financial support from the Academy 
of Finland (project:8120503). 

\end{acknowledgements}
\bibliographystyle{aa}
\bibliography{database}

\begin{thebibliography}{88}
\expandafter\ifx\csname natexlab\endcsname\relax\def\natexlab#1{#1}\fi

\bibitem[{{Alard}(2000)}]{2000A&AS..144..363A}
{Alard}, C. 2000, \aaps, 144, 363, (A00)

\bibitem[{{Alard} \& {Lupton}(1998)}]{1998ApJ...503..325A}
{Alard}, C. \& {Lupton}, R.~H. 1998, \apj, 503, 325

\bibitem[{{Amanullah} {et~al.}(2010){Amanullah}, {Lidman}, {Rubin}, {Aldering},
  {Astier}, {Barbary}, {Burns}, {Conley}, {Dawson}, {Deustua}, {Doi}, {Fabbro},
  {Faccioli}, {Fakhouri}, {Folatelli}, {Fruchter}, {Furusawa}, {Garavini},
  {Goldhaber}, {Goobar}, {Groom}, {Hook}, {Howell}, {Kashikawa}, {Kim}, {Knop},
  {Kowalski}, {Linder}, {Meyers}, {Morokuma}, {Nobili}, {Nordin}, {Nugent},
  {{\"O}stman}, {Pain}, {Panagia}, {Perlmutter}, {Raux}, {Ruiz-Lapuente},
  {Spadafora}, {Strovink}, {Suzuki}, {Wang}, {Wood-Vasey}, {Yasuda}, \&
  {Supernova Cosmology Project}}]{2010ApJ...716..712A}
{Amanullah}, R., {Lidman}, C., {Rubin}, D., {et~al.} 2010, \apj, 716, 712

\bibitem[{{Astier} {et~al.}(2006){Astier}, {Guy}, {Regnault}, {Pain},
  {Aubourg}, {Balam}, {Basa}, {Carlberg}, {Fabbro}, {Fouchez}, {Hook},
  {Howell}, {Lafoux}, {Neill}, {Palanque-Delabrouille}, {Perrett}, {Pritchet},
  {Rich}, {Sullivan}, {Taillet}, {Aldering}, {Antilogus}, {Arsenijevic},
  {Balland}, {Baumont}, {Bronder}, {Courtois}, {Ellis}, {Filiol}, {Gon{\c
  c}alves}, {Goobar}, {Guide}, {Hardin}, {Lusset}, {Lidman}, {McMahon},
  {Mouchet}, {Mourao}, {Perlmutter}, {Ripoche}, {Tao}, \&
  {Walton}}]{2006A&A...447...31A}
{Astier}, P., {Guy}, J., {Regnault}, N., {et~al.} 2006, \aap, 447, 31

\bibitem[{{Barbary} {et~al.}(2010){Barbary}, {Aldering}, {Amanullah},
  {Brodwin}, {Connolly}, {Dawson}, {Doi}, {Eisenhardt}, {Faccioli}, {Fadeyev},
  {Fakhouri}, {Fruchter}, {Gilbank}, {Gladders}, {Goldhaber}, {Goobar},
  {Hattori}, {Hsiao}, {Huang}, {Ihara}, {Kashikawa}, {Koester}, {Konishi},
  {Kowalski}, {Lidman}, {Lubin}, {Meyers}, {Morokuma}, {Oda}, {Panagia},
  {Perlmutter}, {Postman}, {Ripoche}, {Rosati}, {Rubin}, {Schlegel},
  {Spadafora}, {Stanford}, {Strovink}, {Suzuki}, {Takanashi}, {Tokita},
  {Yasuda}, \& {Supernova Cosmology Project}}]{2010arXiv1010.5786B}
{Barbary}, K., {Aldering}, G., {Amanullah}, R., {et~al.} 2010, ArXiv e-prints

\bibitem[{{Baron} {et~al.}(2004){Baron}, {Nugent}, {Branch}, \&
  {Hauschildt}}]{2004ApJ...616L..91B}
{Baron}, E., {Nugent}, P.~E., {Branch}, D., \& {Hauschildt}, P.~H. 2004, \apjl,
  616, L91

\bibitem[{{Barris} \& {Tonry}(2004)}]{2004ApJ...613L..21B}
{Barris}, B.~J. \& {Tonry}, J.~L. 2004, \apjl, 613, L21

\bibitem[{{Bazin} {et~al.}(2009){Bazin}, {Palanque-Delabrouille}, {Rich},
  {Ruhlmann-Kleider}, {Aubourg}, {Le Guillou}, {Astier}, {Balland}, {Basa},
  {Carlberg}, {Conley}, {Fouchez}, {Guy}, {Hardin}, {Hook}, {Howell}, {Pain},
  {Perrett}, {Pritchet}, {Regnault}, {Sullivan}, {Antilogus}, {Arsenijevic},
  {Baumont}, {Fabbro}, {Le Du}, {Lidman}, {Mouchet}, {Mour{\~a}o}, \&
  {Walker}}]{2009A&A...499..653B}
{Bazin}, G., {Palanque-Delabrouille}, N., {Rich}, J., {et~al.} 2009, \aap, 499,
  653

\bibitem[{{Berta} {et~al.}(2008){Berta}, {Rubele}, {Franceschini}, {Held},
  {Rizzi}, {Rodighiero}, {Cimatti}, {Dias}, {Feruglio}, {La Franca},
  {Lonsdale}, {Maiolino}, {Matute}, {Rowan-Robinson}, {Sacchi}, \&
  {Zamorani}}]{2008A&A...488..533B}
{Berta}, S., {Rubele}, S., {Franceschini}, A., {et~al.} 2008, \aap, 488, 533

\bibitem[{{Bertin} \& {Arnouts}(1996)}]{1996A&AS..117..393B}
{Bertin}, E. \& {Arnouts}, S. 1996, \aaps, 117, 393

\bibitem[{{Botticella} {et~al.}(2008){Botticella}, {Riello}, {Cappellaro},
  {Benetti}, {Altavilla}, {Pastorello}, {Turatto}, {Greggio}, {Patat},
  {Valenti}, {Zampieri}, {Harutyunyan}, {Pignata}, \&
  {Taubenberger}}]{2008A&A...479...49B}
{Botticella}, M.~T., {Riello}, M., {Cappellaro}, E., {et~al.} 2008, \aap, 479,
  49

\bibitem[{{Brown} {et~al.}(2007){Brown}, {Dessart}, {Holland}, {Immler},
  {Landsman}, {Blondin}, {Blustin}, {Breeveld}, {Dewangan}, {Gehrels},
  {Hutchins}, {Kirshner}, {Mason}, {Mazzali}, {Milne}, {Modjaz}, \&
  {Roming}}]{2007ApJ...659.1488B}
{Brown}, P.~J., {Dessart}, L., {Holland}, S.~T., {et~al.} 2007, \apj, 659, 1488

\bibitem[{{Calzetti} {et~al.}(2000){Calzetti}, {Armus}, {Bohlin}, {Kinney},
  {Koornneef}, \& {Storchi-Bergmann}}]{2000ApJ...533..682C}
{Calzetti}, D., {Armus}, L., {Bohlin}, R.~C., {et~al.} 2000, \apj, 533, 682

\bibitem[{{Cappellaro} {et~al.}(1997){Cappellaro}, {Turatto}, {Tsvetkov},
  {Bartunov}, {Pollas}, {Evans}, \& {Hamuy}}]{1997A&A...322..431C}
{Cappellaro}, E., {Turatto}, M., {Tsvetkov}, D.~Y., {et~al.} 1997, \aap, 322,
  431

\bibitem[{{Cardelli} {et~al.}(1989){Cardelli}, {Clayton}, \&
  {Mathis}}]{1989ApJ...345..245C}
{Cardelli}, J.~A., {Clayton}, G.~C., \& {Mathis}, J.~S. 1989, \apj, 345, 245

\bibitem[{{Coleman} {et~al.}(1980){Coleman}, {Wu}, \&
  {Weedman}}]{1980ApJS...43..393C}
{Coleman}, G.~D., {Wu}, C., \& {Weedman}, D.~W. 1980, \apjs, 43, 393

\bibitem[{{Dahl{\'e}n} \& {Fransson}(1999)}]{1999A&A...350..349D}
{Dahl{\'e}n}, T. \& {Fransson}, C. 1999, \aap, 350, 349

\bibitem[{{Dahl{\'e}n} {et~al.}(2004){Dahl{\'e}n}, {Fransson}, {{\"O}stlin}, \&
  {N{\"a}slund}}]{2004MNRAS.350..253D}
{Dahl{\'e}n}, T., {Fransson}, C., {{\"O}stlin}, G., \& {N{\"a}slund}, M. 2004,
  \mnras, 350, 253

\bibitem[{{Dahlen} {et~al.}(2010){Dahlen}, {Mobasher}, {Dickinson}, {Ferguson},
  {Giavalisco}, {Grogin}, {Guo}, {Koekemoer}, {Lee}, {Lee}, {Nonino}, {Riess},
  \& {Salimbeni}}]{2010ApJ...724..425D}
{Dahlen}, T., {Mobasher}, B., {Dickinson}, M., {et~al.} 2010, \apj, 724, 425

\bibitem[{{Dahlen} {et~al.}(2008){Dahlen}, {Strolger}, \&
  {Riess}}]{2008ApJ...681..462D}
{Dahlen}, T., {Strolger}, L.-G., \& {Riess}, A.~G. 2008, \apj, 681, 462

\bibitem[{{Dahlen} {et~al.}(2004){Dahlen}, {Strolger}, {Riess}, {Mobasher},
  {Chary}, {Conselice}, {Ferguson}, {Fruchter}, {Giavalisco}, {Livio}, {Madau},
  {Panagia}, \& {Tonry}}]{2004ApJ...613..189D}
{Dahlen}, T., {Strolger}, L.-G., {Riess}, A.~G., {et~al.} 2004, \apj, 613, 189

\bibitem[{{D'Andrea} {et~al.}(2010){D'Andrea}, {Sako}, {Dilday}, {Frieman},
  {Holtzman}, {Kessler}, {Konishi}, {Schneider}, {Sollerman}, {Wheeler},
  {Yasuda}, {Cinabro}, {Jha}, {Nichol}, {Lampeitl}, {Smith}, {Atlee},
  {Bassett}, {Castander}, {Goobar}, {Miquel}, {Nordin}, {{\"O}stman}, {Prieto},
  {Quimby}, {Riess}, \& {Stritzinger}}]{2010ApJ...708..661D}
{D'Andrea}, C.~B., {Sako}, M., {Dilday}, B., {et~al.} 2010, \apj, 708, 661

\bibitem[{{Dessart} {et~al.}(2008){Dessart}, {Blondin}, {Brown}, {Hicken},
  {Hillier}, {Holland}, {Immler}, {Kirshner}, {Milne}, {Modjaz}, \&
  {Roming}}]{2008ApJ...675..644D}
{Dessart}, L., {Blondin}, S., {Brown}, P.~J., {et~al.} 2008, \apj, 675, 644

\bibitem[{{Di Carlo} {et~al.}(2002){Di Carlo}, {Massi}, {Valentini}, {Di
  Paola}, {D'Alessio}, {Brocato}, {Guidubaldi}, {Dolci}, {Pedichini},
  {Speziali}, {Li Causi}, {Caratti o Garatti}, {Cappellaro}, {Turatto},
  {Arkharov}, {Gnedin}, {Larionov}, {Benetti}, {Pastorello}, {Aretxaga},
  {Chavushyan}, {Vega}, {Danziger}, \& {Tornamb{\'e}}}]{2002ApJ...573..144D}
{Di Carlo}, E., {Massi}, F., {Valentini}, G., {et~al.} 2002, \apj, 573, 144

\bibitem[{{Dilday} {et~al.}(2010){Dilday}, {Smith}, {Bassett}, {Becker},
  {Bender}, {Castander}, {Cinabro}, {Filippenko}, {Frieman}, {Galbany},
  {Garnavich}, {Goobar}, {Hopp}, {Ihara}, {Jha}, {Kessler}, {Lampeitl},
  {Marriner}, {Miquel}, {Moll{\'a}}, {Nichol}, {Nordin}, {Riess}, {Sako},
  {Schneider}, {Sollerman}, {Wheeler}, {{\"O}stman}, {Bizyaev}, {Brewington},
  {Malanushenko}, {Malanushenko}, {Oravetz}, {Pan}, {Simmons}, \&
  {Snedden}}]{2010ApJ...713.1026D}
{Dilday}, B., {Smith}, M., {Bassett}, B., {et~al.} 2010, \apj, 713, 1026

\bibitem[{{Elmhamdi} {et~al.}(2003){Elmhamdi}, {Danziger}, {Chugai},
  {Pastorello}, {Turatto}, {Cappellaro}, {Altavilla}, {Benetti}, {Patat}, \&
  {Salvo}}]{2003MNRAS.338..939E}
{Elmhamdi}, A., {Danziger}, I.~J., {Chugai}, N., {et~al.} 2003, \mnras, 338,
  939

\bibitem[{{Frieman} {et~al.}(2008){Frieman}, {Bassett}, {Becker}, {Choi},
  {Cinabro}, {DeJongh}, {Depoy}, {Dilday}, {Doi}, {Garnavich}, {Hogan},
  {Holtzman}, {Im}, {Jha}, {Kessler}, {Konishi}, {Lampeitl}, {Marriner},
  {Marshall}, {McGinnis}, {Miknaitis}, {Nichol}, {Prieto}, {Riess}, {Richmond},
  {Romani}, {Sako}, {Schneider}, {Smith}, {Takanashi}, {Tokita}, {van der
  Heyden}, {Yasuda}, {Zheng}, {Adelman-McCarthy}, {Annis}, {Assef},
  {Barentine}, {Bender}, {Blandford}, {Boroski}, {Bremer}, {Brewington},
  {Collins}, {Crotts}, {Dembicky}, {Eastman}, {Edge}, {Edmondson}, {Elson},
  {Eyler}, {Filippenko}, {Foley}, {Frank}, {Goobar}, {Gueth}, {Gunn},
  {Harvanek}, {Hopp}, {Ihara}, {Ivezi{\'c}}, {Kahn}, {Kaplan}, {Kent},
  {Ketzeback}, {Kleinman}, {Kollatschny}, {Kron}, {Krzesi{\'n}ski}, {Lamenti},
  {Leloudas}, {Lin}, {Long}, {Lucey}, {Lupton}, {Malanushenko}, {Malanushenko},
  {McMillan}, {Mendez}, {Morgan}, {Morokuma}, {Nitta}, {Ostman}, {Pan},
  {Rockosi}, {Romer}, {Ruiz-Lapuente}, {Saurage}, {Schlesinger}, {Snedden},
  {Sollerman}, {Stoughton}, {Stritzinger}, {Subba Rao}, {Tucker}, {Vaisanen},
  {Watson}, {Watters}, {Wheeler}, {Yanny}, \& {York}}]{2008AJ....135..338F}
{Frieman}, J.~A., {Bassett}, B., {Becker}, A., {et~al.} 2008, \aj, 135, 338

\bibitem[{{Gal-Yam} {et~al.}(2009){Gal-Yam}, {Mazzali}, {Ofek}, {Nugent},
  {Kulkarni}, {Kasliwal}, {Quimby}, {Filippenko}, {Cenko}, {Chornock},
  {Waldman}, {Kasen}, {Sullivan}, {Beshore}, {Drake}, {Thomas}, {Bloom},
  {Poznanski}, {Miller}, {Foley}, {Silverman}, {Arcavi}, {Ellis}, \&
  {Deng}}]{2009Natur.462..624G}
{Gal-Yam}, A., {Mazzali}, P., {Ofek}, E.~O., {et~al.} 2009, \nat, 462, 624

\bibitem[{{Giacconi} {et~al.}(2001){Giacconi}, {Rosati}, {Tozzi}, {Nonino},
  {Hasinger}, {Norman}, {Bergeron}, {Borgani}, {Gilli}, {Gilmozzi}, \&
  {Zheng}}]{2001ApJ...551..624G}
{Giacconi}, R., {Rosati}, P., {Tozzi}, P., {et~al.} 2001, \apj, 551, 624

\bibitem[{{Gilliland} {et~al.}(1999){Gilliland}, {Nugent}, \&
  {Phillips}}]{1999ApJ...521...30G}
{Gilliland}, R.~L., {Nugent}, P.~E., \& {Phillips}, M.~M. 1999, \apj, 521, 30

\bibitem[{{Glazebrook} {et~al.}(2006){Glazebrook}, {Verma}, {Boyle}, {Oliver},
  {Mann}, \& {Monbleau}}]{2006AJ....131.2383G}
{Glazebrook}, K., {Verma}, A., {Boyle}, B., {et~al.} 2006, \aj, 131, 2383

\bibitem[{{Goobar}(2008)}]{2008ApJ...686L.103G}
{Goobar}, A. 2008, \apjl, 686, L103

\bibitem[{{Hamuy} {et~al.}(2002){Hamuy}, {Maza}, {Pinto}, {Phillips},
  {Suntzeff}, {Blum}, {Olsen}, {Pinfield}, {Ivanov}, {Augusteijn}, {Brillant},
  {Chadid}, {Cuby}, {Doublier}, {Hainaut}, {Le Floc'h}, {Lidman},
  {Petr-Gotzens}, {Pompei}, \& {Vanzi}}]{2002AJ....124..417H}
{Hamuy}, M., {Maza}, J., {Pinto}, P.~A., {et~al.} 2002, \aj, 124, 417

\bibitem[{{Hendry} {et~al.}(2005){Hendry}, {Smartt}, {Maund}, {Pastorello},
  {Zampieri}, {Benetti}, {Turatto}, {Cappellaro}, {Meikle}, {Kotak}, {Irwin},
  {Jonker}, {Vermaas}, {Peletier}, {van Woerden}, {Exter}, {Pollacco}, {Leon},
  {Verley}, {Benn}, \& {Pignata}}]{2005MNRAS.359..906H}
{Hendry}, M.~A., {Smartt}, S.~J., {Maund}, J.~R., {et~al.} 2005, \mnras, 359,
  906

\bibitem[{{Immler} {et~al.}(2007){Immler}, {Brown}, {Milne}, {Dessart},
  {Mazzali}, {Landsman}, {Gehrels}, {Petre}, {Burrows}, {Nousek}, {Chevalier},
  {Williams}, {Koss}, {Stockdale}, {Kelley}, {Weiler}, {Holland}, {Pian},
  {Roming}, {Pooley}, {Nomoto}, {Greiner}, {Campana}, \&
  {Soderberg}}]{2007ApJ...664..435I}
{Immler}, S., {Brown}, P.~J., {Milne}, P., {et~al.} 2007, \apj, 664, 435

\bibitem[{{Jha} {et~al.}(2006){Jha}, {Kirshner}, {Challis}, {Garnavich},
  {Matheson}, {Soderberg}, {Graves}, {Hicken}, {Alves}, {Arce}, {Balog},
  {Barmby}, {Barton}, {Berlind}, {Bragg}, {Brice{\~n}o}, {Brown}, {Buckley},
  {Caldwell}, {Calkins}, {Carter}, {Concannon}, {Donnelly}, {Eriksen},
  {Fabricant}, {Falco}, {Fiore}, {Garcia}, {G{\'o}mez}, {Grogin}, {Groner},
  {Groot}, {Haisch}, {Hartmann}, {Hergenrother}, {Holman}, {Huchra},
  {Jayawardhana}, {Jerius}, {Kannappan}, {Kim}, {Kleyna}, {Kochanek},
  {Koranyi}, {Krockenberger}, {Lada}, {Luhman}, {Luu}, {Macri}, {Mader},
  {Mahdavi}, {Marengo}, {Marsden}, {McLeod}, {McNamara}, {Megeath}, {Moraru},
  {Mossman}, {Muench}, {Mu{\~n}oz}, {Muzerolle}, {Naranjo}, {Nelson-Patel},
  {Pahre}, {Patten}, {Peters}, {Peters}, {Raymond}, {Rines}, {Schild},
  {Sobczak}, {Spahr}, {Stauffer}, {Stefanik}, {Szentgyorgyi}, {Tollestrup},
  {V{\"a}is{\"a}nen}, {Vikhlinin}, {Wang}, {Willner}, {Wolk}, {Zajac}, {Zhao},
  \& {Stanek}}]{2006AJ....131..527J}
{Jha}, S., {Kirshner}, R.~P., {Challis}, P., {et~al.} 2006, \aj, 131, 527

\bibitem[{{Johnson} \& {Crotts}(2006)}]{2006AJ....132..756J}
{Johnson}, B.~D. \& {Crotts}, A.~P.~S. 2006, \aj, 132, 756

\bibitem[{{Kankare} {et~al.}(2008){Kankare}, {Mattila}, {Ryder},
  {P{\'e}rez-Torres}, {Alberdi}, {Romero-Canizales}, {D{\'{\i}}az-Santos},
  {V{\"a}is{\"a}nen}, {Efstathiou}, {Alonso-Herrero}, {Colina}, \&
  {Kotilainen}}]{2008ApJ...689L..97K}
{Kankare}, E., {Mattila}, S., {Ryder}, S., {et~al.} 2008, \apjl, 689, L97

\bibitem[{{Kessler} {et~al.}(2010{\natexlab{a}}){Kessler}, {Bassett}, {Belov},
  {Bhatnagar}, {Campbell}, {Conley}, {Frieman}, {Glazov}, {Hlozek}, {Jha},
  {Kuhlmann}, {Kunz}, {Lampeitl}, {Mahabal}, {Newling}, {Nichol}, {Parkinson},
  {Sajeeth Philip}, {Poznanski}, {Richards}, {Rodney}, {Sako}, {Schneider},
  {Smith}, {Stritzinger}, \& {Varughese}}]{2010arXiv1008.1024K}
{Kessler}, R., {Bassett}, B., {Belov}, P., {et~al.} 2010{\natexlab{a}}, ArXiv
  e-prints

\bibitem[{{Kessler} {et~al.}(2009){Kessler}, {Becker}, {Cinabro}, {Vanderplas},
  {Frieman}, {Marriner}, {Davis}, {Dilday}, {Holtzman}, {Jha}, {Lampeitl},
  {Sako}, {Smith}, {Zheng}, {Nichol}, {Bassett}, {Bender}, {Depoy}, {Doi},
  {Elson}, {Filippenko}, {Foley}, {Garnavich}, {Hopp}, {Ihara}, {Ketzeback},
  {Kollatschny}, {Konishi}, {Marshall}, {McMillan}, {Miknaitis}, {Morokuma},
  {M{\"o}rtsell}, {Pan}, {Prieto}, {Richmond}, {Riess}, {Romani}, {Schneider},
  {Sollerman}, {Takanashi}, {Tokita}, {van der Heyden}, {Wheeler}, {Yasuda}, \&
  {York}}]{2009ApJS..185...32K}
{Kessler}, R., {Becker}, A.~C., {Cinabro}, D., {et~al.} 2009, \apjs, 185, 32

\bibitem[{{Kessler} {et~al.}(2010{\natexlab{b}}){Kessler}, {Cinabro},
  {Bassett}, {Dilday}, {Frieman}, {Garnavich}, {Jha}, {Marriner}, {Nichol},
  {Sako}, {Smith}, {Bernstein}, {Bizyaev}, {Goobar}, {Kuhlmann}, {Schneider},
  \& {Stritzinger}}]{2010ApJ...717...40K}
{Kessler}, R., {Cinabro}, D., {Bassett}, B., {et~al.} 2010{\natexlab{b}}, \apj,
  717, 40

\bibitem[{{Kim} {et~al.}(1996){Kim}, {Goobar}, \&
  {Perlmutter}}]{1996PASP..108..190K}
{Kim}, A., {Goobar}, A., \& {Perlmutter}, S. 1996, \pasp, 108, 190

\bibitem[{{Kinney} {et~al.}(1996){Kinney}, {Calzetti}, {Bohlin}, {McQuade},
  {Storchi-Bergmann}, \& {Schmitt}}]{1996ApJ...467...38K}
{Kinney}, A.~L., {Calzetti}, D., {Bohlin}, R.~C., {et~al.} 1996, \apj, 467, 38

\bibitem[{{Kuznetsova} \& {Connolly}(2007)}]{2007ApJ...659..530K}
{Kuznetsova}, N.~V. \& {Connolly}, B.~M. 2007, \apj, 659, 530

\bibitem[{{La Franca} {et~al.}(2004){La Franca}, {Gruppioni}, {Matute},
  {Pozzi}, {Lari}, {Mignoli}, {Zamorani}, {Alexander}, {Cocchia}, {Danese},
  {Franceschini}, {H{\'e}raudeau}, {Kotilainen}, {Linden-V{\o}rnle}, {Oliver},
  {Rowan-Robinson}, {Serjeant}, {Spinoglio}, \& {Verma}}]{2004AJ....127.3075L}
{La Franca}, F., {Gruppioni}, C., {Matute}, I., {et~al.} 2004, \aj, 127, 3075

\bibitem[{{LeFevre} {et~al.}(2003){LeFevre}, {Saisse}, {Mancini}, {Brau-Nogue},
  {Caputi}, {Castinel}, {D'Odorico}, {Garilli}, {Kissler-Patig}, {Lucuix},
  {Mancini}, {Pauget}, {Sciarretta}, {Scodeggio}, {Tresse}, \&
  {Vettolani}}]{2003SPIE.4841.1670L}
{LeFevre}, O., {Saisse}, M., {Mancini}, D., {et~al.} 2003, in Presented at the
  Society of Photo-Optical Instrumentation Engineers (SPIE) Conference, Vol.
  4841, Instrument Design and Performance for Optical/Infrared Ground-based
  Telescopes. Edited by Iye, Masanori; Moorwood, Alan F. M. Proceedings of the
  SPIE, Volume 4841, pp. 1670-1681 (2003)., ed. M.~{Iye} \& A.~F.~M.
  {Moorwood}, 1670--1681

\bibitem[{{Leonard} {et~al.}(2002){Leonard}, {Filippenko}, {Li}, {Matheson},
  {Kirshner}, {Chornock}, {Van Dyk}, {Berlind}, {Calkins}, {Challis},
  {Garnavich}, {Jha}, \& {Mahdavi}}]{2002AJ....124.2490L}
{Leonard}, D.~C., {Filippenko}, A.~V., {Li}, W., {et~al.} 2002, \aj, 124, 2490

\bibitem[{{Levan} {et~al.}(2005){Levan}, {Nugent}, {Fruchter}, {Burud},
  {Branch}, {Rhoads}, {Castro-Tirado}, {Gorosabel}, {Castro Cer{\'o}n},
  {Thorsett}, {Kouveliotou}, {Golenetskii}, {Fynbo}, {Garnavich}, {Holland},
  {Hjorth}, {M{\o}ller}, {Pian}, {Tanvir}, {Ulanov}, {Wijers}, \&
  {Woosley}}]{2005ApJ...624..880L}
{Levan}, A., {Nugent}, P., {Fruchter}, A., {et~al.} 2005, \apj, 624, 880

\bibitem[{{Li} {et~al.}(2011{\natexlab{a}}){Li}, {Chornock}, {Leaman},
  {Filippenko}, {Poznanski}, {Wang}, {Ganeshalingam}, \&
  {Mannucci}}]{2011MNRAS.tmp..317L}
{Li}, W., {Chornock}, R., {Leaman}, J., {et~al.} 2011{\natexlab{a}}, \mnras,
  317

\bibitem[{{Li} {et~al.}(2011{\natexlab{b}}){Li}, {Leaman}, {Chornock},
  {Filippenko}, {Poznanski}, {Ganeshalingam}, {Wang}, {Modjaz}, {Jha}, {Foley},
  \& {Smith}}]{2011MNRAS.tmp..413L}
{Li}, W., {Leaman}, J., {Chornock}, R., {et~al.} 2011{\natexlab{b}}, \mnras,
  413

\bibitem[{{Magnelli} {et~al.}(2009){Magnelli}, {Elbaz}, {Chary}, {Dickinson},
  {Le Borgne}, {Frayer}, \& {Willmer}}]{2009A&A...496...57M}
{Magnelli}, B., {Elbaz}, D., {Chary}, R.~R., {et~al.} 2009, \aap, 496, 57

\bibitem[{{Maoz} \& {Badenes}(2010)}]{2010MNRAS.407.1314M}
{Maoz}, D. \& {Badenes}, C. 2010, \mnras, 407, 1314

\bibitem[{{Mattila} {et~al.}(2007){Mattila}, {V{\"a}is{\"a}nen}, {Farrah},
  {Efstathiou}, {Meikle}, {Dahlen}, {Fransson}, {Lira}, {Lundqvist},
  {{\"O}stlin}, {Ryder}, \& {Sollerman}}]{2007ApJ...659L...9M}
{Mattila}, S., {V{\"a}is{\"a}nen}, P., {Farrah}, D., {et~al.} 2007, \apjl, 659,
  L9

\bibitem[{{Melinder} {et~al.}(2008){Melinder}, {Mattila}, {{\"O}stlin},
  {Menc{\'{\i}}a Trinchant}, \& {Fransson}}]{2008A&A...490..419M}
{Melinder}, J., {Mattila}, S., {{\"O}stlin}, G., {Menc{\'{\i}}a Trinchant}, L.,
  \& {Fransson}, C. 2008, \aap, 490, 419

\bibitem[{{Miknaitis} {et~al.}(2007){Miknaitis}, {Pignata}, {Rest},
  {Wood-Vasey}, {Blondin}, {Challis}, {Smith}, {Stubbs}, {Suntzeff}, {Foley},
  {Matheson}, {Tonry}, {Aguilera}, {Blackman}, {Becker}, {Clocchiatti},
  {Covarrubias}, {Davis}, {Filippenko}, {Garg}, {Garnavich}, {Hicken}, {Jha},
  {Krisciunas}, {Kirshner}, {Leibundgut}, {Li}, {Miceli}, {Narayan}, {Prieto},
  {Riess}, {Salvo}, {Schmidt}, {Sollerman}, {Spyromilio}, \&
  {Zenteno}}]{2007ApJ...666..674M}
{Miknaitis}, G., {Pignata}, G., {Rest}, A., {et~al.} 2007, \apj, 666, 674

\bibitem[{{Modjaz} {et~al.}(2009){Modjaz}, {Li}, {Butler}, {Chornock},
  {Perley}, {Blondin}, {Bloom}, {Filippenko}, {Kirshner}, {Kocevski},
  {Poznanski}, {Hicken}, {Foley}, {Stringfellow}, {Berlind}, {Barrado y
  Navascues}, {Blake}, {Bouy}, {Brown}, {Challis}, {Chen}, {de Vries},
  {Dufour}, {Falco}, {Friedman}, {Ganeshalingam}, {Garnavich}, {Holden},
  {Illingworth}, {Lee}, {Liebert}, {Marion}, {Olivier}, {Prochaska},
  {Silverman}, {Smith}, {Starr}, {Steele}, {Stockton}, {Williams}, \&
  {Wood-Vasey}}]{2009ApJ...702..226M}
{Modjaz}, M., {Li}, W., {Butler}, N., {et~al.} 2009, \apj, 702, 226

\bibitem[{{Morokuma} {et~al.}(2008){Morokuma}, {Doi}, {Yasuda}, {Akiyama},
  {Sekiguchi}, {Furusawa}, {Ueda}, {Totani}, {Oda}, {Nagao}, {Kashikawa},
  {Murayama}, {Ouchi}, {Watson}, {Richmond}, {Lidman}, {Perlmutter},
  {Spadafora}, {Aldering}, {Wang}, {Hook}, \& {Knop}}]{2008ApJ...676..163M}
{Morokuma}, T., {Doi}, M., {Yasuda}, N., {et~al.} 2008, \apj, 676, 163

\bibitem[{{Neill} {et~al.}(2006){Neill}, {Sullivan}, {Balam}, {Pritchet},
  {Howell}, {Perrett}, {Astier}, {Aubourg}, {Basa}, {Carlberg}, {Conley},
  {Fabbro}, {Fouchez}, {Guy}, {Hook}, {Pain}, {Palanque-Delabrouille},
  {Regnault}, {Rich}, {Taillet}, {Aldering}, {Antilogus}, {Arsenijevic},
  {Balland}, {Baumont}, {Bronder}, {Ellis}, {Filiol}, {Gon{\c c}alves},
  {Hardin}, {Kowalski}, {Lidman}, {Lusset}, {Mouchet}, {Mourao}, {Perlmutter},
  {Ripoche}, {Schlegel}, \& {Tao}}]{2006AJ....132.1126N}
{Neill}, J.~D., {Sullivan}, M., {Balam}, D., {et~al.} 2006, \aj, 132, 1126

\bibitem[{Nugent(2007)}]{2007NugentMISC}
Nugent, P. 2007, Peter Nugent's Spectral Templates,
  \url{http://supernova.lbl.gov/~nugent/nugent_templates.html}

\bibitem[{{Nugent} {et~al.}(2002){Nugent}, {Kim}, \&
  {Perlmutter}}]{2002PASP..114..803N}
{Nugent}, P., {Kim}, A., \& {Perlmutter}, S. 2002, \pasp, 114, 803

\bibitem[{{Pastorello} {et~al.}(2006){Pastorello}, {Sauer}, {Taubenberger},
  {Mazzali}, {Nomoto}, {Kawabata}, {Benetti}, {Elias-Rosa}, {Harutyunyan},
  {Navasardyan}, {Zampieri}, {Iijima}, {Botticella}, {di Rico}, {Del Principe},
  {Dolci}, {Gagliardi}, {Ragni}, \& {Valentini}}]{2006MNRAS.370.1752P}
{Pastorello}, A., {Sauer}, D., {Taubenberger}, S., {et~al.} 2006, \mnras, 370,
  1752

\bibitem[{{Pastorello} {et~al.}(2010){Pastorello}, {Smartt}, {Botticella},
  {Maguire}, {Fraser}, {Smith}, {Kotak}, {Magill}, {Valenti}, {Young},
  {Gezari}, {Bresolin}, {Kudritzki}, {Howell}, {Rest}, {Metcalfe}, {Mattila},
  {Kankare}, {Huang}, {Urata}, {Burgett}, {Chambers}, {Dombeck}, {Flewelling},
  {Grav}, {Heasley}, {Hodapp}, {Kaiser}, {Luppino}, {Lupton}, {Magnier},
  {Monet}, {Morgan}, {Onaka}, {Price}, {Rhoads}, {Siegmund}, {Stubbs},
  {Sweeney}, {Tonry}, {Wainscoat}, {Waterson}, {Waters}, \&
  {Wynn-Williams}}]{2010arXiv1008.2674P}
{Pastorello}, A., {Smartt}, S.~J., {Botticella}, M.~T., {et~al.} 2010, ArXiv
  e-prints

\bibitem[{{Perlmutter} {et~al.}(1997){Perlmutter}, {Gabi}, {Goldhaber},
  {Goobar}, {Groom}, {Hook}, {Kim}, {Kim}, {Lee}, {Pain}, {Pennypacker},
  {Small}, {Ellis}, {McMahon}, {Boyle}, {Bunclark}, {Carter}, {Irwin},
  {Glazebrook}, {Newberg}, {Filippenko}, {Matheson}, {Dopita}, {Couch}, \& {The
  Supernova Cosmology Project}}]{1997ApJ...483..565P}
{Perlmutter}, S., {Gabi}, S., {Goldhaber}, G., {et~al.} 1997, \apj, 483, 565

\bibitem[{{Phillips} {et~al.}(1999){Phillips}, {Lira}, {Suntzeff}, {Schommer},
  {Hamuy}, \& {Maza}}]{1999AJ....118.1766P}
{Phillips}, M.~M., {Lira}, P., {Suntzeff}, N.~B., {et~al.} 1999, \aj, 118, 1766

\bibitem[{{Poznanski} {et~al.}(2007{\natexlab{a}}){Poznanski}, {Maoz}, \&
  {Gal-Yam}}]{2007AJ....134.1285P}
{Poznanski}, D., {Maoz}, D., \& {Gal-Yam}, A. 2007{\natexlab{a}}, \aj, 134,
  1285

\bibitem[{{Poznanski} {et~al.}(2007{\natexlab{b}}){Poznanski}, {Maoz},
  {Yasuda}, {Foley}, {Doi}, {Filippenko}, {Fukugita}, {Gal-Yam}, {Jannuzi},
  {Morokuma}, {Oda}, {Schweiker}, {Sharon}, {Silverman}, \&
  {Totani}}]{2007MNRAS.382.1169P}
{Poznanski}, D., {Maoz}, D., {Yasuda}, N., {et~al.} 2007{\natexlab{b}}, \mnras,
  382, 1169

\bibitem[{{Quimby} {et~al.}(2007){Quimby}, {Wheeler}, {H{\"o}flich}, {Akerlof},
  {Brown}, \& {Rykoff}}]{2007ApJ...666.1093Q}
{Quimby}, R.~M., {Wheeler}, J.~C., {H{\"o}flich}, P., {et~al.} 2007, \apj, 666,
  1093

\bibitem[{{Rau} {et~al.}(2009){Rau}, {Kulkarni}, {Law}, {Bloom}, {Ciardi},
  {Djorgovski}, {Fox}, {Gal-Yam}, {Grillmair}, {Kasliwal}, {Nugent}, {Ofek},
  {Quimby}, {Reach}, {Shara}, {Bildsten}, {Cenko}, {Drake}, {Filippenko},
  {Helfand}, {Helou}, {Howell}, {Poznanski}, \&
  {Sullivan}}]{2009PASP..121.1334R}
{Rau}, A., {Kulkarni}, S.~R., {Law}, N.~M., {et~al.} 2009, \pasp, 121, 1334

\bibitem[{{Richardson} {et~al.}(2006){Richardson}, {Branch}, \&
  {Baron}}]{2006AJ....131.2233R}
{Richardson}, D., {Branch}, D., \& {Baron}, E. 2006, \aj, 131, 2233

\bibitem[{{Richardson} {et~al.}(2002){Richardson}, {Branch}, {Casebeer},
  {Millard}, {Thomas}, \& {Baron}}]{2002AJ....123..745R}
{Richardson}, D., {Branch}, D., {Casebeer}, D., {et~al.} 2002, \aj, 123, 745

\bibitem[{{Riello} \& {Patat}(2005)}]{2005MNRAS.362..671R}
{Riello}, M. \& {Patat}, F. 2005, \mnras, 362, 671

\bibitem[{{Riess} {et~al.}(2007){Riess}, {Strolger}, {Casertano}, {Ferguson},
  {Mobasher}, {Gold}, {Challis}, {Filippenko}, {Jha}, {Li}, {Tonry}, {Foley},
  {Kirshner}, {Dickinson}, {MacDonald}, {Eisenstein}, {Livio}, {Younger}, {Xu},
  {Dahl{\'e}n}, \& {Stern}}]{2007ApJ...659...98R}
{Riess}, A.~G., {Strolger}, L., {Casertano}, S., {et~al.} 2007, \apj, 659, 98

\bibitem[{{Rigopoulou} {et~al.}(2005){Rigopoulou}, {Vacca}, {Berta},
  {Franceschini}, \& {Aussel}}]{2005A&A...440...61R}
{Rigopoulou}, D., {Vacca}, W.~D., {Berta}, S., {Franceschini}, A., \& {Aussel},
  H. 2005, \aap, 440, 61

\bibitem[{{Rodney} \& {Tonry}(2009)}]{2009ApJ...707.1064R}
{Rodney}, S.~A. \& {Tonry}, J.~L. 2009, \apj, 707, 1064

\bibitem[{{Ruiter} {et~al.}(2009){Ruiter}, {Belczynski}, \&
  {Fryer}}]{2009ApJ...699.2026R}
{Ruiter}, A.~J., {Belczynski}, K., \& {Fryer}, C. 2009, \apj, 699, 2026

\bibitem[{{Sahu} {et~al.}(2006){Sahu}, {Anupama}, {Srividya}, \&
  {Muneer}}]{2006MNRAS.372.1315S}
{Sahu}, D.~K., {Anupama}, G.~C., {Srividya}, S., \& {Muneer}, S. 2006, \mnras,
  372, 1315

\bibitem[{{Sako} {et~al.}(2008){Sako}, {Bassett}, {Becker}, {Cinabro},
  {DeJongh}, {Depoy}, {Dilday}, {Doi}, {Frieman}, {Garnavich}, {Hogan},
  {Holtzman}, {Jha}, {Kessler}, {Konishi}, {Lampeitl}, {Marriner}, {Miknaitis},
  {Nichol}, {Prieto}, {Riess}, {Richmond}, {Romani}, {Schneider}, {Smith},
  {Subba Rao}, {Takanashi}, {Tokita}, {van der Heyden}, {Yasuda}, {Zheng},
  {Barentine}, {Brewington}, {Choi}, {Dembicky}, {Harnavek}, {Ihara}, {Im},
  {Ketzeback}, {Kleinman}, {Krzesi{\'n}ski}, {Long}, {Malanushenko},
  {Malanushenko}, {McMillan}, {Morokuma}, {Nitta}, {Pan}, {Saurage}, \&
  {Snedden}}]{2008AJ....135..348S}
{Sako}, M., {Bassett}, B., {Becker}, A., {et~al.} 2008, \aj, 135, 348

\bibitem[{{Sharon} {et~al.}(2010){Sharon}, {Gal-Yam}, {Maoz}, {Filippenko},
  {Foley}, {Silverman}, {Ebeling}, {Ma}, {Ofek}, {Kneib}, {Donahue}, {Ellis},
  {Freedman}, {Kirshner}, {Mulchaey}, {Sarajedini}, \&
  {Voit}}]{2010ApJ...718..876S}
{Sharon}, K., {Gal-Yam}, A., {Maoz}, D., {et~al.} 2010, \apj, 718, 876

\bibitem[{{Smartt}(2009)}]{2009ARA&A..47...63S}
{Smartt}, S.~J. 2009, \araa, 47, 63

\bibitem[{{Stern} {et~al.}(2004){Stern}, {van Dokkum}, {Nugent}, {Sand},
  {Ellis}, {Sullivan}, {Bloom}, {Frail}, {Kneib}, {Koopmans}, \&
  {Treu}}]{2004ApJ...612..690S}
{Stern}, D., {van Dokkum}, P.~G., {Nugent}, P., {et~al.} 2004, \apj, 612, 690

\bibitem[{{Sullivan} {et~al.}(2006){Sullivan}, {Le Borgne}, {Pritchet},
  {Hodsman}, {Neill}, {Howell}, {Carlberg}, {Astier}, {Aubourg}, {Balam},
  {Basa}, {Conley}, {Fabbro}, {Fouchez}, {Guy}, {Hook}, {Pain},
  {Palanque-Delabrouille}, {Perrett}, {Regnault}, {Rich}, {Taillet}, {Baumont},
  {Bronder}, {Ellis}, {Filiol}, {Lusset}, {Perlmutter}, {Ripoche}, \&
  {Tao}}]{2006ApJ...648..868S}
{Sullivan}, M., {Le Borgne}, D., {Pritchet}, C.~J., {et~al.} 2006, \apj, 648,
  868

\bibitem[{{Tominaga} {et~al.}(2011){Tominaga}, {Morokuma}, {Blinnikov},
  {Baklanov}, {Sorokina}, \& {Nomoto}}]{2011ApJS..193...20T}
{Tominaga}, N., {Morokuma}, T., {Blinnikov}, S.~I., {et~al.} 2011, \apjs, 193,
  20

\bibitem[{{Tonry} {et~al.}(2003){Tonry}, {Schmidt}, {Barris}, {Candia},
  {Challis}, {Clocchiatti}, {Coil}, {Filippenko}, {Garnavich}, {Hogan},
  {Holland}, {Jha}, {Kirshner}, {Krisciunas}, {Leibundgut}, {Li}, {Matheson},
  {Phillips}, {Riess}, {Schommer}, {Smith}, {Sollerman}, {Spyromilio},
  {Stubbs}, \& {Suntzeff}}]{2003ApJ...594....1T}
{Tonry}, J.~L., {Schmidt}, B.~P., {Barris}, B., {et~al.} 2003, \apj, 594, 1

\bibitem[{{Totani} {et~al.}(2008){Totani}, {Morokuma}, {Oda}, {Doi}, \&
  {Yasuda}}]{2008PASJ...60.1327T}
{Totani}, T., {Morokuma}, T., {Oda}, T., {Doi}, M., \& {Yasuda}, N. 2008,
  \pasj, 60, 1327

\bibitem[{{Vanzella} {et~al.}(2002){Vanzella}, {Cristiani}, {Arnouts},
  {Dennefeld}, {Fontana}, {Grazian}, {Nonino}, {Petitjean}, \&
  {Saracco}}]{2002A&A...396..847V}
{Vanzella}, E., {Cristiani}, S., {Arnouts}, S., {et~al.} 2002, \aap, 396, 847

\bibitem[{{Wood-Vasey} {et~al.}(2007){Wood-Vasey}, {Miknaitis}, {Stubbs},
  {Jha}, {Riess}, {Garnavich}, {Kirshner}, {Aguilera}, {Becker}, {Blackman},
  {Blondin}, {Challis}, {Clocchiatti}, {Conley}, {Covarrubias}, {Davis},
  {Filippenko}, {Foley}, {Garg}, {Hicken}, {Krisciunas}, {Leibundgut}, {Li},
  {Matheson}, {Miceli}, {Narayan}, {Pignata}, {Prieto}, {Rest}, {Salvo},
  {Schmidt}, {Smith}, {Sollerman}, {Spyromilio}, {Tonry}, {Suntzeff}, \&
  {Zenteno}}]{2007ApJ...666..694W}
{Wood-Vasey}, W.~M., {Miknaitis}, G., {Stubbs}, C.~W., {et~al.} 2007, \apj,
  666, 694

\bibitem[{{Woosley} {et~al.}(2007){Woosley}, {Blinnikov}, \&
  {Heger}}]{2007Natur.450..390W}
{Woosley}, S.~E., {Blinnikov}, S., \& {Heger}, A. 2007, \nat, 450, 390

\bibitem[{{Young} {et~al.}(2008){Young}, {Smartt}, {Mattila}, {Tanvir},
  {Bersier}, {Chambers}, {Kaiser}, \& {Tonry}}]{2008A&A...489..359Y}
{Young}, D.~R., {Smartt}, S.~J., {Mattila}, S., {et~al.} 2008, \aap, 489, 359

\end{thebibliography}
\Online
\appendix
\section{Observational data for the 31 transient objects}
This appendix contains a table with extended information for the 31
transient objects detected in the SVISS ELAIS-S1 field. The astrometry
is accurate to within 0.4\arcsec (for details on the astrometry see
Menc\'ia-Trinchant et al., in prep.). The photometrical errors are
obtained as described in Section~\ref{sec:phot}. For non-detections the
magnitude is given as a lower limit and the limit is also given as the
error. The limiting magnitudes given in the table are $1\sigma$ limiting
magnitudes derived from the photometric accuracy investigation.  The
host galaxy redshifts given in the table are photometric redshifts
derived from the $UBVRI$ galaxy photometry. Objects that were assumed to be
hostless are given a redshift of N/A in the table.

\onllongtab{1}{
\begin{longtable}{l c c c c c c c c} 
\caption{Extended information on the transient objects}\\     \hline
\hline  
Id & RA (deg.) & Dec (deg.) & Host $z_p$ & Epoch & $m_R$ & $\delta m_R$ & 
$m_I$ & $\delta m_I$\\ 
\hline
\endfirsthead
\caption{Continued.} \\
\hline
Id & RA (deg.) & Dec (deg.) & Host $z_p$ & Epoch & $m_R$ & $\delta m_R$ & 
$m_I$ & $\delta m_I$\\ 
\hline
\endhead
\hline
\endfoot
\hline
\endlastfoot
                                                                 
SVISS-SN43 & 7.9102451 & -44.499318 &  0.46  & 1  &  24.881 &  0.0811 &  24.741 &  0.1405\\   
           &           &            &        & 2  &  25.416 &  0.1149 &  24.797 &  0.1429\\   
           &           &            &        & 3  &  25.615 &  0.1458 &   25.47 &  0.3558\\   
           &           &            &        & 4  &  26.427 &   0.222 &  25.518 &  0.3195\\   
           &           &            &        & 5  &  26.649 &  0.5808 &  26.927 &  0.5354\\   
           &           &            &        & 6  &   27.29 &   0.421 & $>$26.894 & 26.8935\\   
           &           &            &        & 7  &   27.36 &  0.4706 & $>$26.719 & 26.7194\\   
SVISS-SN161& 7.9549322 & -44.484425 &  0.54  & 1  & $>$28.134 & 28.1342 & $>$28.024 & 28.0236\\
           &           &            &        & 2  &  $>$27.94 & 27.9396 & $>$27.746 & 27.7464\\   
           &           &            &        & 3  &  25.115 &  0.0595 &  25.503 &  0.1809\\   
           &           &            &        & 4  &  24.207 &    0.02 &  23.878 &  0.0362\\   
           &           &            &        & 5  &  25.925 &  0.1307 &  25.383 &  0.1815\\   
           &           &            &        & 6  &  27.213 &  0.3706 &  26.709 &  0.4135\\   
           &           &            &        & 7  & $>$28.084 & 28.0837 & $>$27.722 & 27.7219\\   
SVISS-SN115& 8.1434494 & -44.39665  &  0.40  & 1  &   $>$27.3 &    27.3 & $>$27.211 & 27.2112\\
           &           &            &        & 2  &   22.07 &    0.02 &  22.235 &    0.02\\   
           &           &            &        & 3  &  22.884 &    0.02 &  22.262 &    0.02\\   
           &           &            &        & 4  &  24.644 &  0.0676 &  23.416 &    0.02\\   
           &           &            &        & 5  &  24.965 &  0.0839 &  23.982 &  0.0716\\   
           &           &            &        & 6  &  25.221 &  0.1118 &  24.703 &  0.1808\\   
           &           &            &        & 7  &   25.36 &  0.1697 &  25.138 &  0.3596\\   
SVISS-SN116& 8.088403  & -44.47483  &  0.57  & 1  & $>$28.657 & 28.6567 & $>$27.598 & 27.5984\\
           &           &            &        & 2  &  22.895 &    0.02 &  22.941 &    0.02\\   
           &           &            &        & 3  &  24.622 &  0.1311 &   23.45 &    0.02\\   
           &           &            &        & 4  &  26.516 &  0.2035 &  25.441 &  0.1349\\   
           &           &            &        & 5  &  26.335 &  0.3015 &  24.983 &  0.1291\\   
           &           &            &        & 6  &  27.643 &  0.5584 &  25.911 &  0.3007\\   
           &           &            &        & 7  &  26.733 &  0.3231 &  26.434 &  0.3512\\   
SVISS-SN309& 8.0001281 & -44.439905 &  0.67  & 1  & $>$28.657 & 28.6567 & $>$27.598 & 27.5984\\
           &           &            &        & 2  & $>$28.508 & 28.5081 & $>$27.635 &  27.635\\   
           &           &            &        & 3  & $>$28.022 & 28.0224 & $>$27.577 & 27.5774\\   
           &           &            &        & 4  &  24.236 &    0.02 &  23.902 &  0.0532\\   
           &           &            &        & 5  &  23.599 &    0.02 &  23.751 &  0.0761\\   
           &           &            &        & 6  &  26.126 &  0.1866 &  25.278 &  0.1809\\   
           &           &            &        & 7  &  27.559 &  0.5768 &   25.75 &  0.2316\\   
SVISS-SN402& 8.0848318 & -44.405671 &  0.34  & 1  &   $>$27.3 &    27.3 & $>$27.211 & 27.2112\\
           &           &            &        & 2  & $>$26.776 & 26.7764 & $>$27.543 & 27.5435\\   
           &           &            &        & 3  & $>$28.096 & 28.0961 & $>$26.972 & 26.9718\\   
           &           &            &        & 4  & $>$27.703 & 27.7034 & $>$27.298 & 27.2979\\   
           &           &            &        & 5  & $>$27.221 & 27.2206 & $>$25.897 & 25.8973\\   
           &           &            &        & 6  &  24.173 &    0.02 &  23.362 &    0.02\\   
           &           &            &        & 7  &  23.747 &    0.02 &  23.063 &    0.02\\   
SVISS-SN135& 7.8847856 & -44.377629 &  0.90  & 1  & $>$27.253 & 27.2533 & $>$27.877 & 27.8767\\
           &           &            &        & 2  & $>$28.186 & 28.1863 & $>$27.596 &  27.596\\   
           &           &            &        & 3  &  25.344 &  0.0522 &  24.569 &  0.0683\\   
           &           &            &        & 4  &  27.161 &  0.3049 &  25.469 &  0.1887\\   
           &           &            &        & 5  &  27.815 &   0.612 &  26.092 &  0.3977\\   
           &           &            &        & 6  & $>$27.173 & 27.1731 &  26.736 &  0.7313\\   
           &           &            &        & 7  &  28.259 &  0.5435 &  27.158 &  0.5231\\   
SVISS-SN14 & 8.0329574 & -44.514378 &  N/A   & 1  &  26.187 &  0.1873 &  25.843 &  0.1986\\
           &           &            &        & 2  &  26.291 &  0.2177 &  26.025 &  0.2372\\   
           &           &            &        & 3  &  27.031 &  0.3204 &  26.979 &  0.4307\\   
           &           &            &        & 4  &  27.948 &  0.6347 &  27.462 &  0.5716\\   
           &           &            &        & 5  &  27.766 &  0.5392 &  27.096 &  0.6211\\   
           &           &            &        & 6  &   28.03 &  0.5945 &   27.42 &  0.5754\\   
           &           &            &        & 7  & $>$28.084 & 28.0837 &  26.636 &  0.3541\\   
SVISS-SN51 & 7.9625753 & -44.508399 &  0.53  & 1  &  26.042 &  0.2719 &  24.929 &  0.1633\\
           &           &            &        & 2  &  26.663 &  0.3074 &  25.147 &  0.1941\\   
           &           &            &        & 3  &  27.145 &  0.4142 &   25.47 &  0.3558\\   
           &           &            &        & 4  &    26.9 &   0.288 &  25.967 &  0.4471\\   
           &           &            &        & 5  &   27.25 &  0.4775 &  25.418 &    0.14\\   
           &           &            &        & 6  &  $>$28.19 &   28.19 &  25.954 &  0.3464\\   
           &           &            &        & 7  & $>$28.022 & 28.0217 &  26.682 &  0.5789\\   
SVISS-SN54 & 7.9628816 & -44.443032 &  0.77  & 1  &  25.467 &    0.16 &  24.977 &  0.1696\\
           &           &            &        & 2  &  26.722 &  0.3195 &  25.563 &  0.2678\\   
           &           &            &        & 3  & $>$27.967 &  27.967 &  26.066 &  0.6209\\   
           &           &            &        & 4  & $>$28.386 & 28.3863 &  26.671 &  0.6957\\   
           &           &            &        & 5  & $>$28.275 & 28.2753 & $>$28.148 & 28.1477\\   
           &           &            &        & 6  & $>$27.008 &  27.008 & $>$26.489 &  26.489\\   
           &           &            &        & 7  & $>$28.022 & 28.0217 & $>$26.719 & 26.7194\\   
SVISS-SN261& 8.0816154 & -44.476649 &  N/A   & 1  & $>$28.134 & 28.1342 & $>$28.024 & 28.0236\\
           &           &            &        & 2  &  $>$27.94 & 27.9396 & $>$27.746 & 27.7464\\   
           &           &            &        & 3  & $>$28.285 & 28.2846 & $>$27.675 &  27.675\\   
           &           &            &        & 4  & $>$28.171 & 28.1708 & $>$27.643 & 27.6432\\   
           &           &            &        & 5  & $>$28.204 & 28.2036 & $>$27.305 & 27.3051\\   
           &           &            &        & 6  &  24.892 &  0.0416 &    23.8 &  0.0426\\   
           &           &            &        & 7  &  26.361 &  0.2134 &  24.922 &  0.1047\\   
SVISS-SN55 & 7.8047583 & -44.499259 &  0.59  & 1  & $>$27.918 & 27.9184 & $>$27.667 & 27.6672\\
           &           &            &        & 2  &  25.776 &  0.1298 &  25.566 &  0.2291\\   
           &           &            &        & 3  &  25.878 &  0.1712 &  24.981 &  0.1349\\   
           &           &            &        & 4  & $>$27.418 & 27.4183 &  25.545 &  0.3008\\   
           &           &            &        & 5  & $>$28.065 & 28.0654 &  25.115 &  0.2053\\   
           &           &            &        & 6  & $>$28.133 & 28.1333 &  26.083 &  0.3359\\   
           &           &            &        & 7  & $>$27.753 & 27.7532 &  26.291 &  0.4234\\   
SVISS-SN31 & 8.0856037 & -44.462252 &  N/A   & 1  &  24.303 &  0.0459 &  23.958 &  0.0535\\
           &           &            &        & 2  &  25.016 &  0.0526 &  24.313 &  0.0634\\   
           &           &            &        & 3  &  25.671 &  0.1325 &  24.756 &  0.1057\\   
           &           &            &        & 4  &  26.771 &   0.248 &  25.689 &   0.151\\   
           &           &            &        & 5  &  27.031 &  0.5771 &  25.592 &  0.2423\\   
           &           &            &        & 6  &  27.535 &  0.5243 &   26.24 &  0.3787\\   
           &           &            &        & 7  & $>$27.966 & 27.9659 &  26.308 &   0.327\\   
SVISS-SN56 & 8.0617464 & -44.438378 &  0.51  & 1  & $>$26.694 & 26.694 &  $>$26.359 &  26.359\\
           &           &            &        & 2  & $>$26.776 & 26.7764 &  $>$26.333 &  26.333\\   
           &           &            &        & 3  & $>$28.096 & 28.0961 &  $>$25.939 & 25.939\\   
           &           &            &        & 4  &  25.372 &  0.1405 &  24.312 &  0.1262\\   
           &           &            &        & 5  &  25.343 &  0.1372 &  23.897 &  0.0672\\   
           &           &            &        & 6  &  27.093 &  0.5583 &  24.884 &  0.2127\\   
           &           &            &        & 7  & $>$27.322 & 27.3217 & $>$26.176 & 26.1764\\   
SVISS-SN357& 7.9953784 & -44.396938 &  1.36  & 1  &   $>$27.3 &    27.3 & $>$27.211 & 27.2112\\
           &           &            &        & 2  & $>$26.776 & 26.7764 & $>$27.543 & 27.5435\\   
           &           &            &        & 3  & $>$28.096 & 28.0961 & $>$26.972 & 26.9718\\   
           &           &            &        & 4  & $>$27.703 & 27.7034 & $>$27.298 & 27.2979\\   
           &           &            &        & 5  &  24.743 &  0.0709 &  25.001 &  0.3677\\   
           &           &            &        & 6  &  24.922 &  0.0813 &  24.624 &  0.1682\\   
           &           &            &        & 7  &  25.344 &  0.1673 &  24.729 &  0.2138\\   
SVISS-SN24 & 7.9023817 & -44.374341 &  0.77  & 1  &  25.184 &   0.138 &  23.816 &  0.0755\\
           &           &            &        & 2  &  26.489 &  0.2786 &  24.369 &  0.1448\\   
           &           &            &        & 3  & $>$27.226 & 27.2258 &  26.055 &  0.3293\\   
           &           &            &        & 4  & $>$27.836 & 27.8362 &  26.635 &  0.5763\\   
           &           &            &        & 5  & $>$27.153 & 27.1535 & $>$26.362 & 26.3618\\   
           &           &            &        & 6  & $>$27.451 &  27.451 & $>$26.641 & 26.6409\\   
           &           &            &        & 7  & $>$27.503 & 27.5032 & $>$26.398 & 26.3982\\   

SVISS-T99  &  8.0147036 & -44.388741 &  1.50 & 1  &  26.438 &   0.4123 &  24.549&   0.1501\\   
           &           &            &        & 2  &  25.545 &   0.3216 &  24.117&   0.1073\\   
           &           &            &        & 3  & $>$28.096 &  28.0961 &  24.315&   0.1309\\   
           &           &            &        & 4  &   26.49 &    0.329 &  24.091&   0.1019\\   
           &           &            &        & 5  &  25.674 &    0.216 &  24.046&    0.077\\   
           &           &            &        & 6  &  25.861 &   0.2149 &  24.139&   0.1084\\   
           &           &            &        & 7  &  24.644 &   0.1045 &  23.611&   0.1222\\   
SVISS-T104 &  7.9834014 & -44.414223 &  1.52 & 1  &  24.255 &    0.053 &  23.965&   0.0922\\
           &           &            &        & 2  &  24.567 &   0.0901 &   24.03&   0.0987\\   
           &           &            &        & 3  &  23.751 &     0.02 &  24.033&   0.0971\\   
           &           &            &        & 4  &  25.153 &   0.1144 &  24.243&   0.1186\\   
           &           &            &        & 5  &  26.532 &   0.5599 &  25.183&    0.468\\   
           &           &            &        & 6  &   27.25 &   0.6157 &  $>$26.62&  26.6198\\   
           &           &            &        & 7  & $>$27.322 &  27.3217 & $>$26.176&  26.1764\\   
SVISS-T167 &  8.0149816 & -44.361213 &  1.95 & 1  &   $>$27.3 &     27.3 & $>$27.211&  27.2112\\
           &           &            &        & 2  & $>$26.776 &  26.7764 &  25.764&   0.2732\\   
           &           &            &        & 3  &   26.07 &   0.2859 &  25.337&   0.2675\\   
           &           &            &        & 4  &  25.587 &   0.1695 &  24.991&   0.2025\\   
           &           &            &        & 5  &  25.988 &   0.3184 &   25.02&   0.3775\\   
           &           &            &        & 6  &  25.193 &   0.1084 &  24.783&   0.1944\\   
           &           &            &        & 7  &   25.81 &   0.2537 &  25.215&   0.3938\\   
SVISS-T182 &  8.1151531 & -44.431759 &  0.52 & 1  &   $>$27.3 &     27.3 & $>$27.211&  27.2112\\
           &           &            &        & 2  & $>$26.776 &  26.7764 & $>$27.543&  27.5435\\   
           &           &            &        & 3  &    24.4 &   0.2141 &  24.431&   0.1453\\   
           &           &            &        & 4  &  23.198 &     0.02 &  23.078&     0.02\\   
           &           &            &        & 5  &  23.451 &     0.02 &   22.84&     0.02\\   
           &           &            &        & 6  &  23.332 &     0.02 &  23.095&     0.02\\   
           &           &            &        & 7  &   23.92 &     0.02 &  24.134&   0.1091\\   
SVISS-T23  &  7.8220527 & -44.379293 &  0.65 & 1  &  25.619 &    0.254 &  25.926&   0.5165\\
           &           &            &        & 2  &  25.317 &   0.1843 &  25.601&   0.2831\\   
           &           &            &        & 3  &   25.48 &   0.2386 &  25.233&   0.2195\\   
           &           &            &        & 4  &  25.923 &   0.2426 &  26.226&   0.4586\\   
           &           &            &        & 5  &  25.259 &   0.2237 & $>$26.362&  26.3618\\   
           &           &            &        & 6  &  26.155 &   0.3105 &  26.391&    0.446\\   
           &           &            &        & 7  &   26.25 &   0.3229 & $>$26.398&  26.3982\\   
SVISS-T26  &  7.9150698 & -44.334945 &  1.23 & 1  & $>$26.887 &   26.887 &  25.088&    0.234\\
           &           &            &        & 2  &  24.828 &   0.1365 &  24.297&   0.1378\\   
           &           &            &        & 3  &  24.365 &   0.1184 &  23.805&   0.0795\\   
           &           &            &        & 4  &   24.65 &   0.1249 &   24.02&   0.0846\\   
           &           &            &        & 5  &  24.114 &     0.02 &  23.852&   0.0797\\   
           &           &            &        & 6  &  24.643 &    0.111 &  23.905&   0.3598\\   
           &           &            &        & 7  &  24.967 &   0.1364 &  24.367&   0.1622\\   
SVISS-T68  &  7.8542558 & -44.400151 &  0.34 & 1  & $>$27.253 &  27.2533 & $>$27.877&  27.8767\\
           &           &            &        & 2  &   26.33 &    0.193 &  25.546&   0.1742\\   
           &           &            &        & 3  &  26.515 &   0.3176 &  25.981&   0.2312\\   
           &           &            &        & 4  &  26.989 &   0.2703 &  25.697&   0.2381\\   
           &           &            &        & 5  &  26.196 &   0.1796 &  25.744&     0.29\\   
           &           &            &        & 6  &  26.926 &   0.8029 & $>$26.911&  26.9113\\   
           &           &            &        & 7  & $>$28.472 &  28.4721 &  25.381&    0.159\\   
SVISS-T306 &  7.9255526 & -44.517406 &   0.4 & 1  &  23.315 &     0.02 &   23.44&     0.02\\
           &           &            &        & 2  &  23.438 &     0.02 &  23.718&   0.0461\\   
           &           &            &        & 3  &  23.054 &     0.02 &  23.312&     0.02\\   
           &           &            &        & 4  &  23.795 &     0.02 &  24.143&   0.0783\\   
           &           &            &        & 5  &  23.294 &     0.02 &  23.281&     0.02\\   
           &           &            &        & 6  &  24.042 &     0.02 &  24.809&    0.153\\   
           &           &            &        & 7  &  25.272 &   0.0876 & $>$26.719&  26.7194\\   
SVISS-T39  &  7.9069841 & -44.529026 &   1.8 & 1  &  22.849 &     0.02 &  22.691&     0.02\\
           &           &            &        & 2  &  23.087 &     0.02 &  22.806&     0.02\\   
           &           &            &        & 3  &  23.366 &     0.02 &  23.124&     0.02\\   
           &           &            &        & 4  &  23.645 &     0.02 &  23.418&     0.02\\   
           &           &            &        & 5  &  24.422 &    0.082 &  23.565&   0.0596\\   
           &           &            &        & 6  &  24.115 &     0.02 &  23.516&   0.1144\\   
           &           &            &        & 7  &  22.971 &     0.02 &  22.845&     0.02\\   
SVISS-T50  &  7.9607529 & -44.511618 &   0.7 & 1  &  25.762 &   0.2132 &  25.272&   0.2126\\
           &           &            &        & 2  &  26.444 &   0.2649 &  25.635&   0.2819\\   
           &           &            &        & 3  &  27.018 &   0.3854 & $>$26.602&  26.6022\\   
           &           &            &        & 4  &  27.628 &   0.4064 & $>$26.764&   26.764\\   
           &           &            &        & 5  & $>$28.275 &  28.2753 & $>$28.148&  28.1477\\   
           &           &            &        & 6  &  $>$28.19 &    28.19 & $>$26.894&  26.8935\\   
           &           &            &        & 7  &  27.538 &   0.5227 &  26.654&   0.5722\\   
SVISS-T6   &  7.870627 & -44.513955  &  1.64 & 1  &  24.408 &   0.1208 &  25.075&   0.2029\\
           &           &            &        & 2  &  23.841 &     0.02 &   24.69&   0.1199\\   
           &           &            &        & 3  &  25.261 &   0.1689 &  25.071&   0.2404\\   
           &           &            &        & 4  &  23.709 &     0.02 &  24.144&   0.0849\\   
           &           &            &        & 5  &  24.466 &   0.0946 & $>$26.476&  26.4764\\   
           &           &            &        & 6  & $>$27.004 &  27.0039 &  24.518&   0.1963\\   
           &           &            &        & 7  &  24.187 &     0.02 &  24.602&   0.2067\\   
SVISS-T36  &  7.8742713 & -44.409262 &  N/A  & 1  &  23.593 &     0.02 &  22.525&     0.02\\
           &           &            &        & 2  &   22.87 &     0.02 &   22.35&     0.02\\   
           &           &            &        & 3  &  23.355 &     0.02 &  22.776&     0.02\\   
           &           &            &        & 4  &  21.948 &     0.02 &  21.894&     0.02\\   
           &           &            &        & 5  &  23.726 &     0.02 &  21.124&     0.02\\   
           &           &            &        & 6  &  22.609 &     0.02 &  22.253&     0.02\\   
           &           &            &        & 7  &  23.133 &     0.02 &  22.636&     0.02\\   
SVISS-T59  &  7.7726376 & -44.492906 &  0.64 & 1  & $>$27.364 &  27.3641 & $>$26.725&  26.7249\\
           &           &            &        & 2  &  27.552 &   0.5635 &  24.851&   0.1431\\   
           &           &            &        & 3  &  25.557 &   0.2283 &  24.719&   0.1742\\   
           &           &            &        & 4  &   $>$27.3 &  27.2997 &  24.528&   0.1315\\   
           &           &            &        & 5  &  25.371 &    0.193 &  24.148&   0.1936\\   
           &           &            &        & 6  &  24.868 &   0.1861 &  24.741&   0.2644\\   
           &           &            &        & 7  &  26.298 &   0.3412 &  24.968&    0.277\\   
SVISS-T106 &  7.9025451 & -44.494797 &  0.09 & 1  & $>$27.364 &  27.3641 & $>$26.725&  26.7249\\
           &           &            &        & 2  & $>$27.587 &  27.5866 & $>$26.912&   26.912\\   
           &           &            &        & 3  & $>$27.161 &  27.1611 & $>$26.657&   26.657\\   
           &           &            &        & 4  &  24.564 &   0.0862 &  23.261&     0.02\\   
           &           &            &        & 5  &  23.795 &     0.02 &  23.188&     0.02\\   
           &           &            &        & 6  &  22.995 &     0.02 &  22.372&     0.02\\   
           &           &            &        & 7  &  23.411 &     0.02 &  22.305&     0.02\\   
SVISS-T138 &  7.8257044 & -44.441017 &  N/A  & 1  & $>$27.364 &  27.3641 & $>$26.725&  26.7249\\
           &           &            &        & 2  & $>$27.587 &  27.5866 & $>$26.912&   26.912\\   
           &           &            &        & 3  & $>$27.161 &  27.1611 & 126.657&   26.657\\   
           &           &            &        & 4  &  22.897 &     0.02 &  21.947&     0.02\\   
           &           &            &        & 5  &  22.829 &     0.02 &  21.746&     0.02\\   
           &           &            &        & 6  &  22.467 &     0.02 &  21.475&     0.02\\   
           &           &            &        & 7  &  22.393 &     0.02 &  21.575&     0.02\\   
\end{longtable} 
}                                                    
\end{document}